\documentclass[preprint,12pt,3p]{elsarticle}

\usepackage{moreverb}
\usepackage{mathptmx}      
\usepackage[english]{babel} 
\usepackage{graphicx,xspace,float}
\usepackage{listings}
\usepackage[
 colorlinks,
 urlcolor = NavyBlue,
 citecolor = NavyBlue,
 linkcolor = NavyBlue,
]{hyperref}
\usepackage{tabularx}
\usepackage{xcolor}
\usepackage{caption}
\usepackage{subcaption}
\usepackage{xspace}
\usepackage{wrapfig}
\usepackage{amsmath}

\lstdefinelanguage{ABS}{keywords=
{assert, switch, when, newclass,newinterface,update,foreach,null,this,thisDC,dyndelta,new,data,type,def,case,case2,of,cog,class,interface,extends,implements,if,else,await,get,local,Fut,return,skip,while,module,duration,duration2,now,import, export, uses, from,
  suspend,delta,adds,modifies,removes,original,then,productline,features ,corefeatures,optionalfeatures,after,when,product,hasAttribute,hasMethod,root,extension,group,allof,oneof,require,exclude,original,ifin,ifout,opt,spawn,newgroup,acquire,in,except,joins,leaves,as,subtypeOf}, 
sensitive=true, comment=[l]{//}, morecomment=[s]{/*}{*/}, morestring=[b]"}

\definecolor{codegreen}{rgb}{0,0.6,0}
\definecolor{codegray}{rgb}{0.5,0.5,0.5}
\definecolor{codepurple}{rgb}{0.58,0,0.82}
\definecolor{backcolour}{rgb}{0.95,0.95,0.92}

\newcommand{\Abs}[1]{\lstinline[language=ABS,columns=fullflexible,mathescape=true,inputencoding=latin1,extendedchars,keywordstyle=\bf\sffamily,basicstyle=\sffamily]!#1!}

\newenvironment{absinline}[1][]{%
\lstset{language=ABS,
		 showstringspaces=false,
		 columns=fullflexible,mathescape=true,%
         keywordstyle=\bf\sffamily,commentstyle=\sl\sffamily,%
         basicstyle=\normalsize\sffamily,inputencoding=latin1, 
         extendedchars,xleftmargin=2em,#1}}
{}

\lstdefinestyle{absstyle}{
language=ABS,columns=fullflexible,
 		   mathescape=true,%
 		   showstringspaces=false,%
keywordstyle=\bf\sffamily,
numbers=none,
commentstyle=\sl\sffamily,%
basicstyle=\footnotesize\sffamily,
inputencoding=latin1, 
extendedchars,xleftmargin=2pt
}

\lstnewenvironment{absexample}{
\noindent 
\lstset{style=absstyle,
basicstyle=\footnotesize\sffamily,
keywordstyle=\bfseries\color{codepurple},
framerule=1pt,
backgroundcolor=\color{backcolour},
rulecolor=\color{codepurple!80!black},
frame=tblr,
xleftmargin=4pt,
xrightmargin=6pt,
}}
{}

\lstnewenvironment{absmodelcode}{
\noindent 
\lstset{style=absstyle,
basicstyle=\scriptsize\sffamily,
keywordstyle=\bfseries\color{codepurple},
commentstyle=\color{teal},%
framerule=1pt,
backgroundcolor=\color{backcolour},
rulecolor=\color{black},
frame=tblr,
xleftmargin=18pt,
xrightmargin=1pt,
numbers=left,
framexleftmargin=2em,
stepnumber=1,
escapechar=|,
breaklines=true,
}}
{}

\lstnewenvironment{absexamplen}[1][]{
\lstset{style=absstyle,
basicstyle=\footnotesize\sffamily,
keywordstyle=\bfseries\color{violet},
 framerule=1pt,tabsize=2,
backgroundcolor=\color{backcolour},
rulecolor=\color{black},
frame=tblr,
captionpos=b,
xleftmargin=4pt,#1
}}
{}

\lstnewenvironment{absexamplesm}{
\noindent 
\lstset{style=absstyle,
basicstyle=\scriptsize\sffamily,
keywordstyle=\bfseries\color{codepurple},
framerule=1pt,
backgroundcolor=\color{backcolour},
rulecolor=\color{codepurple!80!black},
frame=tblr,
xleftmargin=4pt,
xrightmargin=6pt,
}}
{}

\definecolor{bluegray}{rgb}{0.4, 0.6, 0.8}
\definecolor{calpolypomonagreen}{rgb}{0.12, 0.3, 0.17}
\definecolor{applegreen}{rgb}{0.55, 0.71, 0.0}
\definecolor{goldenbrown}{rgb}{0.6, 0.4, 0.08}

\newcommand{\ABS}{ABS\xspace}
\newcommand{\rtabs}{Real-Time ABS\xspace}

\newcommand{\etal}{\textit{et al.}\@\xspace}
\newcommand\x{}

\newcolumntype{q}{>{\hsize=.05\hsize}X}
\newcolumntype{w}{>{\hsize=.19\hsize}X}
\newcolumntype{e}{>{\hsize=.36\hsize}X}
\newcolumntype{t}{>{\hsize=.36\hsize}X}

\newcolumntype{b}{>{\hsize=.13\hsize}X}
\newcolumntype{m}{>{\hsize=.45\hsize}X}
\newcolumntype{g}{>{\hsize=.4\hsize}X}

\newcommand{\aaanoop}[1]{}

\date{Preprint, May 2023}

\begin{document}

\title{Predicting Resource Consumption of Kubernetes Container Systems
  using Resource Models\tnoteref{tfn1}}

\author[1]{Gianluca Turin\corref{cor}}\ead{gianlutu@uio.no}
\author[2]{Andrea Borgarelli\fnref{fn2}}\ead{aborgare@mpi-sws.org}
\author[2]{Simone Donetti}\ead{simone.donetti@unito.it}

\author[2]{Ferruccio Damiani}\ead{ferruccio.damiani@unito.it}
\author[1]{Einar~Broch~Johnsen\corref{cor}}\ead{einarj@uio.no}
\author[1]{S. Lizeth Tapia Tarifa\corref{cor}}\ead{sltarifa@ifi.uio.no}
\cortext[cor]{Corresponding author}
\fntext[fn2]{Andrea Borgarelli did part of the work after he became a student of the CS@Max Planck doctoral program.}

\tnotetext[tfn1]{This work was supported by the Research Council of Norway through the project 274515 \emph{ADAPt: Exploiting Abstract Data-Access Patterns for Better Data Locality in Parallel Processing} \newline  (\href{https://www.mn.uio.no/ifi/english/research/projects/adapt/}{\sf www.mn.uio.no/ifi/english/research/projects/adapt/}).}

\affiliation[1]{organization={University of Oslo}, addressline={Gaustadalleen 23B},
postcode={NO-0373}, city={Oslo}, country={Norway}}
\affiliation[2]{organization={Universita' degli Studi di Torino}, addressline={Corso Svizzera 185},
postcode={10149}, city={Torino}, country={Italy}}

\begin{abstract}
  Cloud computing has radically changed the way organisations operate
  their software by allowing them to achieve high availability of
  services at affordable cost. Containerized microservices is an
  enabling technology for this change, and advanced container
  orchestration platforms such as Kubernetes are used for service
  management.  Despite the flourishing ecosystem of monitoring tools
  for such orchestration platforms, service management is still mainly
  a manual effort.
  
  The modeling of cloud computing systems is an essential step towards
  automatic management, but the modeling of cloud systems of such
  complexity remains challenging and, as yet, unaddressed. In fact
  modeling resource consumption will be a key to comparing the outcome
  of possible deployment scenarios. This paper considers how to derive
  resource models for cloud systems empirically. We do so based on
  models of deployed services in a formal modeling language with
  explicit CPU and memory resources; once the adherence to the real
  system is good enough, formal properties can be verified in the
  model.

  Targeting a likely microservices application, we present a model of
  Kubernetes developed in Real-Time ABS. We report on leveraging data
  collected empirically from small deployments to simulate the
  execution of higher intensity scenarios on larger deployments. We
  discuss the challenges and limitations that arise from this
  approach, and identify constraints under which we obtain
  satisfactory accuracy.  
 \end{abstract}


\maketitle

\section{Introduction}
\label{intro}
Cloud-native applications are collections of
microservices~\cite{Newman15,balalaie2016}, i.e., small, independent,
and loosely coupled services.  Deploying these applications is
challenging and error-prone.  Container technologies such as
Docker~\cite{dockerpaper} facilitate this deployment process by
addressing the complexity rising from modules dependencies and by
isolating small services in a protected environment. Container
orchestrator systems such as Kubernetes~\cite{burns2016} are used to
organize and deploy containerized services in cloud-native
applications.  The Cloud Native Computing Foundation (CNCF) reports an
increased adoption of containers by 300\% from 2016 to
2021~\cite{cncf20report}. Their most recent user
survey~\cite{cncf21report} shows the adoption of Kubernetes has grown
along: 96\% of organizations are either using or evaluating Kubernetes
in production. In general we can expect the adoption of Cloud
computing to continue to increase, and support for these spreading
technologies will be of paramount importance~\cite{manifesto2018}.

Kubernetes is Google's third generation of container orchestrator
systems~\cite{burns2016} and was open-sourced in 2014.  The system
provides a layer between the cluster operator and the applications
running on the cluster. Applications are implemented as collections of
services, each developed, deployed and scaled individually. It
leverages containerization to handle scaling and failover for the
application, and provides deployment patterns, service definitions,
service discovery and basic load balancing~\cite{Gilly2011}.
Containers are deployed in pods, which are abstractions for groups of
containerized components. Services and automatic scalers let the
application scale, adapt the application to variable demand, restart
or gradually update failing components, accommodating continuous
deployment.

Deploying and running a microservice application in Kubernetes in a proficient way remains a highly technical challenge~\cite{k8sdeployexample}, despite a flourishing ecosystem of open source plugins and documentation. Performance is affected by several steps of the Dev\-Ops toolchain, such as defining requested resources, (anti)affinity and load balancing. The performance outcome of a Kubernetes deployment is strictly affected by the operator decisions, and thus deployment cannot be easily automated. To track performance over costs, the operator needs to decide on a service allocation for the initial deployment and achieve proper load balancing across the cluster nodes while keeping a clear picture of the current cluster settings and demand.  It is difficult to achieve resource-efficient solutions.

Containers are often perceived as lightweight virtual machines, since in some cases they replace virtual machines.  However, this association can be misleading: on the surface containers are independent units of deployment, as they have their own process space, network space and file system, they can orchestrate network ports, and they can safely rely on different kinds of volumes.  Underneath, resources are shared between different containers.  This indirectly affects their resource usage and thereby their availability.  Understanding what goes on under the hood in container orchestration systems is essential in order to reach a proficient deployment.

In this paper, we introduce a modeling framework for cloud-native applications orchestrated using Kubernetes, to predict how CPU and memory resources will be used by multiple containers and how resource consumption will be affected in different cluster settings. The proposed modeling framework can help the system administrator in finding a cluster configuration for a microservice-based system which meets the system's performance requirements.  We aim to facilitate the comparison of different deployment scenarios by means of a highly configurable, executable model. The main contributions of this paper are: 
\begin{enumerate}
\item a framework for modeling the resource-sensitive behavior of cluster configurations for a microservice-based system;
\item a methodology to create formal models of resource consumption for containerized microservices deployed and managed by Kubernetes in this framework; and
\item an evaluation of the proposed methodology on an actual microservice application in Kubernetes, the open-source \emph{Online Boutique} cloud-native application.\footnote{\url{https://github.com/GoogleCloudPlatform/microservices-demo}}
\end{enumerate}
Although the proposed modeling framework abstracts from many aspects of Kubernetes (e.g., rollouts, rollbacks, orchestration of volumes, user roles and authorizations, scalability), once calibrated following our proposed methodology, the derived models already allow system deployment under several configurations to be explored and compared at the modeling level, \emph{before the system is actually deployed in these configurations}.

\paragraph{Paper Overview} Section~\ref{sec:background} introduces
Kubernetes and \rtabs. The developed modeling framework for
cloud-native applications using Kubernetes is presented in
Section~\ref{sec:model}, and the methodology for instantiating the
framework for a specific application, in
Section~\ref{sec:method}. Section~\ref{sec:eval} evaluates the
methodology on different deployments of the \emph{Online Boutique}
application. Section~\ref{sec:discussion} elaborates on the
applicability and extensibility of the presented work.
Section~\ref{sec:related} discusses related work, and
Section~\ref{sec:conclusion} concludes the paper.


\section{Background}
\label{sec:background}

\subsection{Kubernetes}\label{sec:k8s}
\textbf{Kubernetes}~\cite{burns2016} is an open-source system \cite{Kubernetes} for managing containerized applications across multiple hosts. It provides basic mechanisms for deployment, maintenance, and scaling of applications.  The core of Kubernetes includes services running on pods with various components for their management.  We here briefly introduce the main Kubernetes components related to resource management: service allocation and load balancing of service requests, containers, pods, nodes and their capabilities \cite{Kubernetesconcepts,hightower17kubernetes}.

\textbf{Services} represent components that act as basic internal \emph{load balancers} and ambassadors for pods. A service comprises a logical collection of pods (explained below) that perform the same function and presents them as a single entity via a \emph{service endpoint}.  This allows the Kubernetes framework to deploy a service that can keep track of and route requests to the different back-end containers of a particular type. Internal service clients only need to know about the service endpoint. Meanwhile, the service abstraction enables the scaling or replacing of back-end containers as necessary. The address of a service endpoint remains stable regardless of changes to the pods to which it routes requests. By deploying a service, the associated pods gain discoverability, which simplifies container design.  Whenever access to one or more pods needs to be provided to another application or to external service clients, a service can be configured.  Although services, by default, are only available using an internally routable address, they can be made available to the outside of the cluster.

\textbf{Containers}~\cite{pahl2019} facilitate the deployment process by addressing the complexity rising from module dependencies and by isolating small services in protected environments. Container technology enables self-contained, ready-to-deploy parts of applications, including middleware and business logic, to be packaged into binaries and libraries that can be used to run the applications. The processes inside a container share network space, process space, and file system. This means that they can talk to each other through different ports, a process can signal another process, and all files inside the container are available to these processes. Tools like Docker~\cite{dockerpaper} provide engines to package applications into containers. The core of a container engine is leveraged by Kubernetes to run the pods.  When a container is built, its image --- the executable binary package that is produced from the container definition --- is usually pushed to an online container registry and tagged. The URL and tag are then set inside Kubernetes deployments to retrieve the containers and activate the corresponding service.

\textbf{Pods} are the basic scheduling unit in Kubernetes. They are high-level abstractions for groups of containerized components.  A pod consists of one or more containers that are guaranteed to be co-located on a host machine and can share resources.  A pod is deployed on a node (explained below) according to its resource requirements and has its own specified resource limits. For two or more pods to be deployed in the same node, the sum of the pods' minimum amounts of required resources needs to be available in the node.  All pods have unique (IP) addresses, which allows
developers to use ports without the risk of conflict. Within the pod, containers can reference each other directly, but a container in one pod cannot address a container in another pod without passing through a reference to a service; the service then holds a reference to the target pod at the specific pod address. The addresses of pods are ephemeral; i.e., they are reassigned on pod creation and system boot.

In Kubernetes, pods can consist of multiple containers, including additional init containers, sidecars, and helper containers that carry out side tasks such as checking health and replying to health probes. Init containers and sidecars are not considered for an application's resource consumption, since the init container only partakes in the creation of the pod and sidecars generally handle networking tasks that can be separated from the consumption of the pod.  In contrast, helper containers can be counted as part of a pod's resource consumption and of the application logic. Putting multiple service containers in the same pod would be an anti-pattern for a cloud-native application, restricting the flexibility of the microservice architecture. Kubernetes offers several options for communication between endpoints (discussed below), which should be preferred over pods with multiple service containers.

\textbf{The nodes} in a cluster are given different roles in the Kubernetes framework: one node functions as the master node and the others as worker nodes. The master node acts as the primary point of contact with the cluster and is responsible for most of the centralized logic that Kubernetes provides. The master node implements a server that acts as a gateway and controller for the cluster by exposing an API for developers and external traffic. It allocates pods and orchestrates communication between the components of the framework.  In contrast, worker nodes host pods and form the larger part of a Kubernetes cluster. Worker nodes have explicit resource capabilities, CPU and memory, which are known by the system. The memory capability is specified as the ratio between occupied space and free space, and the CPU capability in terms of \emph{cores} or \emph{millicores}, where a single core CPU provides 1000 millicores (i.e., milliseconds of processor activity).  When a pod is deployed on a node, the pod detains an amount of millicores which represents the segment of time within which they are allowed to use the CPU.

\textbf{The scheduler} is in charge of service allocation, and assigns pods to specific nodes in the cluster. The scheduler matches the operating requirements of a pod's workload to the resources that are currently available on the nodes in the Kubernetes framework, and places pods on appropriate nodes.  The scheduler is responsible for monitoring the available capacity on each node to make sure that service containers are not scheduled in excess of the available resources. The scheduler needs to keep track of the total capacity of each node as well as the resources already allocated to existing service pods on the nodes.

\textbf{The load balancing} of service requests across multiple pods is handled by the Kubernetes framework.  This load balancing is \emph{content agnostic}; i.e., this load balancing, which is also called layer-4, has limited capabilities because it operates at the Transport layer (TCP/IP) of the ISO/OSI stack.  Modern applications can suffer load balancing issues in Kubernetes due to newer protocols that bypass layer-4 load balancers \cite{Kuberneteslb}. For example, gRPC (Remote Procedure Call) breaks the standard layer-4 load balancing of Kubernetes because it is built on HTTP/2, which multiplexes requests using a single long-lived TCP connection. Thus, multiple requests can be active on the same connection at any point in time. This reduces the overhead of connection management, but it reduces the usefulness of connection-level load balancing: once the connection is established, there is no load balancing and all requests are routed to a single destination pod.  For this reason, clusters are often equipped with additional load balancers, such as the service meshes Linkerd \cite{Linkerd} and Istio \cite{istiopage}. These provide \emph{content aware} layer-7 load balancing, which operates at the Application layer, rerouting requests at the highest level of the network protocol stack by means of a High Availability proxy (HA proxy) for each pod. This type of load balancing is much more expensive than just layer-4, in terms of consumed computational resources and latency. In fact, redistributing requests consume resources on distributed proxies. When layer-4 load balancing fails, a few proxies receive the major part of the requests, and redirecting all that traffic consumes a significant amount of resource only on few nodes.


\subsection{Real-Time ABS}
\label{sec:rtabs}
\ABS \cite{abswebsite, johnsen10fmco} is an executable modeling
language which targets the design and verification of concurrent and
distributed systems.  \ABS is an actor-based, object-oriented,
executable modeling language with a Java-like syntax and a real-time
operational semantics \cite{bjork13isse}.  Its concurrency model is
based on active objects \cite{deboer17csur}, which decouple
communication and synchronization to support very flexible
orchestration of parallel activities within and between active
objects. ABS has previously been used to model and analyze cloud
deployments of resource-aware virtualized systems
\cite{johnsen16ifmcloud,schlatte21coordination}, including workflow
processing \cite{johnsen17softcom}, AWS deployment decisions
\cite{johnsen16isola}, Hadoop \cite{lin16fase}, Spark Streaming
\cite{lin20ijguc} and industrial cloud applications~\cite{ABHJSTW13}.
Therefore, ABS is an adequate match for exploring resource usage
analysis in Kubernetes.  We can understand ABS in terms of layers.

\emph{The functional layer} of \ABS is used to model computations on
the internal data of objects. It allows designers to abstract from the
implementation details of imperative data structures at an early stage
in the software design.  The functional layer combines parametric
algebraic data types (ADTs) and a simple functional language with case
distinction and pattern matching. \ABS includes a library with
predefined datatypes such as \Abs{Bool}, \Abs{Int}, \Abs{String},
\Abs{Rat}, \Abs{Float}, \Abs{Unit}, etc. It also has parametric
datatypes such as lists, sets and maps.  All other types and functions
are user-defined.

\emph{The imperative layer} of \ABS allows designers to express
communication and synchronization between active objects, which
encapsulate threads~\cite{schaefer10ecoop,johnsen10fmco}. Threads are
created automatically at the reception of a method call and terminated
after the execution of the method call is finished.  \ABS combines
active (with a \Abs{run} method which is automatically activated) and
reactive behavior of objects by means of cooperative scheduling:
Inside the active objects, threads may suspend at explicitly defined
release points, after which control may be transferred to another
thread.  Suspension allows other pending threads to be
activated. However, the suspending thread does not signal any other
particular thread, instead the selection of the next thread to be
executed is left to the thread scheduler. In between these explicit
release points, only one thread is active inside an active object,
which means that race conditions are avoided.

\emph{The temporal layer} of \ABS, called
\rtabs \cite{bjork13isse,schlatte21coordination}, develops a real-time
operational semantics for active objects which allows the logical
execution time to be captured during the execution of methods inside
active objects.  To express dense time in the models, \rtabs considers
two types \Abs{Time} and \Abs{Duration}.  Time values represent points
in time as reflected on a global, logical clock during execution. In
contrast, finite duration values represent the passage of time as
local timers over time intervals. Thus, the local passage of time is
expressed in terms of \Abs{duration} statements which capture how long
the local execution is delayed (similar to, e.g., UPPAAL
\cite{larsen97sttt}).

\ABS is an open-source research project \cite{absrepo,schlatte22scp}
supported by a range of analysis tools (see, e.g., the \ABS tool
survey~\cite{ABHJSTW13}); for the analysis results in this paper, we
are using the \ABS simulation tool \cite{schlatte21coordination},
which is implemented in Erlang \cite{armstrong07erlang}.


\section{A Modeling Framework for Resource Consumption in Kubernetes}
\label{sec:model}
\begin{figure}[t]
\centering
\includegraphics[trim={0pt 0 0 0},clip,width=0.7\linewidth]{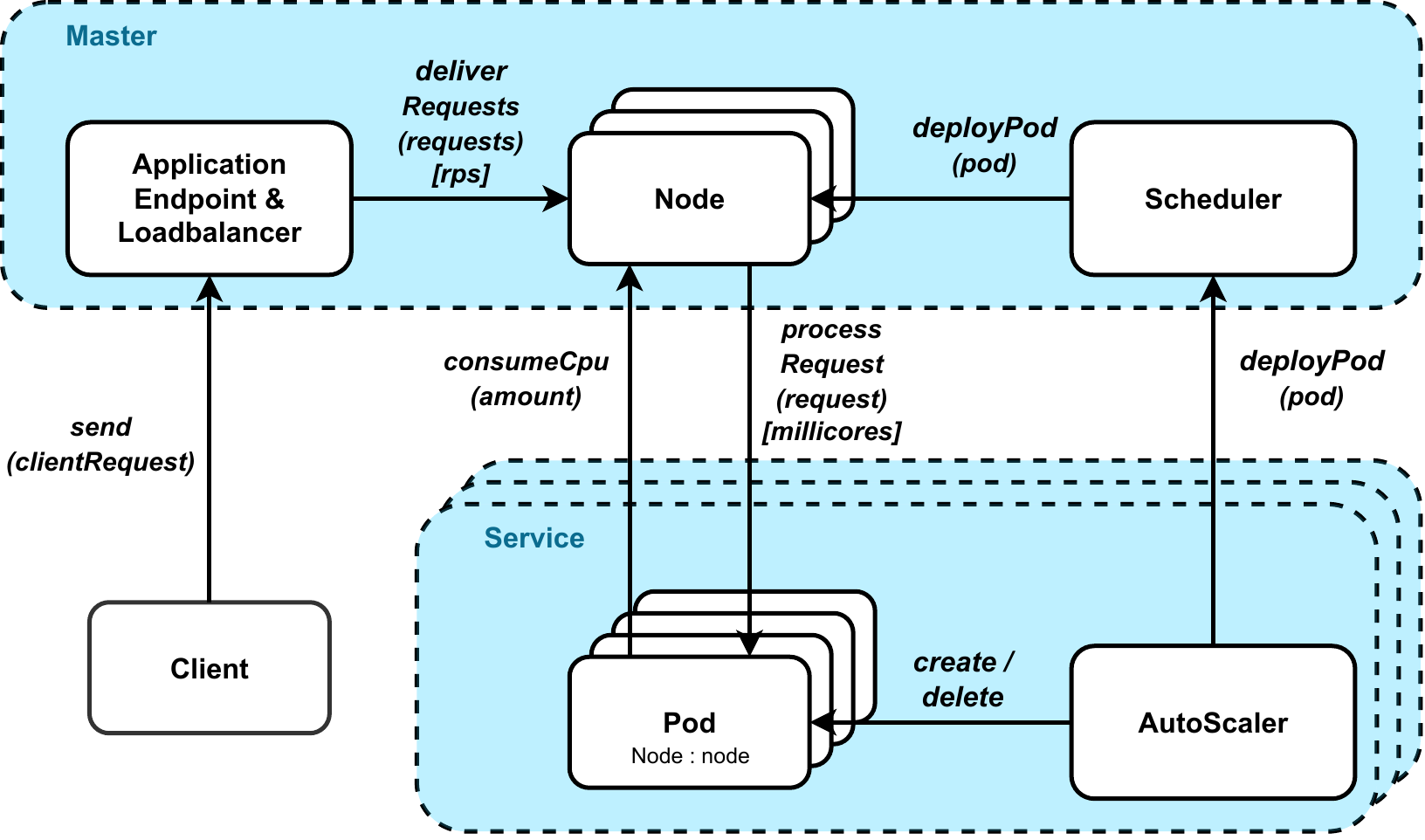}
\caption{\label{fig:modelCD}The architecture of the modeled Kubernetes cluster.}
\end{figure}

In this section, we present a general resource modeling framework for
systems orchestrated using Kubernetes, and discuss how the framework
can be instantiated for a specific cloud-native application. The focus
of the modeling framework is on resource management and load
distribution. We distinguish two main categories of resources:
resources that are temporarily available and periodically recharged
(e.g., CPU or energy), and resources that are acquired and released
(e.g., memory or storage). In this framework we explore both
categories concretely, via CPU and memory resources. Further
generalization of the resource analysis framework is discussed in
Section~\ref{sec:discussion}. The framework is executable and
developed in \ABS; i.e., it provides a simulation environment that can
be used to make model-based load predictions. In particular, the
framework can be used to predict costs for different scenarios under
stress and to compare CPU and memory usage between different system
configurations of Kubernetes nodes, at the \emph{modeling level}.  The
precision of the model will determine the predictive capabilities of
these simulations for real world cluster production
scenarios. Figure~\ref{fig:modelCD} shows the structure of modeled
clusters in the framework.

Clients invoke a service by sending requests to the service endpoint,
requests are distributed among the worker nodes using the load
balancer. The amount of requests a node receives is determined by the
type and number of the pods it hosts and measured in Request Per
Second (RPS). A pod is deployed on a node and consumes its resources
while processing requests.  The autoscaler manages the number of pods
for the service, and calls the scheduler to deploy new pods.

\paragraph{Model Input/Output}

To instantiate the outlined framework on a concrete
cluster of services, we need to specify:
\begin{enumerate}
\item \textbf{application settings,} which include \emph{pod
    configurations} (placement, required resources), \emph{service
    configurations} (specifically, the load balancing policy) and
  \emph{workflows} that are supported by the services of the cluster;
  and
\item \textbf{cost tables,} which specify the resource consumption of
  services for different workflows, at different intensities and for
  different node configurations.
\end{enumerate}
Observe that the isolation properties may vary between container
systems. This is reflected in our modeling framework by the specified
cost tables; for perfectly isolated containers a simpler specification
of cost would suffice \cite{ABHJSTW13,johnsen15jlamp}.

In the remainder of this section, we discuss aspects of the modeling
framework of particular relevance for resource management and load
balancing. The complete modeling framework is open source and
available
online.\footnote{\href{https://github.com/giaku/abs-k8s-model/tree/ld-fixed-nodes}{\tt https://github.com/giaku/abs-k8s-model/tree/ld-fixed-nodes}}

\subsection{Modeling of Requests and Workflows}\label{sec:wf}

In our modeling framework, the workload of services is abstracted into
batches of requests.  A batch of requests has a \emph{size} that
specifies the actual number of requests, while its processing
\emph{cost} is determined by the hosting node according to its cost
table and the total amount of requests that the node is handling.  To
recover a finer granularity for load balancing, batch requests from
clients are partitioned into smaller batches by the load balancer, as
explained in Section~\ref{ssec:3}.
These smaller batches are received by a node and transformed into
resource consumption by the pods hosted on that node.

Figure~\ref{fig:requests} shows how workflows, client requests, and
clients are modeled.
Workflows are compositions of activated services; the \Abs{workflow}
datatype in \ABS defines a workflow to have a name and include a
collection of services.
The \Abs{ClientRequest} datatype in \ABS defines a batch of requests
of a given size to a named workflow.
We assume that the different services of a workflow can be executed in
parallel and abstract from the activation order of the services,
because of the pipeline effect for the multiple requests contained in
each batch when we consider batches of requests.
Clients are implemented by the \Abs{ClientObject} class in \ABS, which
fires batches of requests of a given size of a given workflow to a
service endpoint.

\begin{figure}[t]
\begin{absmodelcode}
data WorkflowData = WorkflowData(String wfName, List<String> services); 
data ClientRequest = ClientRequest(WorkflowData wf, Int rps); 

WorkflowData wf3Data = WorkflowData( // workflow definition with activated services
     "workflow3",
     list["frontend","service1","service2","service3", ...]);

ClientRequest cr1 = ClientRequest(wf3Data, 150); // batch of requests to be sent by a client

Client c1 = // client instantiation
     new ClientObject(cr1,loadBalancer,80,rat(1.0)); // client req, target lb, num of batches, delay between batches
c1!start() // client activation, it will start to asynchronously send its batches over time
\end{absmodelcode}
\vspace{-11pt}
\caption{\label{fig:requests} Request declaration and client hatching.}
\end{figure}

%
%

%

\subsection{Modeling of Services and Pods} \label{ssec:3}
\label{sec:serviceModel}\label{sec:modelPod}
A service in our modeling framework is configurable, and carried out
by a number of pods.  Services are configured by passing parameters to
the service when it is instantiated, including the parameters
\Abs{PodConfig} and \Abs{ServiceConfig}, as shown in
Figure~\ref{fig:podserviceConfig}:
\begin{itemize}
\item The datatype \Abs{PodConfig} specifies the requested CPU
  resources and the CPU limit for processing tasks on the pods, the
  memory cool-down time for insufficient memory and the cost
  granularity.  The \emph{requested CPU resources} specify the minimum
  amount of CPU resources required for the pod to execute and the
  \emph{CPU limit} specifies the maximum amount of CPU resources that
  the pod will consume.  The \emph{memory cool-down} is the time delay
  before rescheduling a pod in case the node lacks sufficient memory
  for the pod to execute.  The \emph{cost granularity} captures the
  number of resource consumption steps the pod will use to process a
  request. The amount of CPU consumed per step is then obtained by
  dividing the cost of the request by the cost granularity.
\item The datatype \Abs{ServiceConfig} specifies the initial number
  of pods, the minimum and maximum number of pods for the service and
  the configuration of the autoscaler.
\end{itemize}

For simplicity in the modeling framework, pods are assumed to consist
of a single container. (A pod with many containers can be modeled by a
pod running one container which consumes the sum of their total
resources.) The pods are deployed onto nodes and consume resources
when they process requests, as shown in
Figure~\ref{fig:processRequest}: memory is allocated at the beginning
and released at the end of the \Abs{processRequest} method, while CPU
resources are consumed gradually according to the
\Abs{costGranularity} parameter of the model. The nature and number of
pods on a node determine the type and number of requests that the node
receives from the service endpoint.

\begin{figure}[t]
\begin{absmodelcode}
data PodConfig = PodConfig(Rat monitorCycle, Rat memoryCooldown, Rat cpuRequest,
	Rat cpuLimit, Int costGranularity);
data ServiceConfig = ServiceConfig(String name, Int startingPods, Int minPods,
	Int maxPods, Rat scalerCycle, Rat upscaleThreshold, Rat downscaleThreshold,
	Int downscalePeriod);
\end{absmodelcode}
\vspace{-11pt}
\caption{\label{fig:podserviceConfig} The datatypes \Abs{PodConfig} and \Abs{ServiceConfig}.}
\end{figure}

\begin{figure}[t]
\begin{absmodelcode}
Unit processRequest(Request request) {
	Rat cost = requestCost(request); Rat requiredMemory = memory(request);
	this.allocateMemory(requiredMemory); // memory allocation |\label{line:alloc-mem}|
	monitor!consumedMemoryUpdate(requiredMemory);
	
	Rat computationStep = (cost / costGranularity) // compute step size
	
	while (cost > 0){ // until fully processed |\label{line:start-while}|
		...
		if (cost > computationStep){ // consume a step size amount
			await this.availableCpu > 0; // check on the pod limit	|\label{line:avail-cpu}|		
			await node!consumeCpu(computationStep,this); // consume node CPU			
			await !this.blocked; // refresh sync			
			availableCpu = availableCpu - computationStep; // reduce available CPU (pod limit)
			monitor!consumeCpu(computationStep);
			cost = cost - computationStep; // cost decreases		
		} else if (cost > 0){ ... } // consume remaining amount, if less than step size
		... suspend; // release control at the end of each consumption step, allows interleaving
	} |\label{line:end-while}|
	
	this.releaseMemory(requiredMemory); // memory release |\label{line:release-mem}|
	monitor!consumedMemoryUpdate(-requiredMemory);
}
\end{absmodelcode}
\vspace{-11pt}
\caption{\label{fig:processRequest} The \Abs{processRequest} method of
  the pods.}
\end{figure}

\subsection{Modeling of Nodes}
\label{sec:nodesModel}

\begin{figure}[t]
\begin{absmodelcode}
Unit convertRps(){
    // 'Service' -> Map<'Workflow',Rps>
    Map<String,Map<String,Int>> serviceWfRpsMap = map[];
    Map<String,Int> totRpsMap = map[];
    
    // scan node requests from master load balancer, retrieve RPS per each service and workflow $\label{fig:line6}$
    foreach (nodeReq in this.nodeReqsQueue){ ... }

    List<String> services = elements(keys(serviceWfRpsMap));

    // create map with <service,total rps>, total amount of RPS for each service
    foreach (service in services){ ... }

    // convert total rps and workflow info in millicores costs, using the cost tables
    Map<String,Rat> serviceConsumptionMap = 
       this.buildServWfConsumption(serviceWfRpsMap,totRpsMap); $\label{fig:line16}$
    
    // generate and send the requests to pods
    foreach (service in services){ $\label{fig:line19}$
      Int totServRps = lookupDefault(totRpsMap,service,0); // total RPS
      Int nOfServPods = lookupDefault(podsMap,service,0); // number of service pods
      Rat serviceCost = lookupDefault(serviceConsumptionMap,service,0); // total cost
      Rat costPerRps = serviceCost / totServRps; // cost per request

      List<Pod> servicePods = lookupDefault(this.servPodsReferenceMap,service,list[]);

      List<Int> rpsQuotas = this.generatePodsRequestsSizes(totServRps,nOfServPods); // split total RPS amount}

      Int index = 0;
      while (index < nOfServPods){ $\label{fig:line30}$ // send each pod its batch of requests
        Int rpsQuota = nth(rpsQuotas,index);
        Rat requestCost = costPerRps * rpsQuota; // compute the cost
        Request req = Request(service,requestCost,1,rpsQuota); // generate the request
        Pod p = nth(servicePods,index);
        // send request to pod with millicores information
        p!processRequest(req); 
        index = index + 1;
      }}}
\end{absmodelcode}
\caption{\label{fig:convertRps} The \Abs{Node convertRps} method
of the the class \Abs{Node}.}
\end{figure}

The Kubernetes master node is not explicitly modeled, its
functionalities are implemented in the model logic. The \Abs{Node}
class in \ABS models the Kubernetes worker node, which has a given
amount of resources available for consumption by its running pods. In
addition to its capacities, the modeled nodes include information about
resource consumption in their cost table, both the capacities and cost
table are specified upon node creation:

\begin{itemize}
\item \textbf{CPU capacity.} CPU resources are \emph{time
    dependent}. They are replenished at every time interval, the total
  amount of computed costs on a node in the time interval cannot
  exceed the node's CPU capacity.
\item \textbf{Memory capacity.} Memory is \emph{time
    independent}. Memory can be acquired and released. The node
  decreases its available memory when a pod starts the processing of a
  request and allocates memory cost on the node memory. The memory
  stays decreased for the whole computation time and the allocated
  amount is restored on request completion.  In case the free memory
  is insufficient, the request remains pending until enough memory is
  available.
\item \textbf{Cost table.} The cost table is used to capture the
  resource consumption on a node for specific configurations.  The
  cost table stores information about resource consumption for each
  workflow and service for different RPS entries.  The table maps
  triplets (workflow, serviceName, RPS) to their known resource
  consumption.
\end{itemize}

The modeling framework considers cluster of nodes from fixed
\emph{images}; i.e., nodes are fixed to be of given configurations
during model instantiation. However, this is not a major limitation as
many different images can be modeled in the framework and nodes can
change between images during a simulation.

Cost tables are introduced to address the problem of dependencies
between pod consumption in container systems without perfect
isolation between pods.  When a pod processes batch requests, it needs
to acquire resources from its node. As the CPU costs associated to RPS
are stored in the node's cost table, the node calculates the actual
resource consumption that each pod will require. If the exact amount
of RPS for a service is not found in the cost table, the model
interpolates between the values of the two closest table entries.  The
number of pods and load balancing between pods affect the number of
requests that a pod receives.

Figure~\ref{fig:convertRps} shows how the the node parses its share of
requests and calculates the total amount of RPS from the queue, for
each service and for each workflow
(Lines~\ref{fig:line6}--\ref{fig:line16}). In the final loop
(Line~\ref{fig:line19}), the amounts of RPS are converted into
millicores by means of the
\Abs{serviceConsumption}\-\Abs{Map}. Finally, the node notifies each
pod about its total consumption that will start consuming (the
while-loop at Line~\ref{fig:line30}). The pods will start to execute
once they know the allocated amount of millicores for this time unit.

\subsection{Modeling of Load Balancers}

In the modeling framework, we have implemented a round-robin load
balancing policy for batches of requests. Other load balancing
policies can be implemented in a similar way, such as random, ring, or
hash. The modeling framework can also accommodate different policies
for different services. Since the modeling framework focuses on
resource consumption, policies that are based on non-functional
properties are difficult to capture within our framework. These
policies, which direct more requests to the best-performing endpoints,
can be based on, e.g., the latency and error rates of the endpoints.
Implementing these policies would require encoding a statistical
distribution of loads in the \Abs{Master LoadBalancer}, which goes
beyond our current modeling framework.

The \Abs{Master LoadBalancer} (\Abs{MLB}) converts batches of workflow
requests sent by the clients into smaller batches of service requests
directed towards the pods.  Figure~\ref{fig:balanceClientRequest}
shows the method \Abs{balanceClientRequest} of the \Abs{MLB} class,
invoked by the clients.  The \Abs{ClientRequest} method refers to a
specific workflow, including the size of the workflow and the list of
services that will be activated (see Section~\ref{sec:wf}). Based on
this information, the \Abs{MLB} generates proper batches of requests
for each service and for each pod of that service.  In the method, the
first cycle (Line~\ref{fig6line6}) goes through the services of the
workflow and divides the RPS with a round-robin policy among the
pods. The second for-loop (Line~\ref{fig6line10}) generates and
forwards the requests to the nodes.

\begin{figure}[t]
\begin{absmodelcode}
Unit balanceClientRequest(ClientRequest request){
    WorkflowData wfd = wf(request);
    String wfName = wfName(wfd);
    List<String> services = services(wfd);
    //here big batches of requests from the client have been transformed into small batches for each pod
    foreach (service in services){ $\label{fig6line6}$
      // group small batches by the node hosting the pods
      Map<Node,Int> servicePodMap = lookupDefault(perServicePodsNumber,service,map[]);
      List<Pair<Node,Int>> quotas = this.generateSubrequestsSizes(rps(request),servicePodMap);

      foreach (quota in quotas){ $\label{fig6line10}$ // send each node its group of small batches
        Node node = fst(quota);
        Int size = snd(quota);
        NodeRequest nodeReq = NodeRequest(service,wfName,size); 
        // create the node request (service,wf,rps)
        node!processRequest(nodeReq);
      }}}
\end{absmodelcode}
\vspace{-11pt}
\caption{\label{fig:balanceClientRequest} The
  \Abs{balanceClientRequest} method of the class
  \Abs{MasterLoadBalancer}.}
\end{figure}

\subsection{Modeling of Schedulers}

The \Abs{Scheduler} class  finds places for pods on nodes.  The
default deployment strategy is to compare the pod's requested CPU
resources to the available CPU resources in the least busy node.
If there are enough CPU resources available on a given node, the pod
is scheduled on that node. If no node has enough resources available
the pod remains pending, to be scheduled in another time interval.
This strategy is implemented in the \Abs{deployPod} method of the
\Abs{Scheduler} class, which is shown in Figure~\ref{fig:deployPod}.

\begin{figure}[t]
\begin{absmodelcode}
Node deployPod(Pod p, ResourcesMonitor rm){
  Bool deployed = False;
  Rat requestedCpu = await rm!getCpuRequest();
  Node result = null;
  
  List<Int> nodeIds = lookupDefault(rulesMap,serviceName,list[]); // check rule
  
  // get the list of candidate nodes
  if (!isEmpty(nodeIds)){ // rule defined
  	Node selected = lookupDefault(nodesMap,head(nodeIds),null);
  	nodesToCheck = list[selected];
  	... // cycle the rule list
  } else { // default behavior
	nodesToCheck = activeNodes
  }
	
  while (!deployed){ // try to deploy on the least loaded node
    Rat maxCpu = -1
    List$\langle$Node$\rangle$ nodesToCheck = tail(activeNodes);
    foreach ( n in nodesToCheck){ ... } // get the node with maximum available CPU
    if (maxCpu >= requestedCpu){await result!addPod(p,rm); deployed = True;} // deploy if enough CPU
    else{await duration(1,1);} // if not successful, wait before retrying
  }
  return result;}
\end{absmodelcode}
\vspace{-11pt}
\caption{\label{fig:deployPod} The \Abs{deployPod} method of the class
  \Abs{Scheduler}.}
\end{figure}

To support the definition of fixed scheduling of pods onto nodes, the
scheduling can also be guided by a map specifying an ordered list of
nodes for each service.
If the list is exhausted, the scheduler will start again from the
beginning of the list.
For example, Figure~\ref{fig:fixedsched} shows a rule defined for the
service \Abs{frontend}. If eight pods are to be deployed, they will
end up two per node; if ten pods are to be deployed, 
three pods will be in node 1 and 2 and two pods in node 3 and 4.

\begin{figure}[t]
\begin{absmodelcode}
Map<String,List<Int>> rulesMap = map[];
rulesMap = put(rulesMap,"frontend",list[1,2,3,4]);
\end{absmodelcode}
\vspace{-11pt}
\caption{\label{fig:fixedsched} A deploy rule for the service
  \Abs{Frontend} in the class \Abs{Scheduler}.}
\end{figure}


\section{A Methodology for Modeling Specific Kubernetes Deployments}
\label{sec:method}
In this section, we propose a methodology to instantiate the modeling
framework in order to make model-based predictions of resource
consumption for a concrete cloud-native application. To this aim, the
cluster administrator needs to know the relevant node configurations
(that specify how pods are deployed on different nodes), the
\emph{workflows} that the application exposes to the endusers (i.e.,
which services are involved in a high-level action such as ``load the
homepage''), and the different \emph{service and pod configurations}
chosen for Kubernetes (i.e., the load balancing choices for services
and the required and maximum amount of resources of each pod). Since
the modeling framework considers deployments of nodes from fixed
images (cf.\ Section~\ref{sec:nodesModel}), we need to derive
execution costs for the images that we consider.

The methodology is used to calibrate the modeling framework to the
targeted cloud-native application by deriving cost tables for the node
images of the model from experiments on the corresponding node
configurations of the application. In the experiments, each node
configuration of the application is monitored while the node is
subjected to an increasing demand for each workflow. Thus,
instantiating the modeling framework for the Kubernetes deployment of
a specific cloud-native application with multiple node images,
requires a few steps:

\begin{enumerate}
\item instrument the cluster,
\item identify suitable workflows,
\item identify node configurations,
\item define a sampling strategy for service loads to derive cost tables, and
\item perform model-based predictions by means of simulation.
\end{enumerate}
We now detail the process for  each step.

\paragraph{Step 1: Instrument the Cluster}
We need to instrument the cluster for monitoring the targeted
cloud-native application.\footnote{In Kubernetes the most widely
  adopted open-source tools for this purpose are Prometheus
  (\url{https://prometheus.io/}) and Grafana
  (\url{https://grafana.com/}), but their integration currently
  requires additional work that the package manager Helm
  (\url{https://helm.sh/}) can significantly relieve. The full
  Prometheus stack chart for Helm is available from the Prometheus
  community.} We deploy the application and isolate the resource
consumption of Kubernetes system pods and monitoring plugins that
perform periodic jobs that would otherwise interfere.  For example, we
used a dedicated worker node to host these system pods; such node,
just like the master node, has not been modeled since it constitutes a
very low amount of resource usage (about 1\% of the total cluster
consumption in our experiments).

Note that many stress test tools send requests synchronously; i.e.,
the thread sending requests will always wait for the previous response
to return. Consequently, when approaching the maximum load capability
of the cluster, the response time grows and yields an RPS rate
drop. This drop cannot be avoided and should not be reproduced in the
model. To run meaningful stress tests, we need a tool capable of
maintaining a fixed number of requests per second, independent of the
response time.\footnote{In our experiments, we have used the
  open-source tool Vegeta, which is available on GitHub,
  \url{https://github.com/tsenart/vegeta}}

\paragraph{Step 2: Identify Suitable Workflows}
We now identify the workflows that are relevant for the resource
consumption of the cloud-native application. We are primarily
interested in workflows that significantly impact the resource needs
of the target application. These workflows can be identified by the
application owner, from the specifications, or by their resource
consumption.  A workflow with low impact in terms of the number of
services involved and the load on those services, would be of limited
interest when we instantiate the modeling framework. A suitable
workflow is static; i.e., the workflow never changes the set of
activated services despite randomised parameters (e.g. `set a
different currency').

\paragraph{Step 3: Identify Node Configurations}
We now identify the node configurations that we want to capture as
node images in the model. We consider nodes configured with a set of
pods that will not change while profiling the resource consumption of
the node. These nodes would typically correspond to the nodes used by
the administrator when scaling the cloud-native application.

\paragraph{Step 4: Define a Sampling Strategy for Service Loads to Derive Cost Tables}
In order to construct cost tables for each modeled worker node image
that reflect the resource usage on the cluster, we need to define a
sampling strategy. We run experiments on the cluster to measure the
resource consumption of every service of any workflow for different
RPS entries; i.e., we run experiments on the cluster to derive the
resource consumption $Y$ for each node configuration $A$, workflow
$\text{\it wf1}$ with its set of activated services, and level of
service requests $r$ (e.g., $25, 50, ..., 150$ RPS). The experiments
result in entries such as
$\text{Cost}_{A} \left( \text{\it wf1}, 25, \text{service1} \right)
\mapsto Y$, where $Y$ is specified in millicores, in the derived cost
table for node $A$. A cost table containing all workflows, their
services and corresponding resource consumption for each node
image. The derived cost tables will be used to calculate the resource
consumption for the pods. Different sampling strategies will provide
more or less entries; the more entries a cost table contains, the more
accurate will be the resulting model.

When our target node reaches its capabilities, the success rate for
the requests drops, the latency of the pods increases and resource
consumption becomes less predictable. Therefore we will let the
maximum RPS in the cost table be such that latencies never exceed a
reasonable amount (e.g., ten times a low demand latency), and success
rate never drops below 90\% during the sampling process.

\paragraph{Step 5: Perform Model-based Predictions by Means of Simulation}

Using an instance of the modeling framework with the derived cost
tables for node images, we can compare the outcomes of running
simulations of different configurations of the modeled cloud-native
application.  Having sampled service loads for multiple node images
allows us to simulate different scheduling choices in the cluster.

Note that Kubernetes systems often scale up and down at the level of
individual pods, thereby changing node compositions by adding or
removing single pods. This may introduce a higher variability for
possible node configurations than what we have considered in the
proposed methodology, where we considered scaling at the abstraction
level of node images to keep the number of node configurations for the
sampling process fairly low. The methodology can be extended to cover
the scaling of individual pods by increasing the number of node
configurations in the sampling process.


\section{Evaluating the Methodology}
\label{sec:eval}

With the proposed methodology, we aim to configure the modeling
framework to predict resource consumption for real cloud-native
applications.  Therefore, we evaluate the predictions of resource
consumption that we obtain for models derived by following the
methodology introduced in Section~\ref{sec:method}.  The proposed
methodology aims to derive cost tables for node images for
cloud-native applications with different workflows.  In order to
evaluate the accuracy of the resulting model for predicting the
resource consumption on the nodes, we apply the methodology to a
cloud-native application with mixed workflows, focus on CPU resources
and consider different node images and load distributions.  In detail,
we investigate the following research questions:

\begin{itemize}
\item[\textit{RQ1}] How accurate is the prediction of CPU consumption
  under a mixed workflow scenario on a cluster with \emph{homogeneous}
  nodes?
\item[\textit{RQ2}] How accurate is the prediction of CPU consumption
  for a mixed workflow scenario, on different cluster setups with
  \emph{heterogeneous} nodes?
\item[\textit{RQ3}] Is there a correlation between fair load
  distribution and response time for microservices application
  deployed in Kubernetes?
\end{itemize}

By a cluster with \emph{homogeneous nodes}, we mean that all nodes in
the cluster contain the same set of pods. In contrast, by a cluster
with \emph{heterogeneous nodes}, we mean that the cluster consists of
different nodes that contain different sets of pods.

\subsection{Experimental Design and Subject}
\label{ssec:casestudies}

To answer these research questions, we performed experiments with a
cloud-native application managed by Kubernetes, running on a cluster.
There are no standard benchmarks for cloud-native
applications. Therefore, we constructed a set of experiments based on
a microservices demo
application\footnote{\url{https://github.com/GoogleCloudPlatform/microservices-demo}}
from Google: an online shop where customers can browse and buy
products.  This application has previously been used to demonstrate
the functionalities and scaling capabilities of several Kubernetes
plugins.  In our experiments, the load generator component of the demo
application was not used since the stress tests have been implemented
differently.

\begin{figure}[t]
\includegraphics[width=\linewidth]{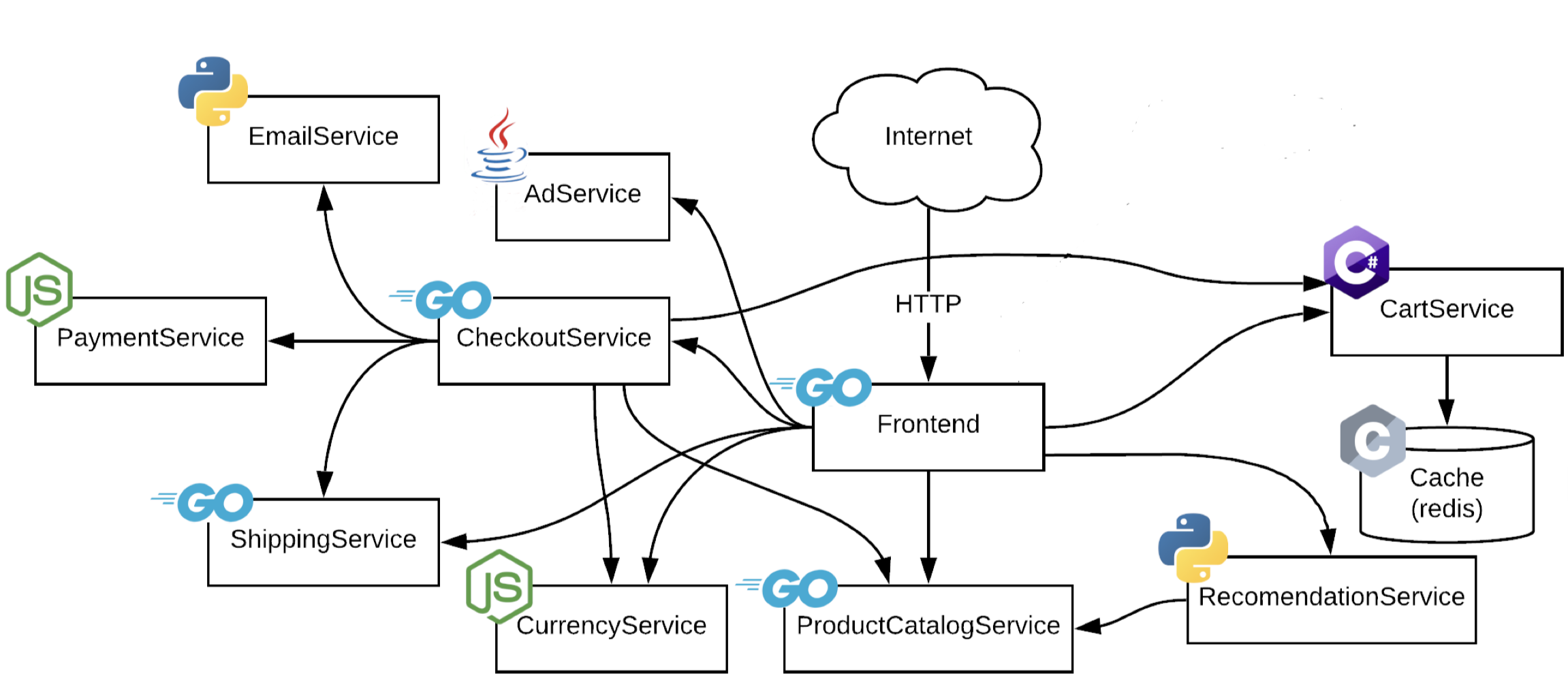}
\caption{Online shop service mesh.\label{fig:md-arch}}
\end{figure}

\paragraph{The Architecture of the Microservice Application}
The microservice architecture of the online boutique, their interactions and the relevant language technologies are shown in Figure~\ref{fig:md-arch}. The communication between the services of the online boutique are mostly based on gRPC calls, an HTTP/2 based protocol which keeps connections alive by bypassing the de facto Kubernetes layer-4 load balancing.

The application was deployed using the load balancer
Istio,\footnote{\url{https://istio.io/}} a service mesh platform working on top of Kubernetes. Istio was chosen for its high availability (HA) proxies based on the open source project Envoy, which provides every service pod with a sidecar container and redirects all requests, achieving layer-7 load balancing also on HTTP/2 based communication protocols. To change the load balancing strategy, we defined Istio Destination Rules for every service with the random policy.

The main workflows that are relevant for users of the online boutique
were identified by browsing the web application. A workflow can be
\emph{get the index page}, \emph{change currency} and \emph{view a
  random product}. Despite their simplicity, tasks like these are
already perceived as workflows and do in fact activate a variety of
services. For example, viewing the index page activates the services
\emph{frontend}, \emph{currencyservice}, \emph{cartservice},
\emph{adservice}, \emph{productcatalogservice} and
\emph{redis-cart}. It is common in this type of applications that a
frontend service builds the frontend page upon information retrieval
from other services. These relations can be seen as service
dependencies, though with our notion of workflows, we separate
external services that can be activated by the users from internal
services that implement the backend of the application.

A service can generally implement multiple functions: In simple cases
(like the online boutique), static workflows activate internal
services in a deterministic way; i.e. no parameter values can be used
to influence the set of services that are activated by a given
workflow (by activation we mean calls of the form
ServiceA.MethodA). Indeed, randomising parameter values for the online
boutique made no difference from the point of view of resource
consumption. In more complex cases, variable workflows can be modeled
as different workflows by fixing different parameters for the same
workflow.

\paragraph{The Configuration of the Cluster}
The online boutique application was deployed on a cluster of nodes
provided by Norwegian Research and Education
Cloud.\footnote{\url{https://www.nrec.no/}} We used five large nodes,
with 4 cores and 16GB of memory, divided into one master and four
worker nodes. In addition, two small nodes, with a single core and 4GB
of memory, were used to host system services, such as the Kubernetes
cluster DNS, the Kubernetes metrics monitor, the monitoring services
(Grafana and Prometheus stack), and the HA proxy services (Istio
components and Kiali for Istio communications monitoring). Finally, a
separate node, with 2 cores and 8GB of memory, was created as an
attacker using Vegeta. Such an additional node is necessary to prevent
that the CPU load generated by the stress tests affects the tailoring
of the target application's performance.

The services needed for the sampling process could easily fit on a single node.  Two services were particularly resource consuming in the first two workflows: \emph{frontend} and \emph{currencyservice}; these two services were deployed in two pods. The third workflow was mostly hitting \emph{frontend}, which has already two pods, and \emph{productcatalogservice} and \emph{recommendationservice}, which we deployed in two pods as well. To instantiate the modeling framework, we focus on the workflows of the online boutique listed in Table~\ref{table:workflows}.

\begin{table}[t]\centering
\begin{tabularx}{0.95\linewidth}{qwet}
\textbf{Name} & \textbf{Description} & \textbf{HTTP request} & \textbf{Services} \\
\hline
\vspace*{.003em} WF1 & \vspace*{.003em} View homepage & \vspace*{.003em} GET http://ip:port/ & \textbf{frontend}, \textbf{currency}, ad, cart, product\-catalog, redis-cart \\\hline
\vspace*{.003em} WF2 & Change \newline currency & POST http://ip:port/\x setCurrency & \textbf{frontend}, \textbf{currency}, ad, cart, productcatalog, redis-cart \\\hline
\vspace*{.003em} WF3 & View \newline product details & GET http://ip:port/product/\x product-code & \textbf{frontend}, currency, ad, cart, \textbf{productcatalog}, redis-cart, \textbf{recommendation}
\end{tabularx}
\caption{An overview of the considered workflows, with a brief description, the corresponding HTTP requests and a listing of the involved services. Services in bold denote the most stressed, hence important, services for the corresponding workflow.\label{table:workflows}}
\end{table}

\begin{table}[t]\centering
\begin{tabularx}{0.95\linewidth}{bmg}
\textbf{Node name} & \textbf{Purpose} & \textbf{Pods} \\
\hline
Node \newline type A & Nodes which implement the workflows and handle a reasonable amount of request throughput & 2$\times$frontend, 2$\times$currency, 1$\times$ad, 1$\times$cart,  2$\times$productcatalog, \hfill \newline 1$\times$redis-cart, 2$\times$recommendation \\\hline
Node \newline type B & \vspace*{.005em}Nodes which favor WF1 and WF2 & 4$\times$frontend, 3$\times$currency, \hfill \newline 1$\times$productcatalog,\newline 1$\times$recommendation \\\hline
Node \newline type C &\vspace*{.005em}Nodes which favor WF3 &  3$\times$frontend, 2$\times$currency, \hfill \newline 2$\times$productcatalog, 3$\times$recommendation
\end{tabularx}
\caption{An overview of the node images, with a brief description and the associated pods.\label{table:nodes}}
\end{table}

We configured a node image (Node type A) to implement the considered
workflows and handle a reasonable amount of request throughput,
another (Node type B) which favors WF1 and WF2 and a third (Node type
C) which favors WF3. The pods on the node images are listed in
Table~\ref{table:nodes}.  An excerpt of the resulting sampling process
outcome for Node type A can be seen in Figure~\ref{fig:calib-a}, the
created maps will then form a higher level map containing the full
calibration data. Next, we identified two other types of nodes, both
meant to extend this deployment.  The two last node types are not
sufficient to implement the workflows of the application by
themselves, so they have been deployed with a helper node hosting the
missing pods for the sampling process.

\begin{figure}[t]
\begin{absmodelcode}
Map<String,Rat> a_wf1_25 = // node type A, workflow 1 with 25 RPS
  map[
    Pair("frontend",526),Pair("currencyservice",434),Pair("adservice",72),
    Pair("cartservice",72),Pair("productcatalogservice",57),Pair("redis-cart",12)];	
//... up to 150 RPS

Map<String,Rat> a_wf2_25 = // same for workflow 2
  map[
    Pair("frontend",558),Pair("currencyservice",436),Pair("adservice",72),
    Pair("cartservice",73),Pair("productcatalogservice",57),Pair("redis-cart",12)];
//... up to 150 RPS

Map<String,Rat> a_wf3_25 = // same for workflow 3
 map[
   Pair("frontend",467),Pair("currencyservice",114),Pair("adservice",73),
   Pair("cartservice",73),Pair("productcatalogservice",276),
   Pair("recommendationservice",135),Pair("redis-cart",12)];
//... up to 150 RPS
\end{absmodelcode}
\vspace{-11pt}
\caption{\label{fig:calib-a}Data from the sampling process integrated
  in the model for node type A.}
\end{figure}

\paragraph{The Configuration of the Experiments}

To investigate research question RQ1, we need to compare model
predictions to the measured CPU consumption under mixed workflow
scenarios on homogeneous nodes.  For this purpose, we performed a
series of mixed workflow stress tests on the cluster with Node type~A.
The mixed workflows varied between two and three workflows, chosen to
cover different scenarios where the workflows are balanced as well as
scenarios where one workflow dominates the service requests.  The
stress tests spanned from 200 to over 500 RPS in
total. Table~\ref{tab:rq1} shows the composition of workflows for each
stress test; stress tests P1--P4 consider two workflows and stress
tests T1--T4 consider three workflows with different RPS. Each stress
test lasted 15 minutes and was executed five times to detect possible
fluctuations in resource consumption.

\begin{table}[t]\centering
\begin{minipage}{0.44\textwidth}\centering
\begin{tabular}{c | c | c | c }
\textbf{Workflow\,mix} & \textbf{WF1} & \textbf{WF2} & \textbf{WF3} \\
\hline
P1 & 100 & & 100 \\
P2 & 300 & 100 & \\
P3 & 150 & 350 & \\
P4 & 150 &  & 350 \\
T1 & 425 & 75 & 75 \\
T2 & 175 & 200 & 175 \\
T3 & 125 & 375 & 100 \\
T4 & 75 & 50 & 400
\end{tabular}
\captionof{table}{\label{tab:rq1}Mixed workflow RPS profiles for RQ1 (workflow pairs are denoted by P and triplets by T).}
\end{minipage}
\qquad \quad 
\begin{minipage}{0.44\textwidth}\centering
\begin{tabular}{c | c | c | c }
\textbf{Workflow\,mix} & \textbf{WF1} & \textbf{WF2} & \textbf{WF3} \\
\hline
P1 & 100 & & 100 \\
P2 & 175 & 100 & \\
P3 & 100 &  & 200 \\
P4 &  & 150 & 150 \\
T1 & 175 & 75 & 25 \\
T2 & 50 & 125 & 125 \\
T3 & 50 & 50 & 200 \\
T4 & 25 & 250 & 25
\end{tabular}
\captionof{table}{\label{tab:rq2}Mixed workflow RPS profiles for RQ2 (workflow pairs are denoted by P and triplets by T).}
\end{minipage}
\end{table}

To investigate research question RQ2, we need to compare model predictions to the measured CPU consumption under mixed workflow scenarios on heterogeneous nodes.  For this purpose, we performed experiments on the cluster with Node types B and C.  Because the load distribution between nodes is slightly unbalanced between different runs of the same deployment on the cluster (see Section~\ref{sec:k8s}), we need to be more careful in designing the experiments with heterogeneous nodes than with homogeneous nodes (used for RQ1). We address this issue by implementing a turnover of the nodes and run three iterations of each stress test on each cluster configuration, resulting in a total of twelve iterations for each stress test for RQ2. For the first three iterations the node types were instantiated on the four worker nodes, for the next three iterations all node types where shifted such that the first worker node hosted the pods of the second worker in the previous round of stress tests, the second worker node hosted the pods of the third worker node in the previous round, etc.  The mixed workflows used in the experiments for RQ2 are shown in Table~\ref{tab:rq2} and the cluster configurations are shown in Table~\ref{tab:nodes}. They have been reduced in the amount of service requests considered, because some cluster configurations could not handle the same service demand as in RQ1.

\begin{wrapfigure}[8]{r}{0.37\linewidth}
\vspace{-73pt}
\hspace*{-.25cm}\begin{minipage}{0.37\textwidth}\centering
  \begin{tabular}{c | c | c  }
    \multicolumn{3}{l}{\quad}\\
    \multicolumn{3}{l}{\quad}\\
    \multicolumn{3}{l}{\quad}\\
    \multicolumn{3}{l}{\quad}\\
\multicolumn{1}{l|}{\textbf{Cluster}}& \multicolumn{1}{l|}{\textbf{Node}} & \multicolumn{1}{l}{\textbf{Node}} \\
\multicolumn{1}{l|}{\textbf{configuration}}& \multicolumn{1}{l|}{\textbf{type B}} & \multicolumn{1}{l}{\textbf{type C}} \\
\hline
1B3C & 1 & 3 \\ 
2B2C & 2 & 2 \\
3B1C & 3 & 1 \\
\end{tabular}
\captionof{table}{\label{tab:nodes} Cluster configurations\\ combining heterogeneous nodes.}
\end{minipage}
\end{wrapfigure}

To investigate research question RQ3, we need to compare fair load
distribution and response time.  For this purpose, we developed a
resource model to calculate the estimated consumption in different
scenarios.  To demonstrate the usefulness of the model, we need to
show that fairly balanced nodes lead to better performance.  Several
metrics have been used for measuring the performance of cloud
systems~\cite{DUAN2017101}. We focused our evaluation exclusively on
\emph{service response time} (i.e., the latency between service
request and response) and the corresponding \emph{system utilisation}
(i.e., the percentage of system resources that are used for service
provisioning). In the experiments, we report on the response time of
the median request and compare their latencies and resource
consumption. The median request is more meaningful than the average
response time since response time distribution is asymmetric w.r.t.\ its
average.  We can compare their latencies and resource consumption
since the computation capabilities of the cluster were never exceeded
in our experiments, all tests obtained a success rate greater than
95\%.


\subsection{Results and Discussion}\label{sec:results}

This section is organized according to research questions RQ1–-RQ3. The scripts developed to perform the experiments have been made available on GitHub.\footnote{\href{https://github.com/giaku/abs-k8s-experiments}{\tt https://github.com/giaku/abs-k8s-experiments}}

\begin{figure}[t]
\centering\footnotesize
\begin{subfigure}[b]{\linewidth}
    \begin{subfigure}[b]{0.5881\linewidth}
    \includegraphics[width=\linewidth,trim={1.05cm 1.2cm .3cm .3cm},clip]{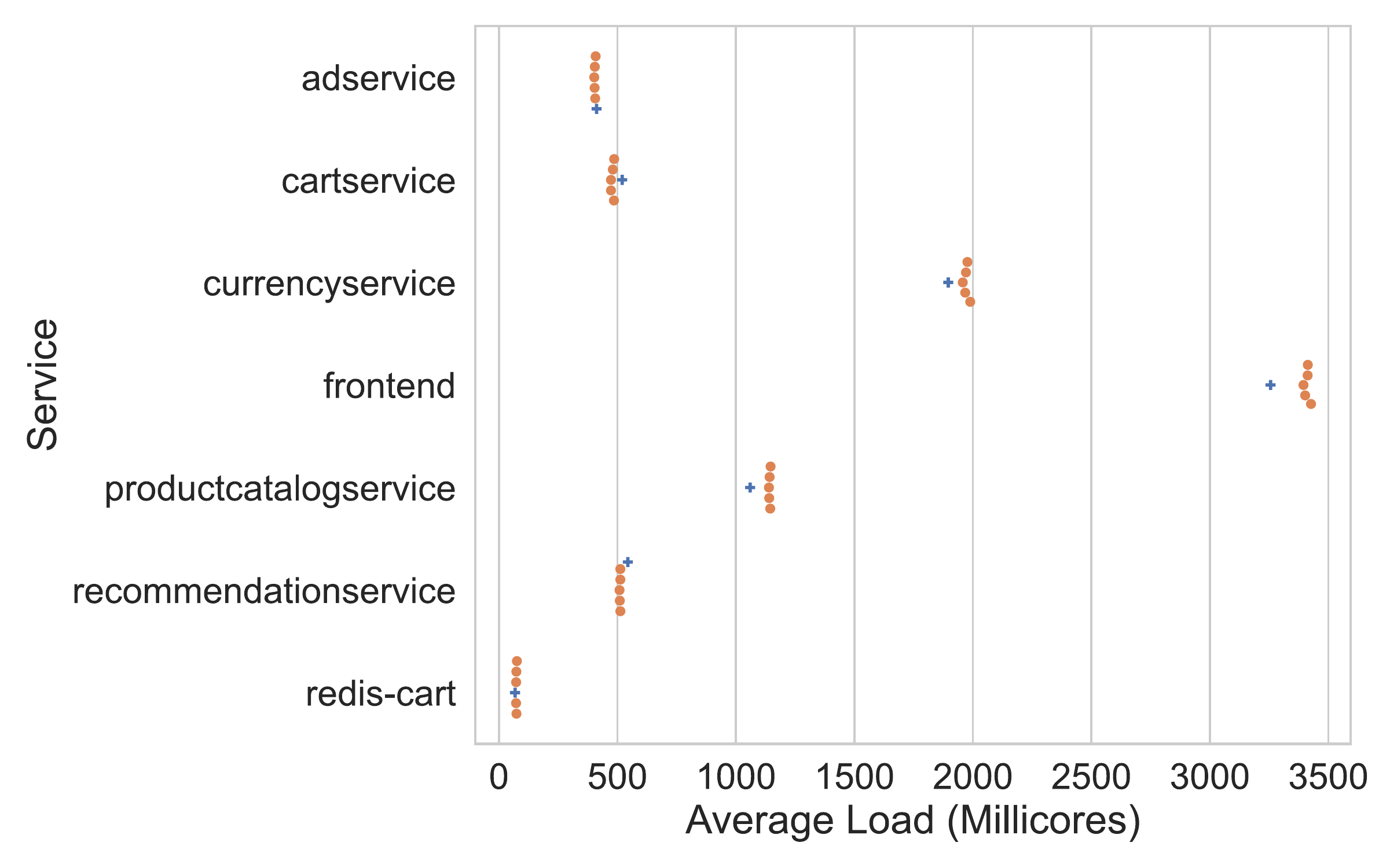}
    \hspace*{.3\linewidth}P1 Average Load (Millicores)
  \end{subfigure}
  \hfill
  \begin{subfigure}[b]{0.404\linewidth}
    \includegraphics[width=\linewidth,trim={8.15cm 1.2cm .3cm .2cm},clip]{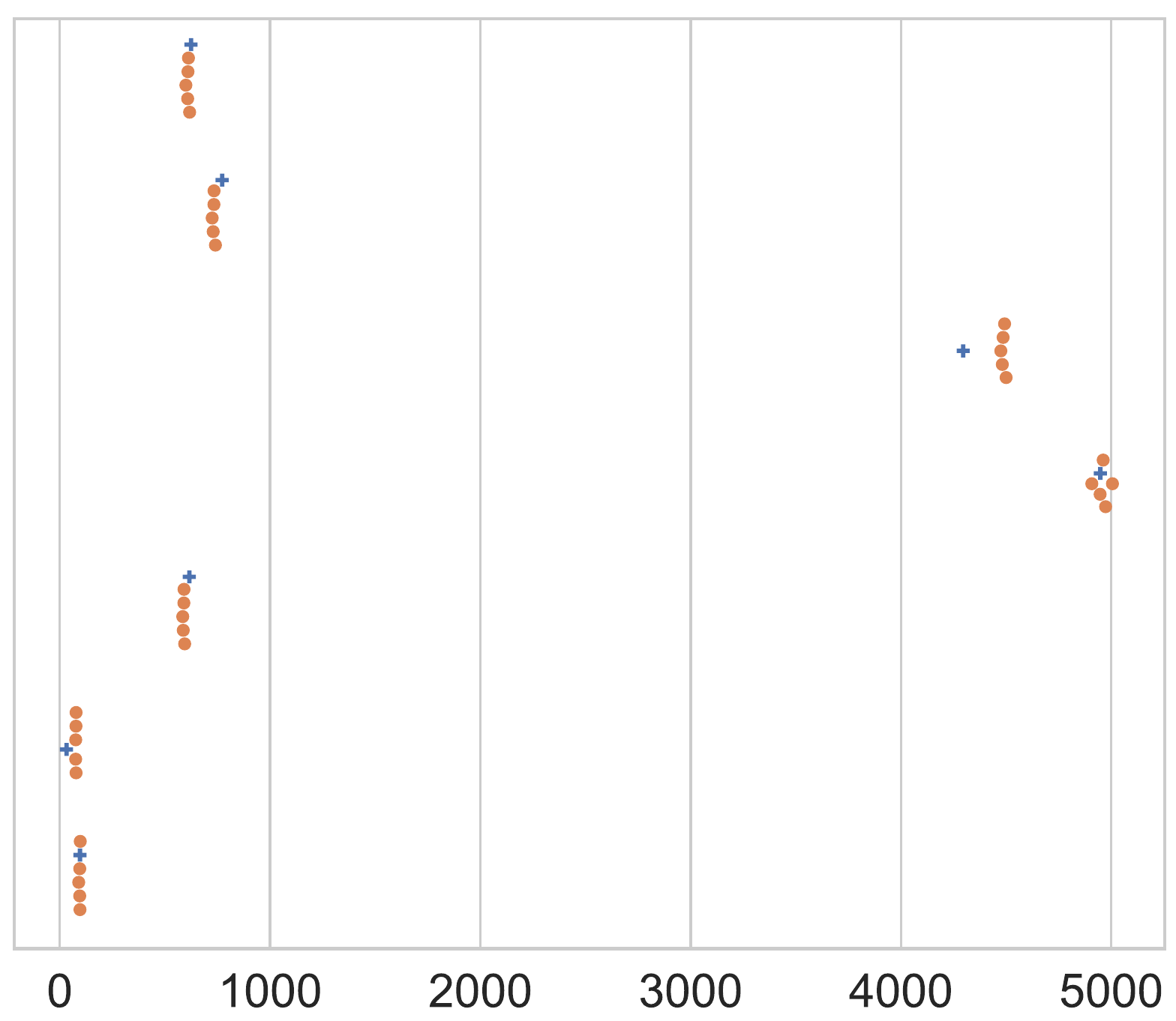}
      \hspace*{.12\linewidth}P2 Average Load (Millicores)
  \end{subfigure}
  \begin{subfigure}[b]{0.528\linewidth}
    \includegraphics[width=\linewidth,trim={1.05cm 1.2cm .45cm .3cm},clip]{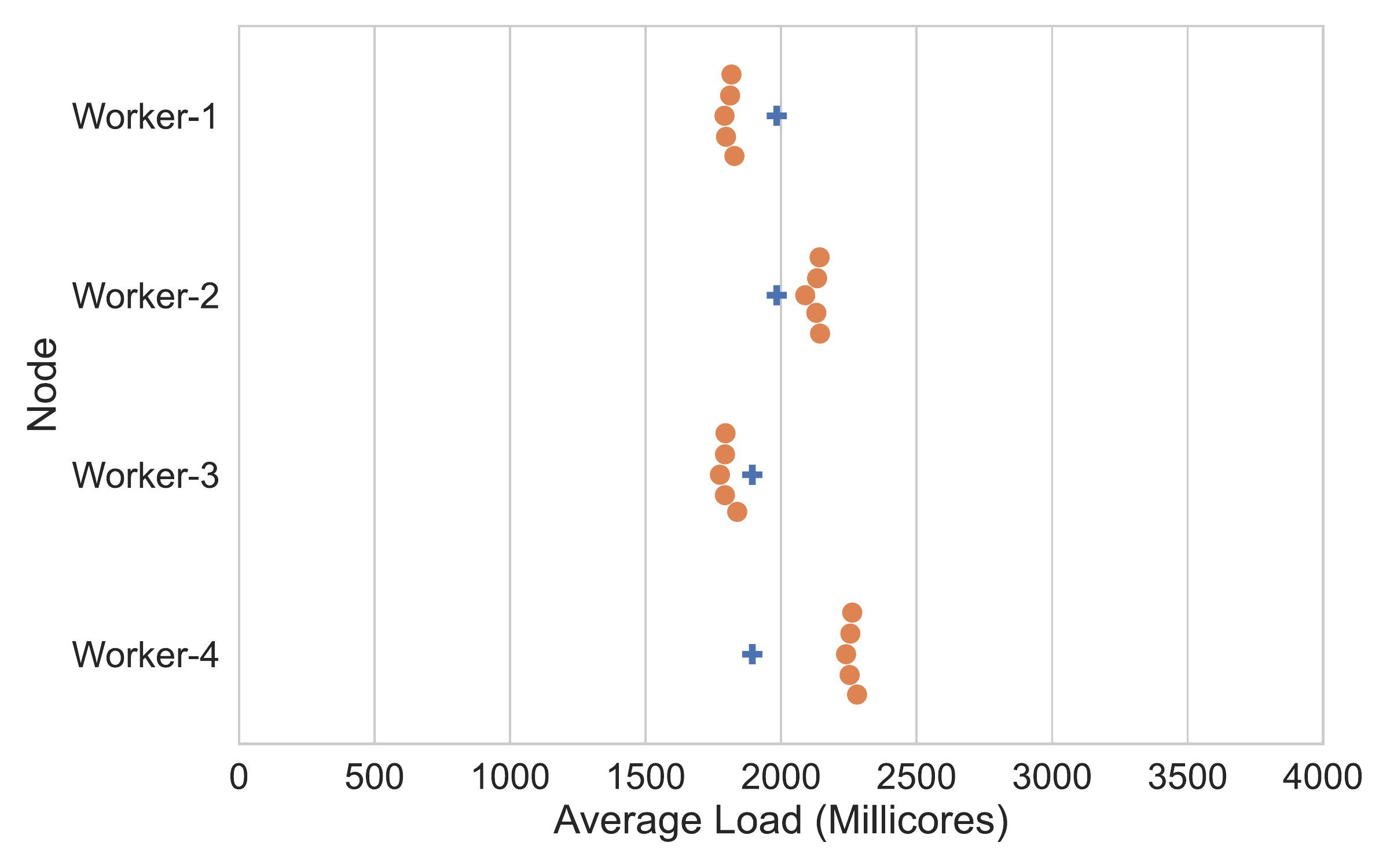}
    \hspace*{.33\linewidth}P1 Average Load (Millicores)
  \end{subfigure}
  \hfill
  \begin{subfigure}[b]{0.4605\linewidth}
    \includegraphics[width=\linewidth,trim={3.95cm 1.2cm .45cm .3cm},clip]{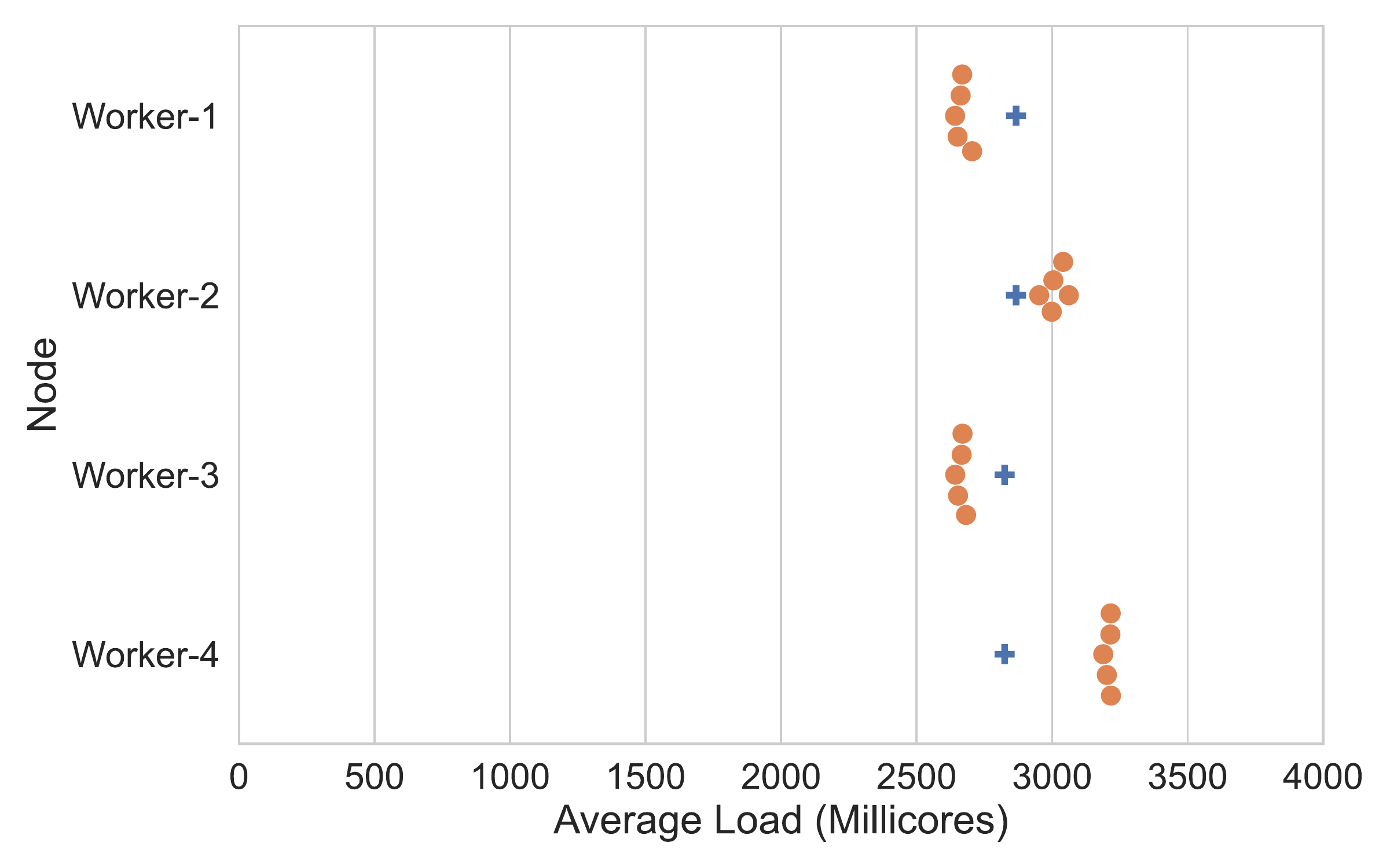}
    \hspace*{.23\linewidth}P2 Average Load (Millicores)
  \end{subfigure}
 \end{subfigure}
 \caption{\label{fig:pairs1}Measurements for the mixed workflows
   P1 and P2, comparing the resource consumption recorded in the
   cluster stress tests (orange dots) with the expected values given by
   the resource model (blue plus). The plots are reported by service
   (top), and by node (bottom).}
\end{figure}

\begin{figure}[t]
\centering\footnotesize
  \begin{subfigure}[b]{\linewidth}
  \hfill
  \begin{subfigure}[b]{0.5881\linewidth}
    \includegraphics[width=\linewidth,trim={1.05cm 1.2cm .3cm .3cm},clip]{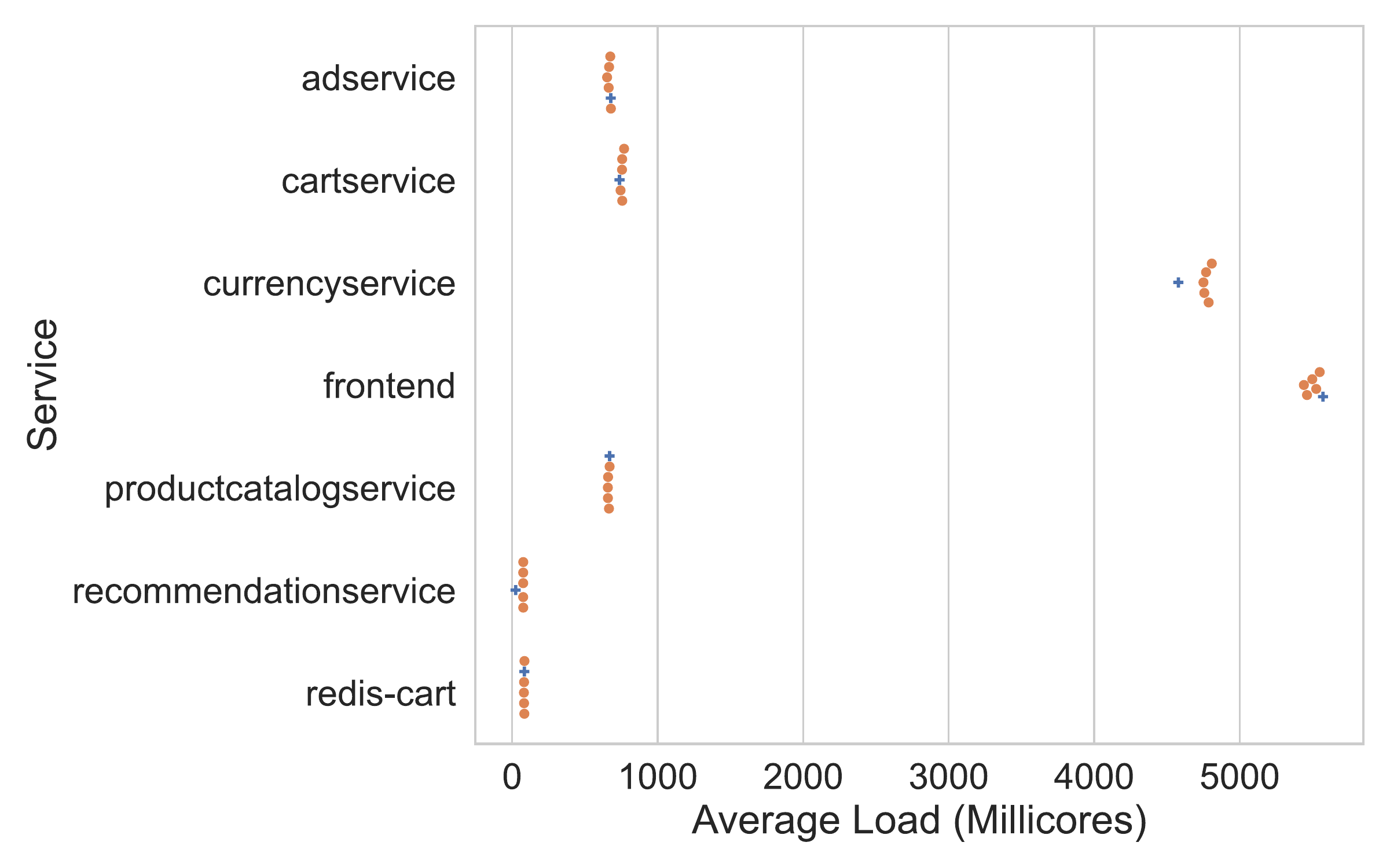}
    \hspace*{.3\linewidth}P3 Average Load (Millicores)
  \end{subfigure}
  \hfill
  \begin{subfigure}[b]{0.404\linewidth}
    \includegraphics[width=\linewidth,trim={8.15cm 1.2cm .3cm .2cm},clip]{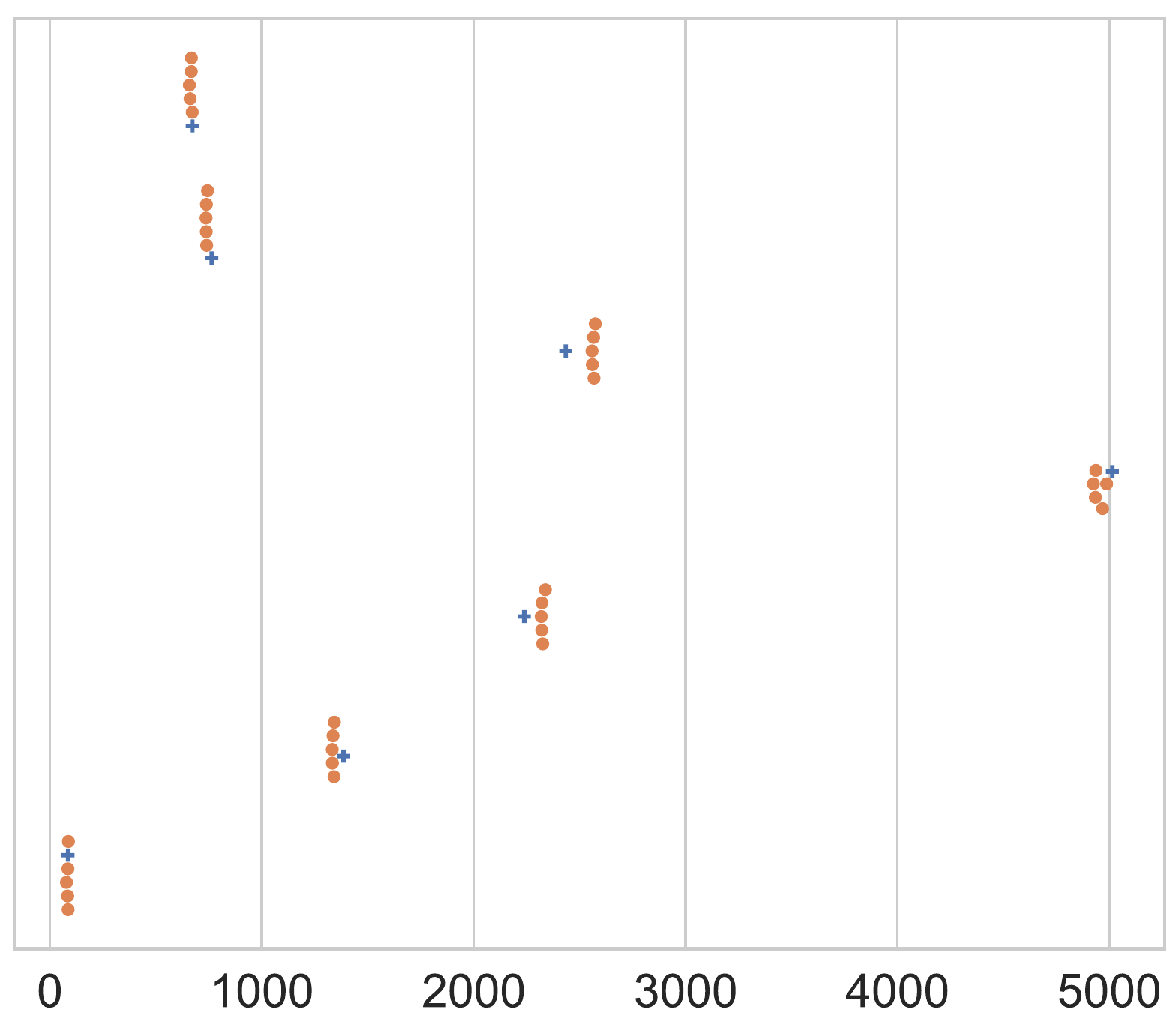}
    \hspace*{.12\linewidth}P4 Average Load (Millicores)
  \end{subfigure}
  \begin{subfigure}[b]{0.528\linewidth}
    \includegraphics[width=\linewidth,trim={1.05cm 1.2cm .45cm .3cm},clip]{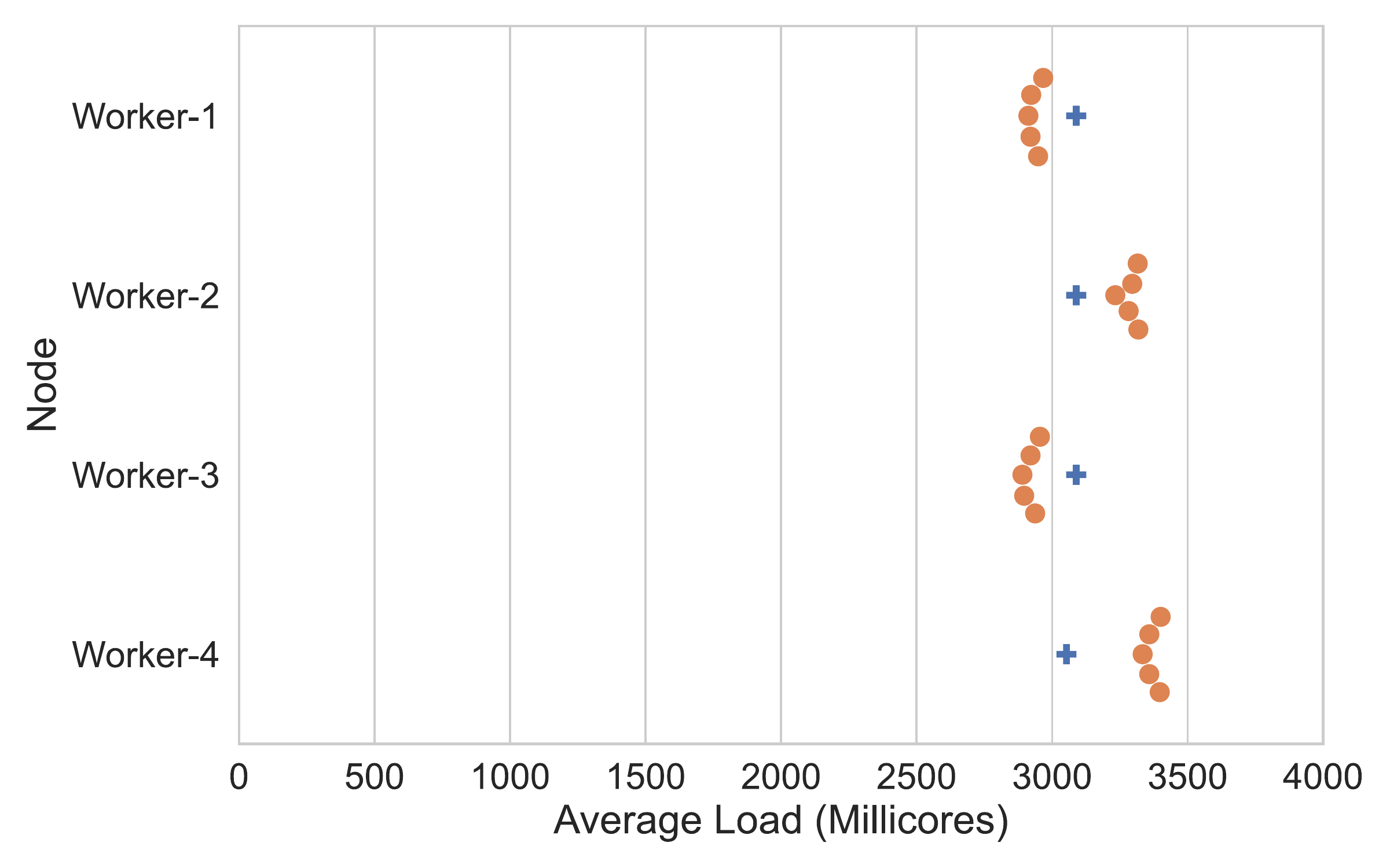}
  \hspace*{.33\linewidth}P3 Average Load (Millicores)
  \end{subfigure}
  \hfill
  \begin{subfigure}[b]{0.4605\linewidth}
    \includegraphics[width=\linewidth,trim={3.95cm 1.2cm .45cm .3cm},clip]{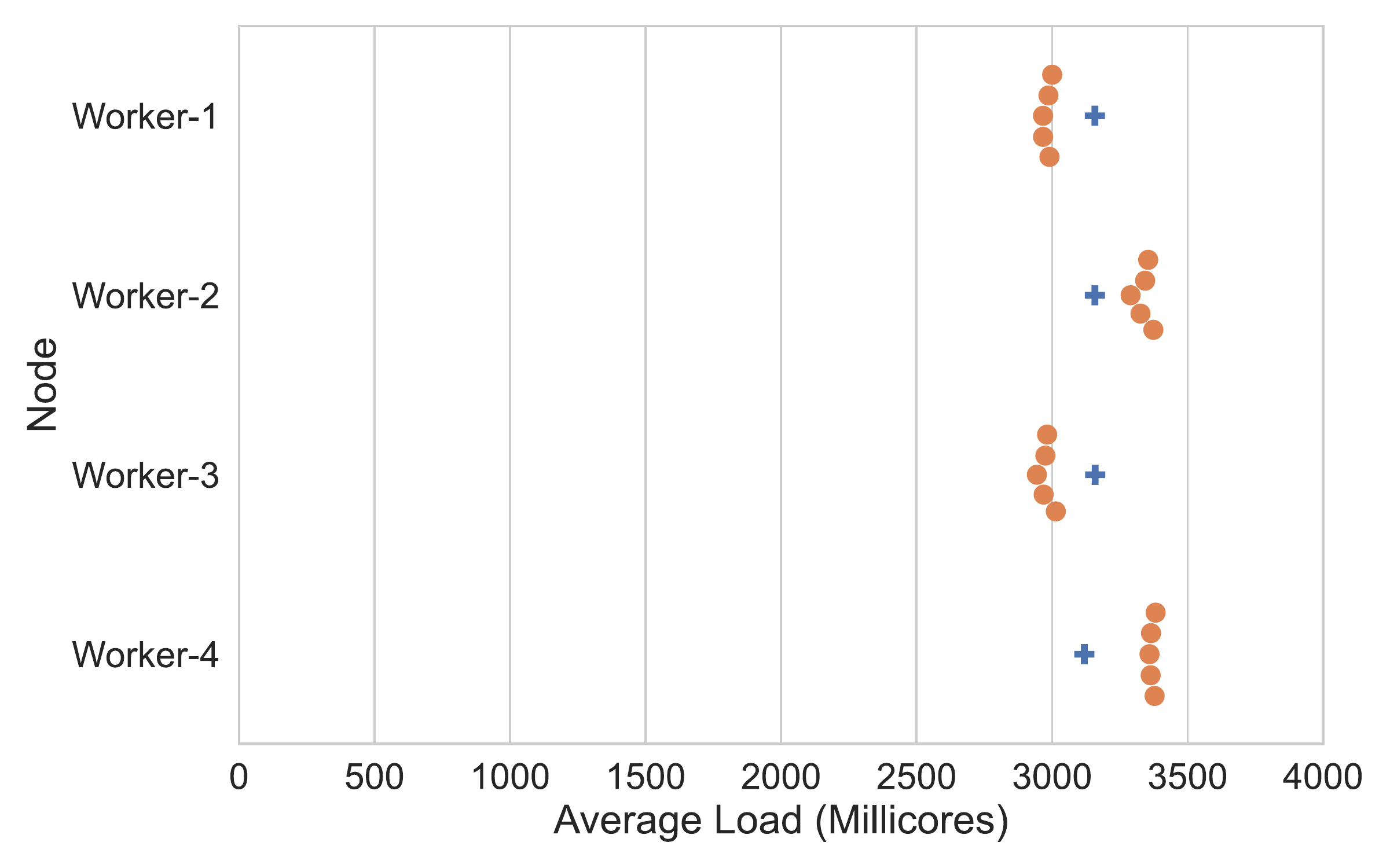}
  \hspace*{.23\linewidth}P4 Average Load (Millicores)
  \end{subfigure}
\end{subfigure}
\caption{\label{fig:pairs2}Measurements for the mixed workflows P3 and
  P4, comparing the resource consumption recorded in the cluster
  stress tests (orange dots) with the expected values given by the
  resource model (blue plus). The plots are reported by service (top),
  and by node (bottom).}
\end{figure}

\begin{figure}[t]
\begin{subfigure}[b]{\linewidth}
\centering\footnotesize
    \begin{subfigure}[b]{0.5881\linewidth}
    \includegraphics[width=\linewidth,trim={1.05cm 1.2cm .3cm .3cm},clip]{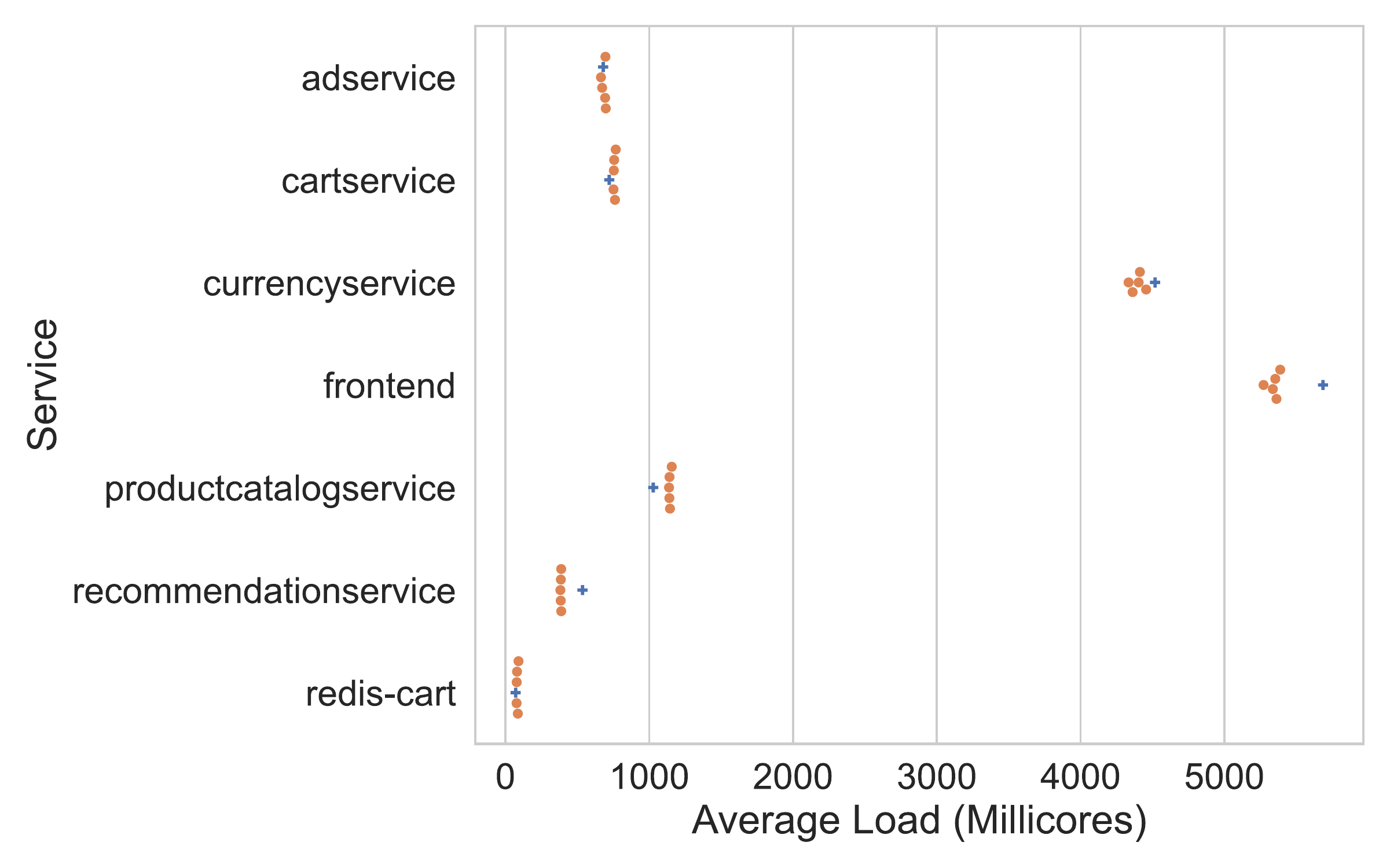}
    \hspace*{.3\linewidth}T1 Average Load (Millicores)
  \end{subfigure}
  \hfill
  \begin{subfigure}[b]{0.404\linewidth}
    \includegraphics[width=\linewidth,trim={8.15cm 1.2cm .3cm .2cm},clip]{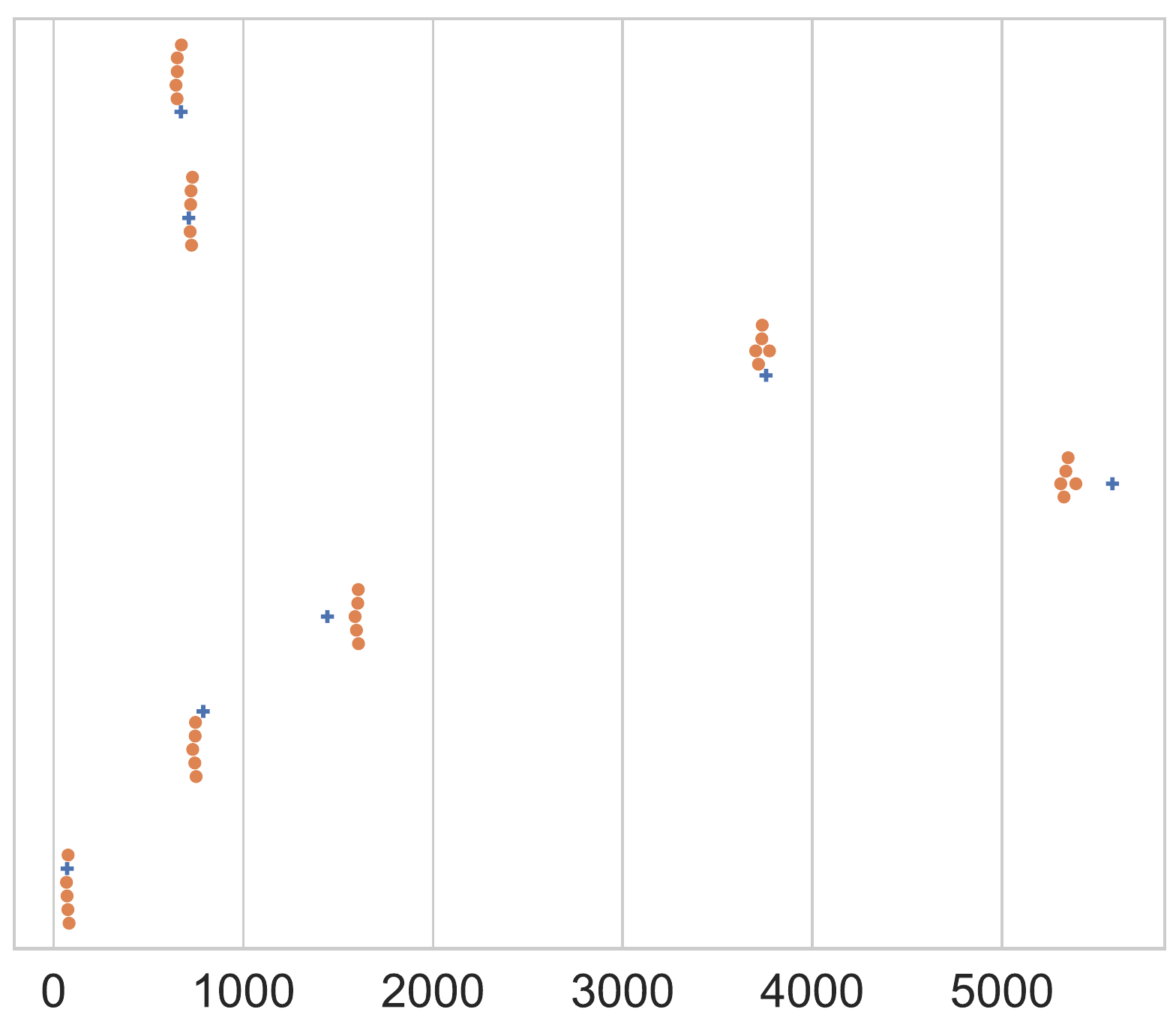}
    \hspace*{.12\linewidth}T2 Average Load (Millicores)
  \end{subfigure}
  \hfill
  \begin{subfigure}[b]{0.528\linewidth}
    \includegraphics[width=\linewidth,trim={1.05cm 1.2cm .45cm .3cm},clip]{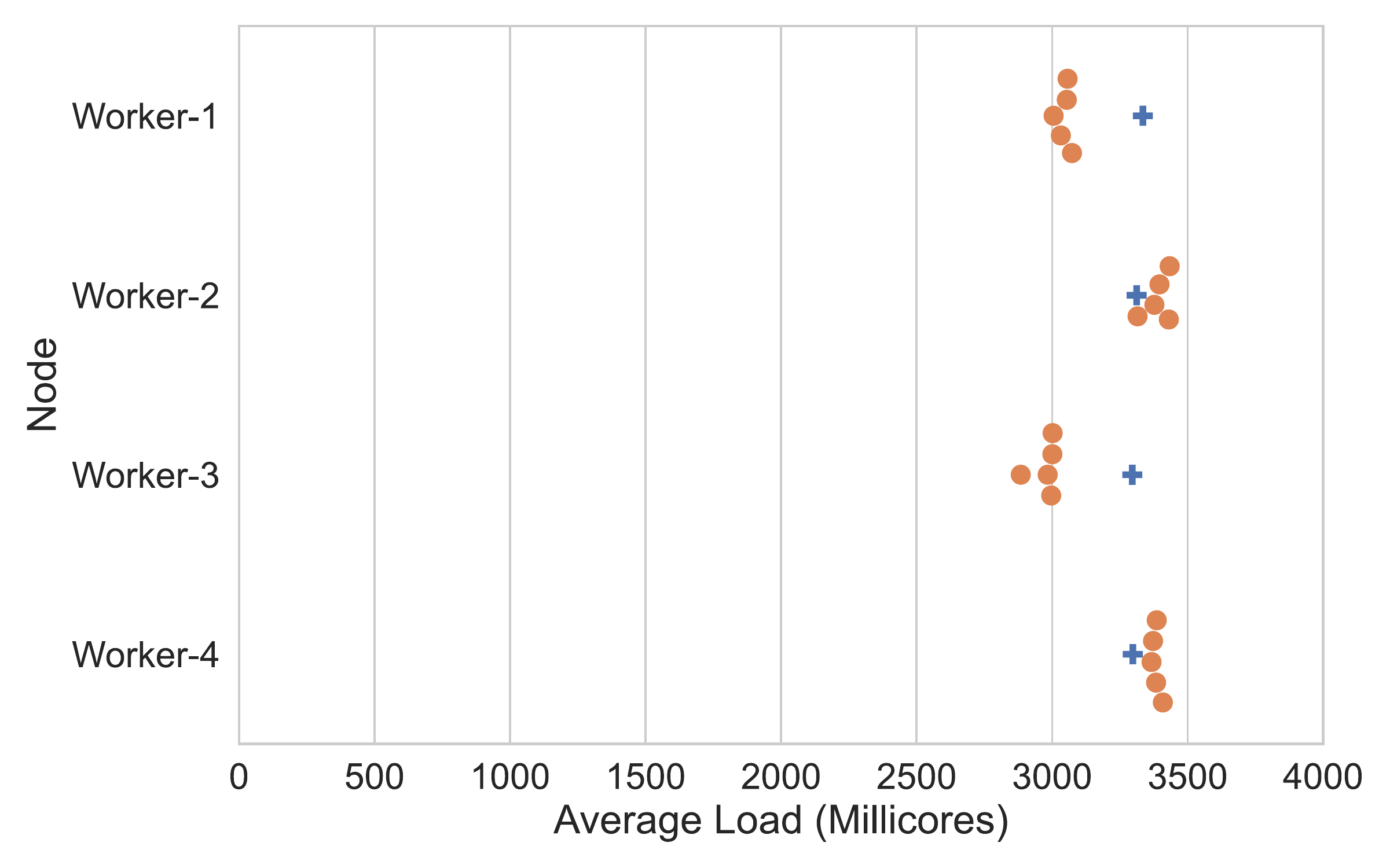}
  \hspace*{.33\linewidth}T1 Average Load (Millicores)
  \end{subfigure} 
  \hfill
  \begin{subfigure}[b]{0.4605\textwidth}
    \includegraphics[width=\linewidth,trim={3.95cm 1.2cm .45cm .3cm},clip]{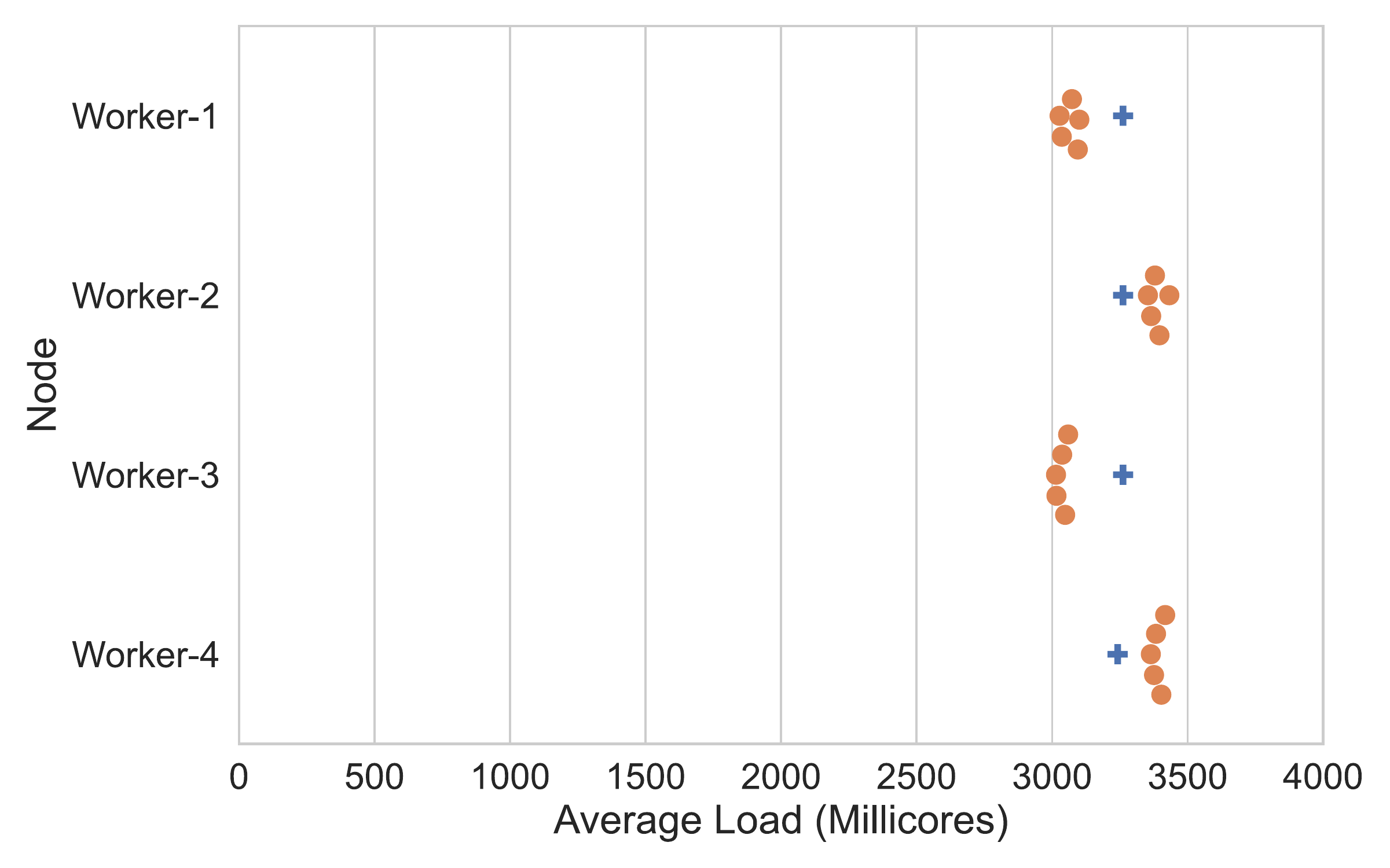}
  \hspace*{.23\linewidth}T2 Average Load (Millicores)
  \end{subfigure}
\end{subfigure}
\caption{\label{fig:triplets1}Measurements for the mixed workflows
  T1 and T2, comparing the resource consumption recorded in the
  cluster stress tests (orange dots) with the expected values given by
  the resource model (blue plus). The plots are reported by service
  (top), and by node (bottom).}
\end{figure}

\begin{figure}[t]
\begin{subfigure}[b]{\linewidth}
\centering\footnotesize
    \begin{subfigure}[b]{0.5881\linewidth}
    \includegraphics[width=\linewidth,trim={1.05cm 1.2cm .3cm .3cm},clip]{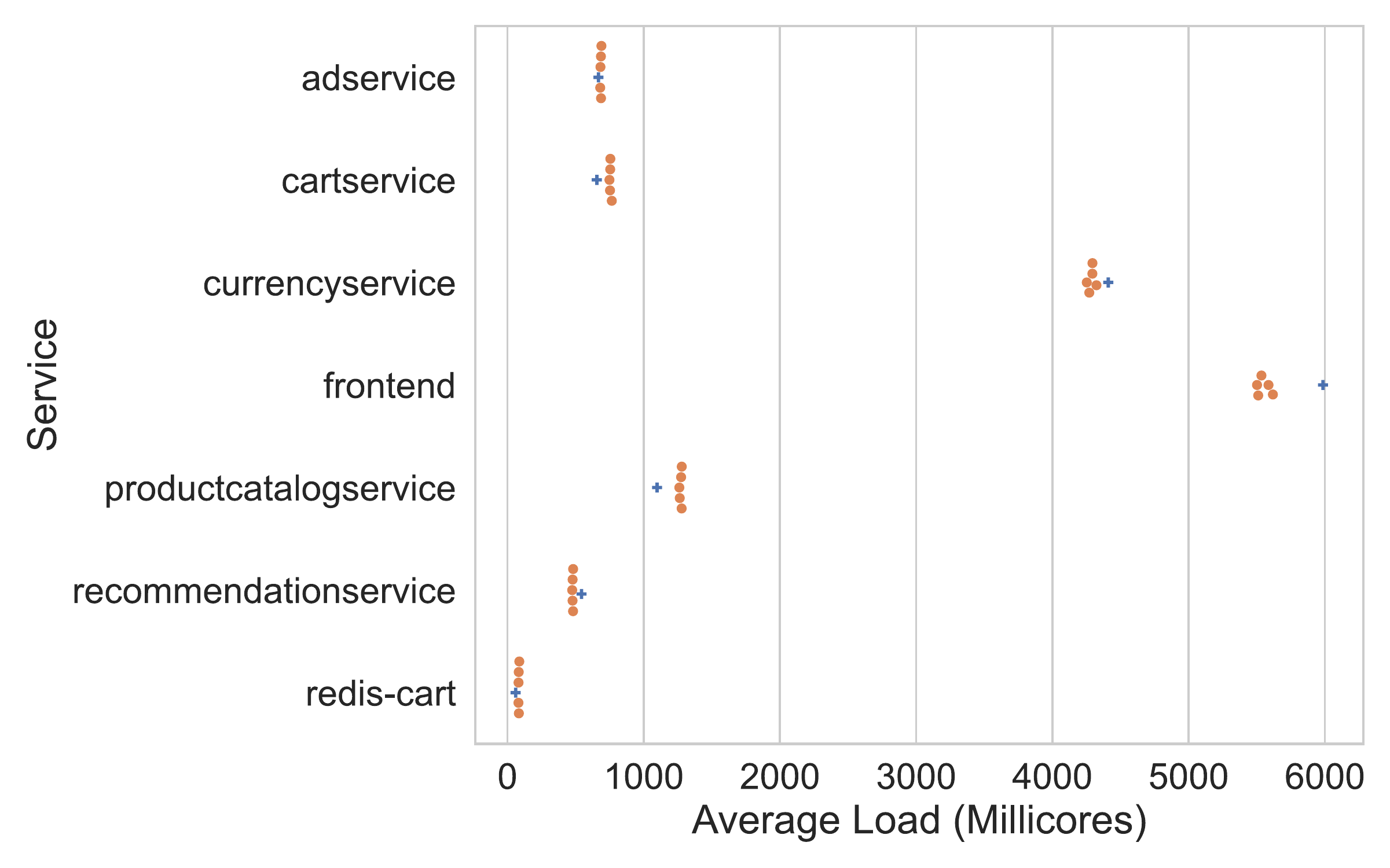}
    \hspace*{.3\linewidth}T3 Average Load (Millicores)
  \end{subfigure}
  \hfill
  \begin{subfigure}[b]{0.404\linewidth}
    \includegraphics[width=\linewidth,trim={8.15cm 1.2cm .3cm .2cm},clip]{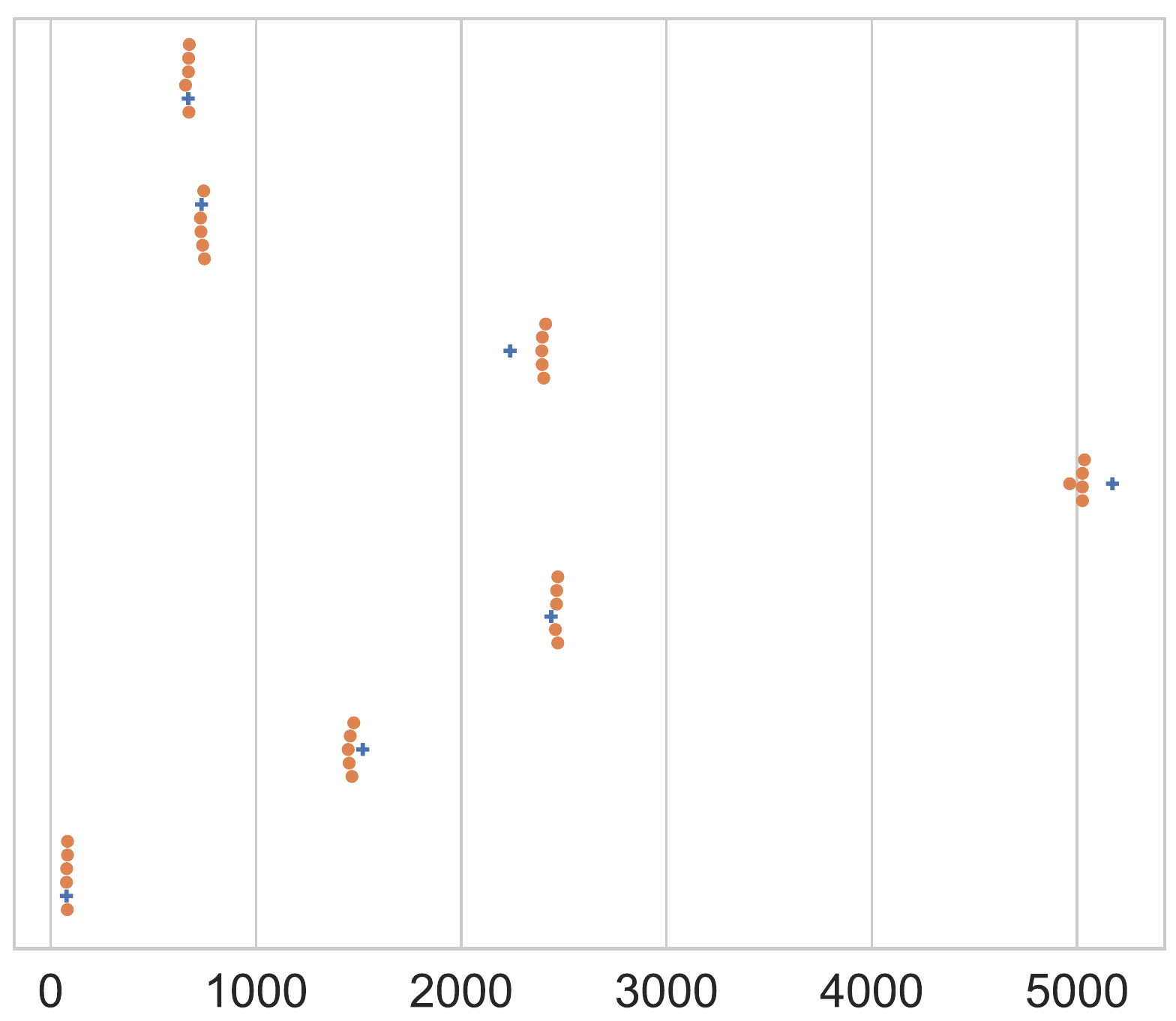}
    \hspace*{.12\linewidth}T4 Average Load (Millicores)
  \end{subfigure}
  \hfill
  \begin{subfigure}[b]{0.528\linewidth}
    \includegraphics[width=\linewidth,trim={1.05cm 1.2cm .45cm .3cm},clip]{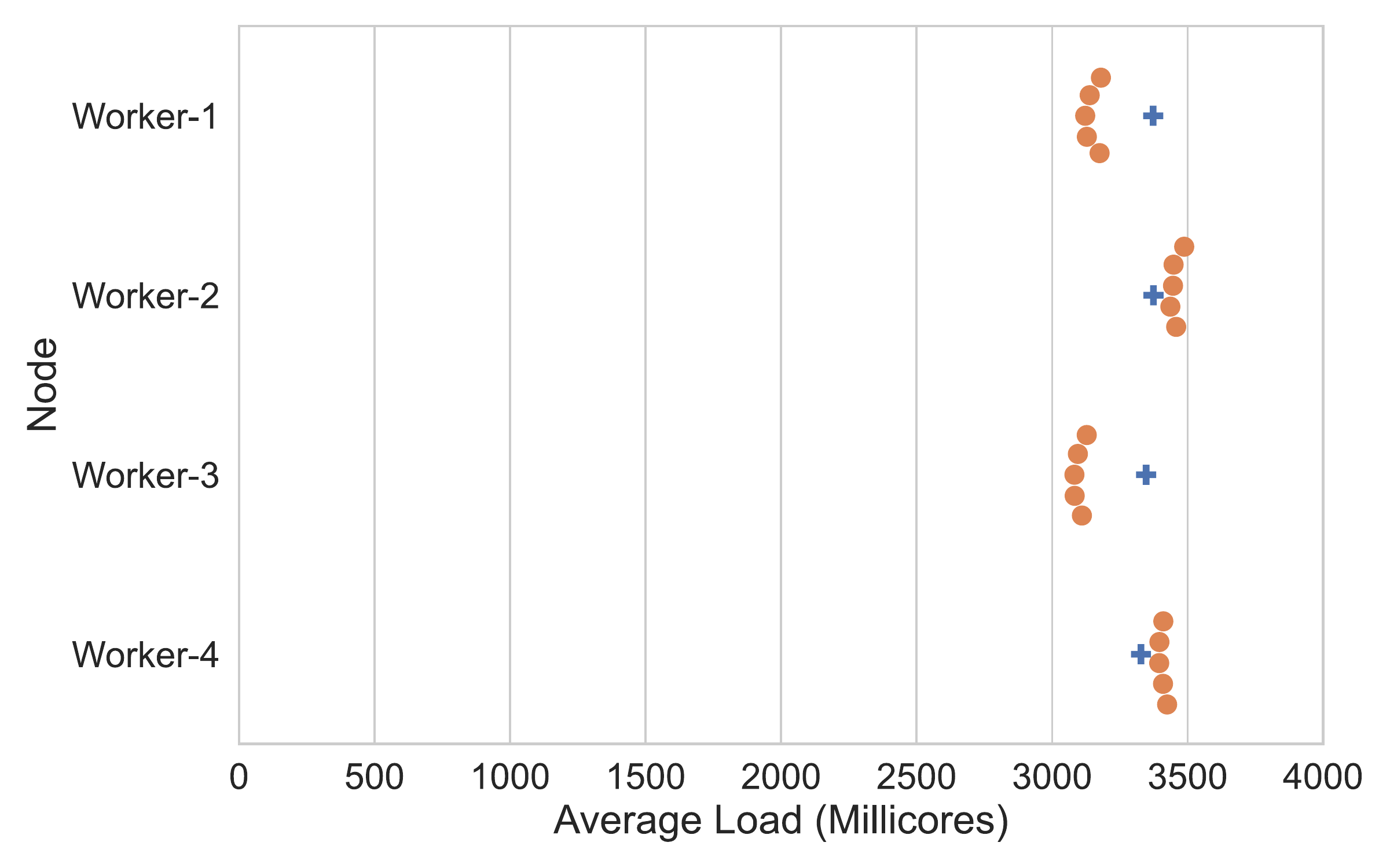}
    \hspace*{.33\linewidth}T3 Average Load (Millicores)
  \end{subfigure}
  \hfill
  \begin{subfigure}[b]{0.4605\linewidth}
    \includegraphics[width=\linewidth,trim={3.95cm 1.2cm .45cm .3cm},clip]{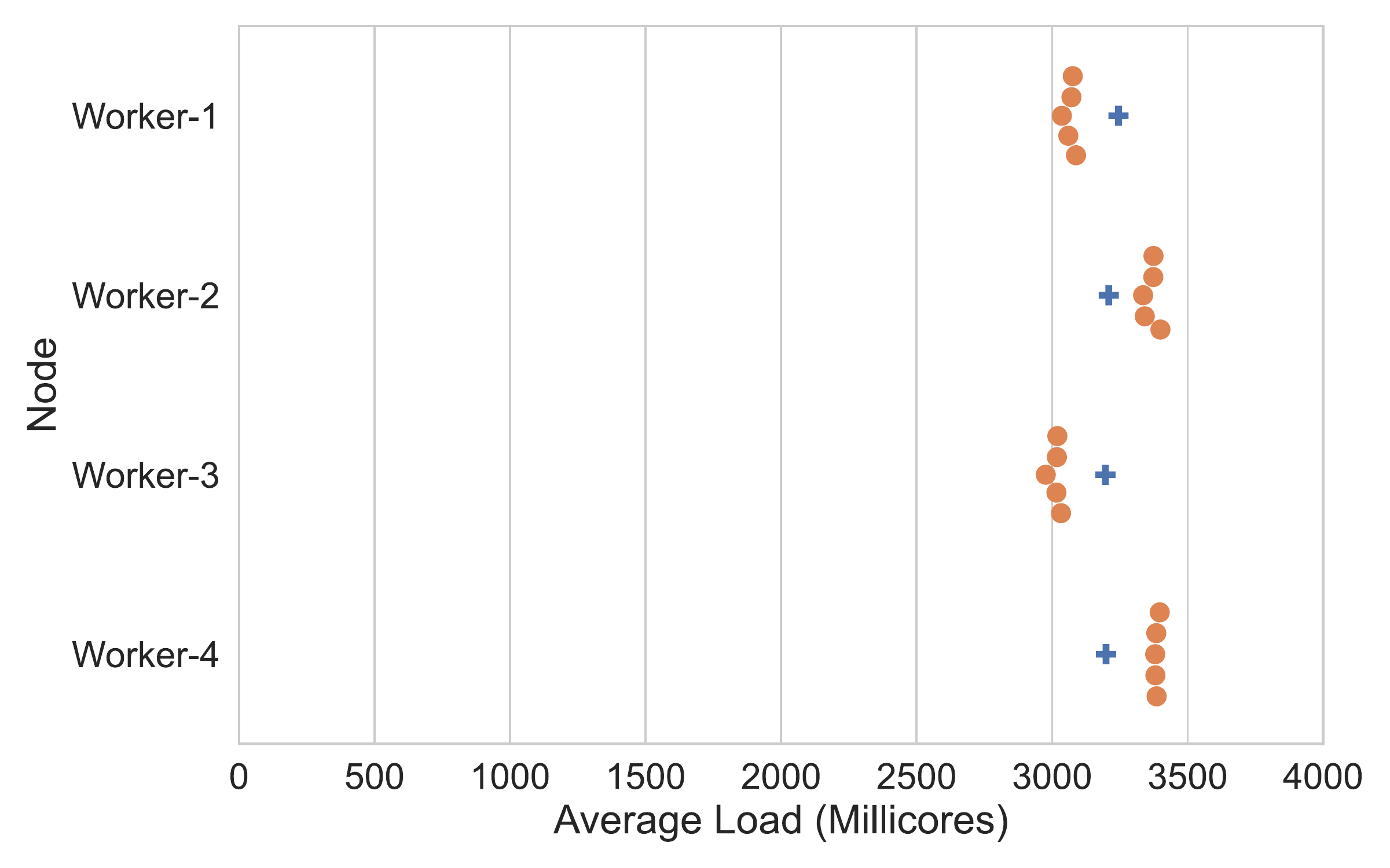}
   \hspace*{.23\linewidth}T4 Average Load (Millicores)
  \end{subfigure}
\end{subfigure}
\caption{\label{fig:triplets2}Measurements for the mixed workflows
  T3 and T4, comparing the resource consumption recorded in the
  cluster stress tests (orange dots) with the expected values given by
  the resource model (blue plus). The plots are reported by service
  (top), and by node (bottom).}
\end{figure}

\paragraph{RQ1}
Figures~\ref{fig:pairs1}--\ref{fig:triplets2} compare measurements for
the five iterations of the stress tests to the corresponding model
predictions for the workflows specified in Table~\ref{tab:rq1}.  In
the figures, consumption is grouped by service in the first row
of plots, and by node in the second row.  Figure~\ref{fig:pairs1}
considers the mixed workflows P1 and P2, Figure~\ref{fig:pairs2}
considers P3 and P4, Figure~\ref{fig:triplets1} considers T1 and T2,
and Figure~\ref{fig:triplets2} considers T3 and T4.

When we look at the consumption by service, the model's prediction is
well aligned with the consumption observed on the cluster. Service
consumption is always very close to the measured outcome.  The model's
largest divergence can be observed for service consumption in workflow
T3 (see Figure~\ref{fig:triplets2}, top left).  For the service
\emph{frontend}, the model predicts a consumption of 6000 millicores
and the system consumes on average 5500, accounting for an
overestimation of 10\%.  This can be explained by the fact that when
the total load on the cluster brings the nodes close to saturation,
the behaviour of the system becomes unpredictable. Resources cannot be
consumed in excess of their availability, and resources are also
needed for the Kubernetes internals.  Furthermore, the real system
degrades performance and some requests fail in order to keep the pace,
while the model can consume every single millicore of CPU.


When we consider the consumption by node, we observe a slightly
unbalanced distribution among the worker nodes in the real system.
Since all nodes are a priori equal, the model predicts the same load
for every node, but the real Kubernetes cluster does not. The reasons
for this difference lie in the load balancing problem discussed in
Section~\ref{sec:k8s}: When the kube-proxies distribute gRPC requests,
they fail to achieve a fair load distribution.  This problem is
addressed by equipping every pod with a sidecar pod that works as an
additional High Availability proxy; these sidecars are provided by
Istio and called Istio-proxies. After the initial distribution of
requests by the Kubernetes system, the Istio-proxies carry out a final
rerouting of the requests.  However, the consumption of these proxies
is not considered part of the system consumption; instead it is part
of the pod consumption.  Consequently, the consumption of a frontend
service is measured together with its sidecar Istio-proxy in the
sampling phase and then replicated in the simulations. However, the
work of the Istio-proxies at different places in the cluster is not
fairly balanced. Some Istio-proxies are redirecting more requests than
others, because the kube-proxies target them more heavily. This lack
of balance does not affect the experiments for RQ1 because the
consumption per service is quite accurate and the consumption per node
is accurate when considering average values.

\paragraph{RQ2}

\begin{figure}[t]
\centering\footnotesize
  \begin{minipage}{0.268\linewidth}\centering
    \includegraphics[width=\linewidth,trim={1.2cm .4cm .3cm .3cm},clip]{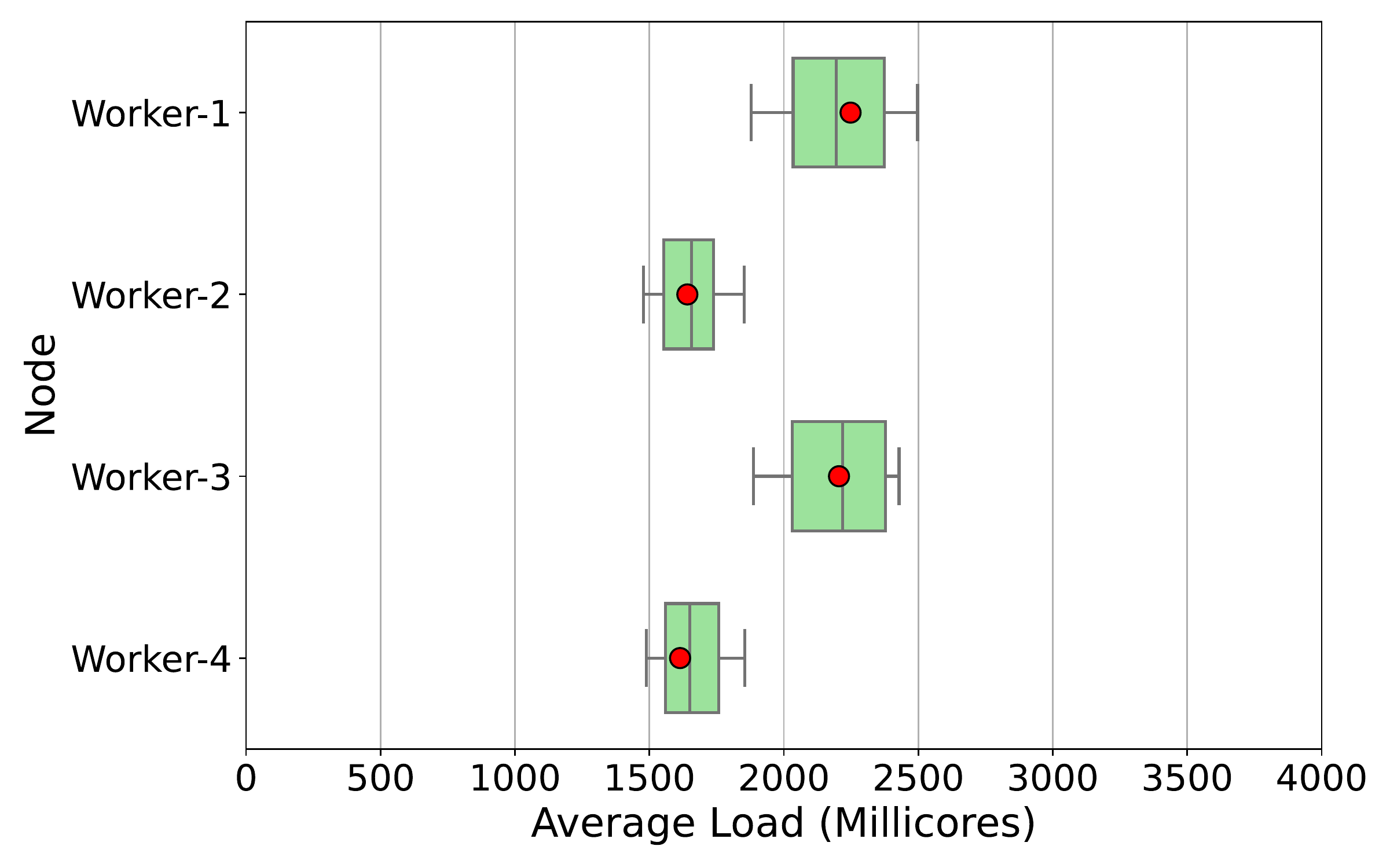}\\
Workflow P1
  \end{minipage}
  \hfill
  \begin{minipage}{0.2344\linewidth}\centering
    \includegraphics[width=\linewidth,trim={4.1cm .4cm .3cm .3cm},clip]{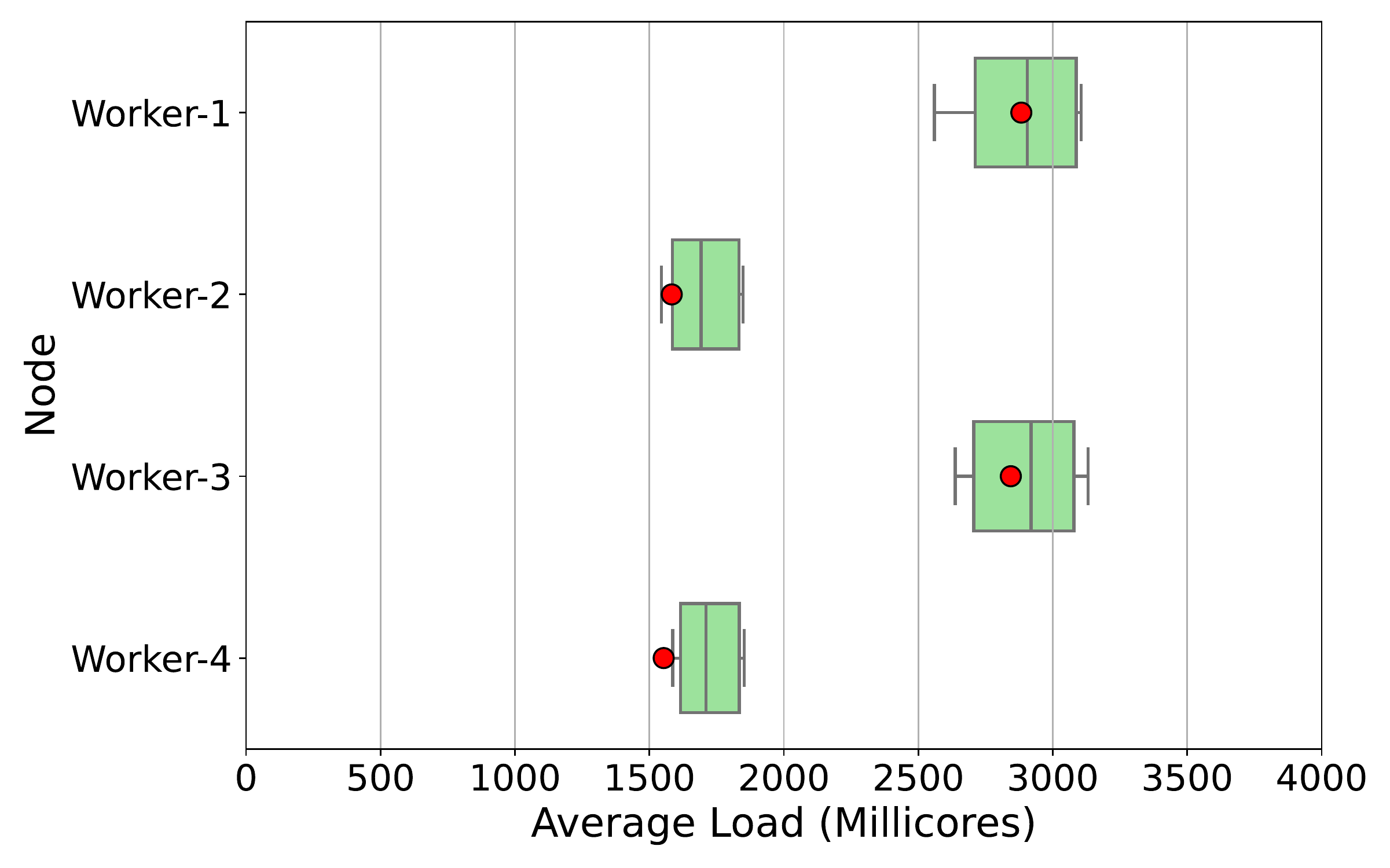}\\
Workflow P2
  \end{minipage}
  \hfill
  \begin{minipage}{0.2344\linewidth}\centering
    \includegraphics[width=\linewidth,trim={4.1cm .4cm .3cm .3cm},clip]{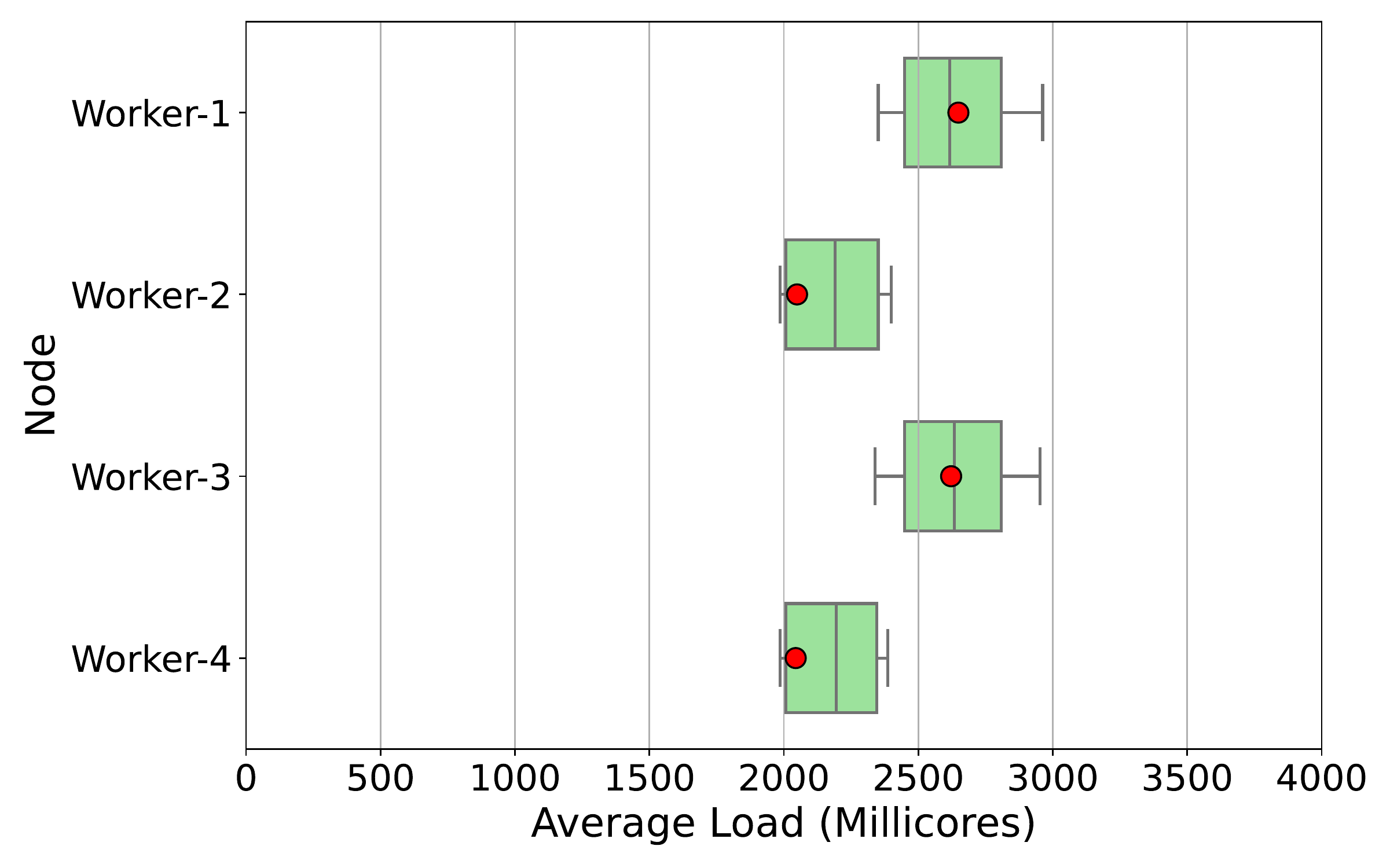}\\
Workflow P3
  \end{minipage}
  \hfill
  \begin{minipage}{0.2344\linewidth}\centering
    \includegraphics[width=\linewidth,trim={4.1cm .4cm .3cm .3cm},clip]{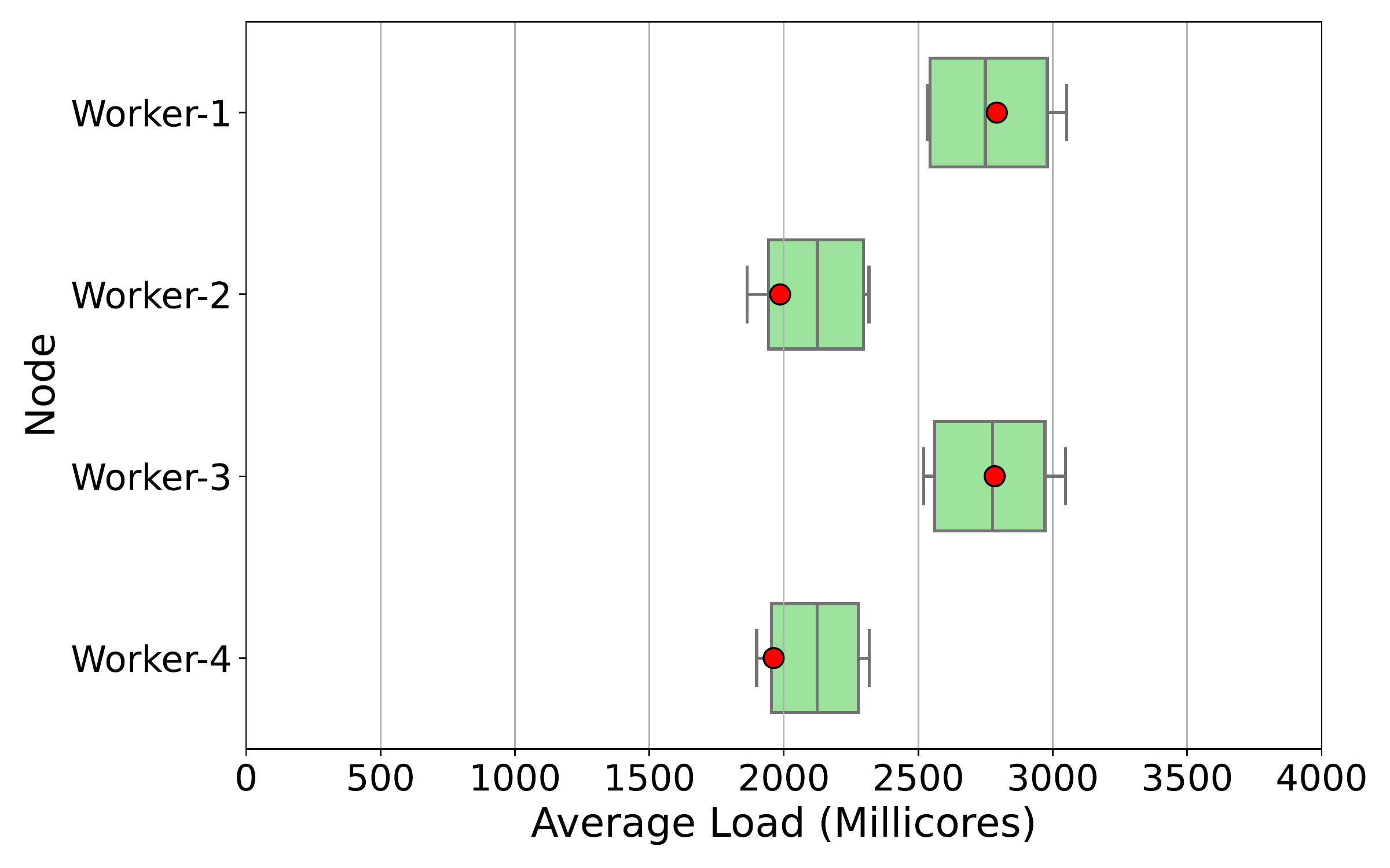}\\
Workflow P4
  \end{minipage}

\medskip

  \begin{minipage}{0.268\linewidth}\centering
    \includegraphics[width=\linewidth,trim={1.2cm .4cm .3cm .3cm},clip]{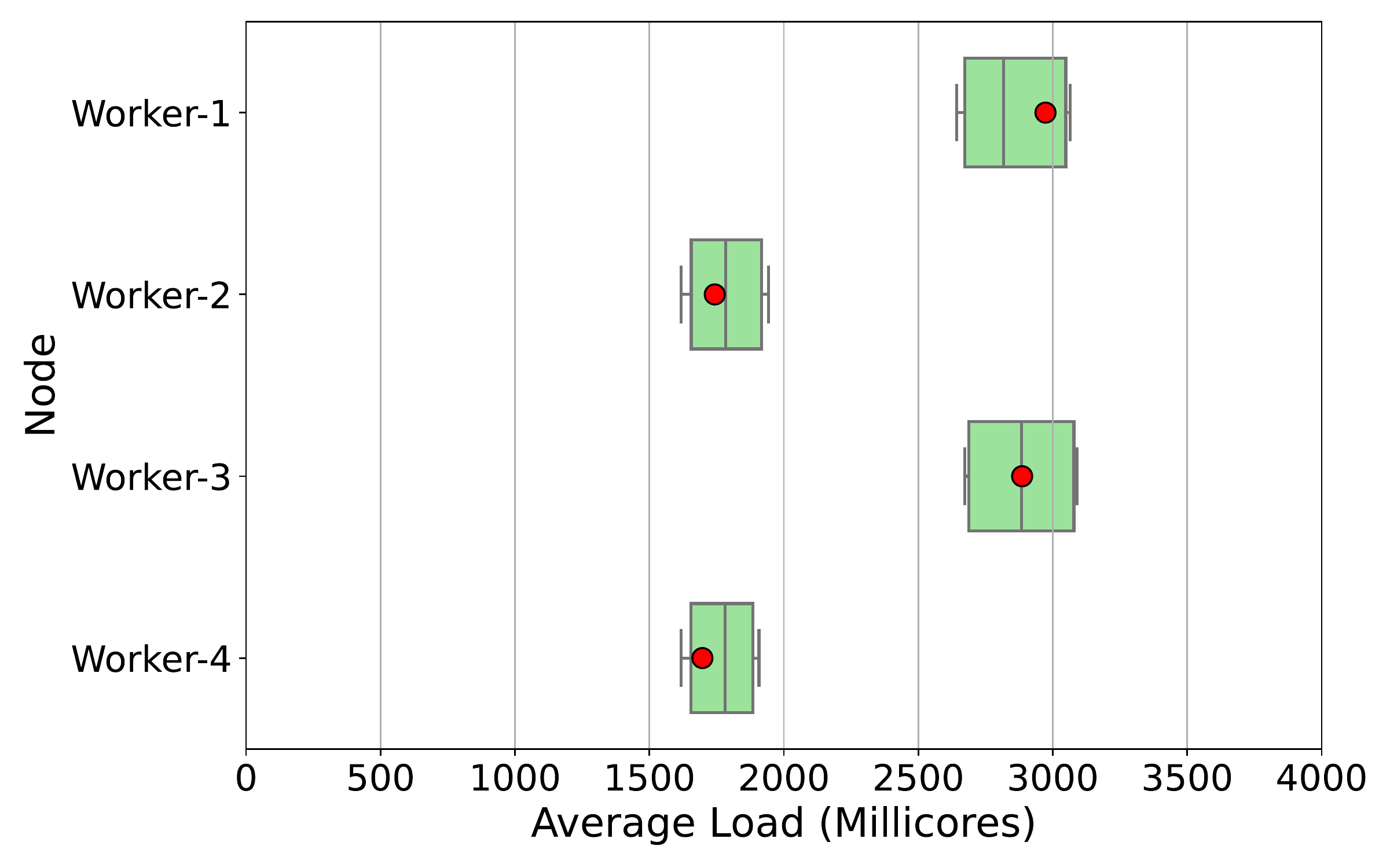}\\
Workflow T1
  \end{minipage}
  \hfill
  \begin{minipage}{0.2344\linewidth}\centering
    \includegraphics[width=\linewidth,trim={4.1cm .4cm .3cm .3cm},clip]{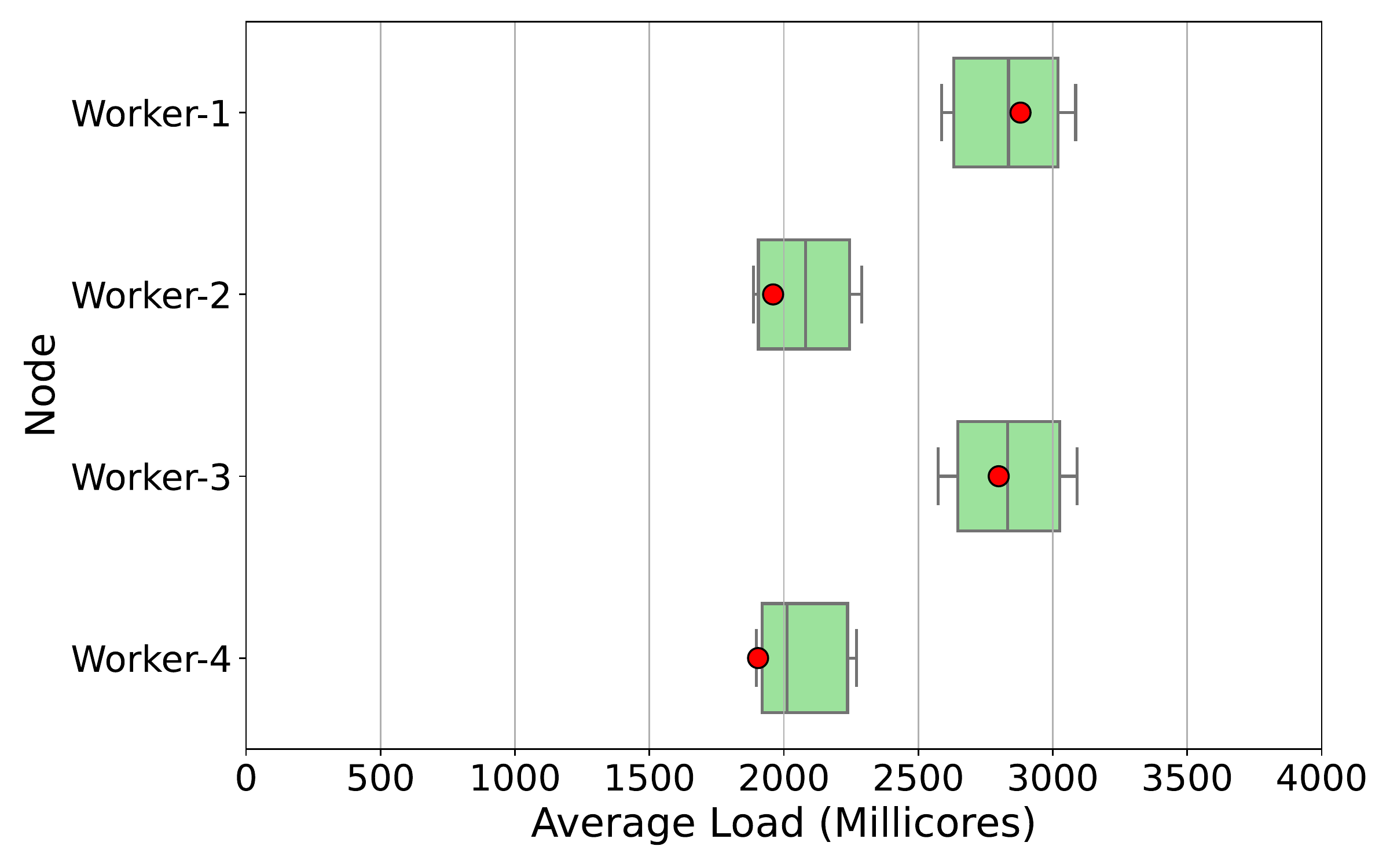}\\
Workflow T2
  \end{minipage}
  \hfill
  \begin{minipage}{0.2344\linewidth}\centering
    \includegraphics[width=\linewidth,trim={4.1cm .4cm .3cm .3cm},clip]{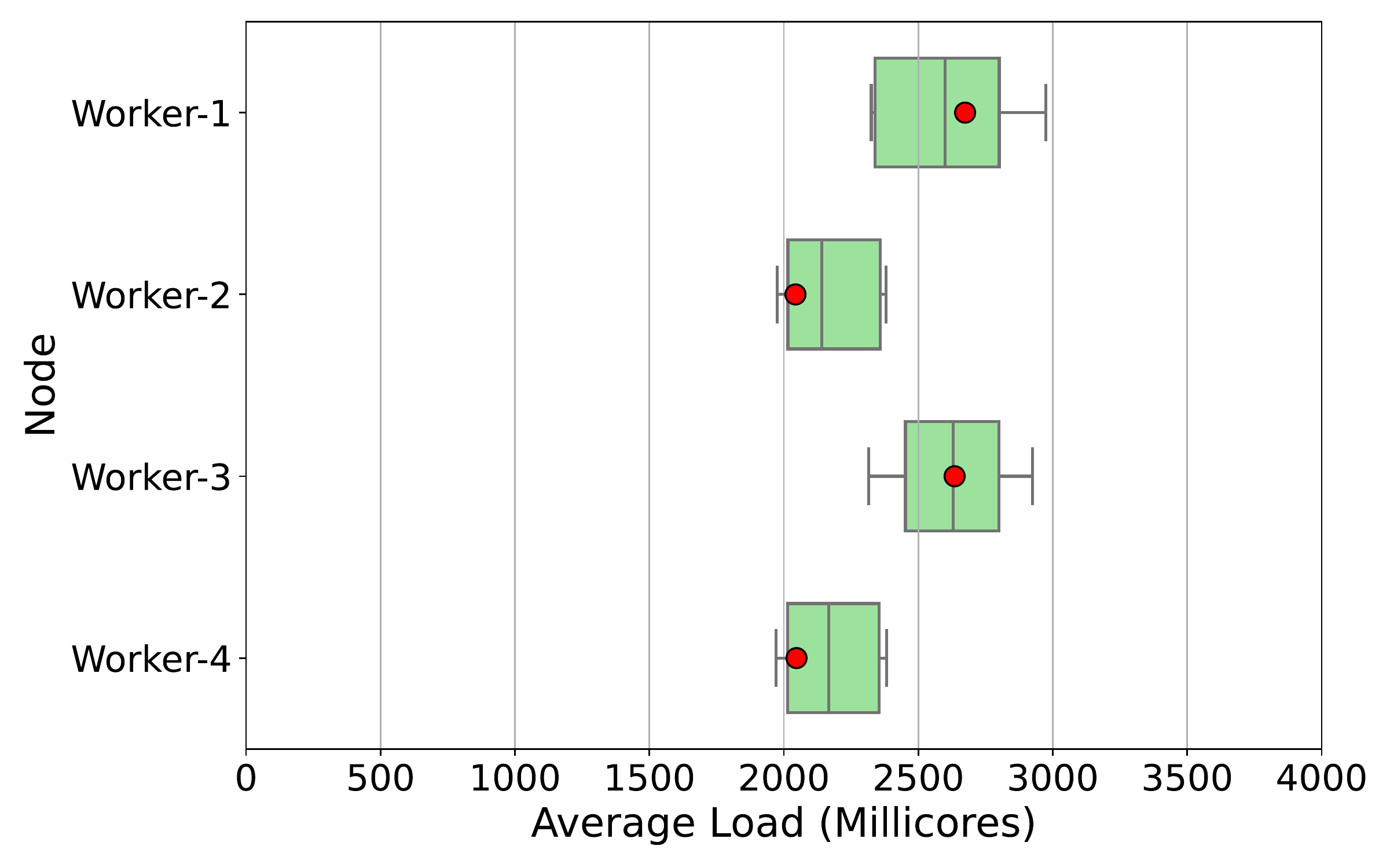}\\
Workflow T3
  \end{minipage}
  \hfill
  \begin{minipage}{0.2344\linewidth}\centering
    \includegraphics[width=\linewidth,trim={4.1cm .4cm .3cm .3cm},clip]{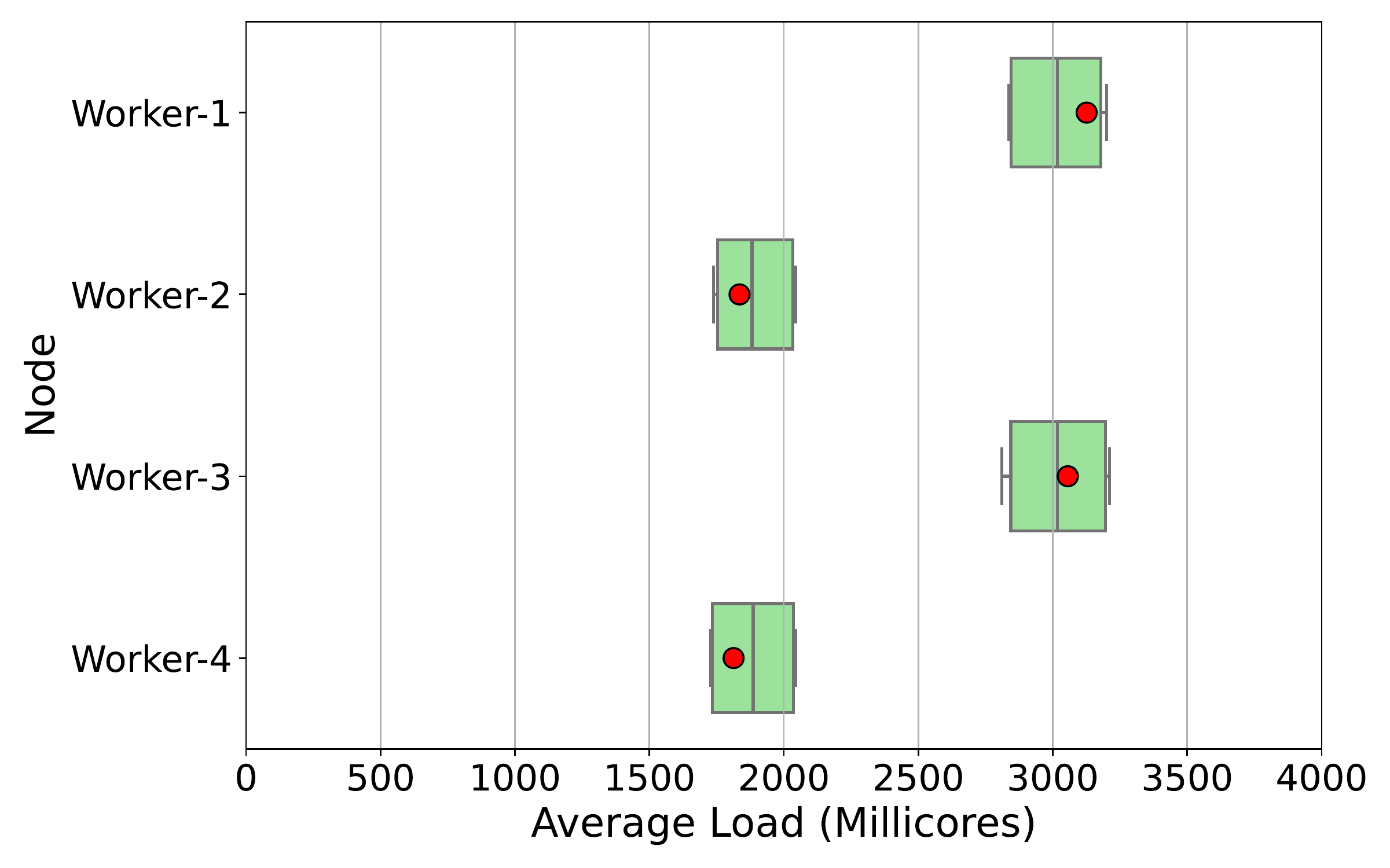}\\
Workflow T4
  \end{minipage}
  \caption{\label{fig:2b2e}Mixed workflow measurements for cluster
    configuration 2B2C. The first line of plots reports consumption by
    service, the second line by node.}
\end{figure}

\begin{figure}
\centering\footnotesize
  \begin{minipage}{0.268\linewidth}\centering
    \includegraphics[width=\linewidth,trim={1.2cm .4cm .3cm .3cm},clip]{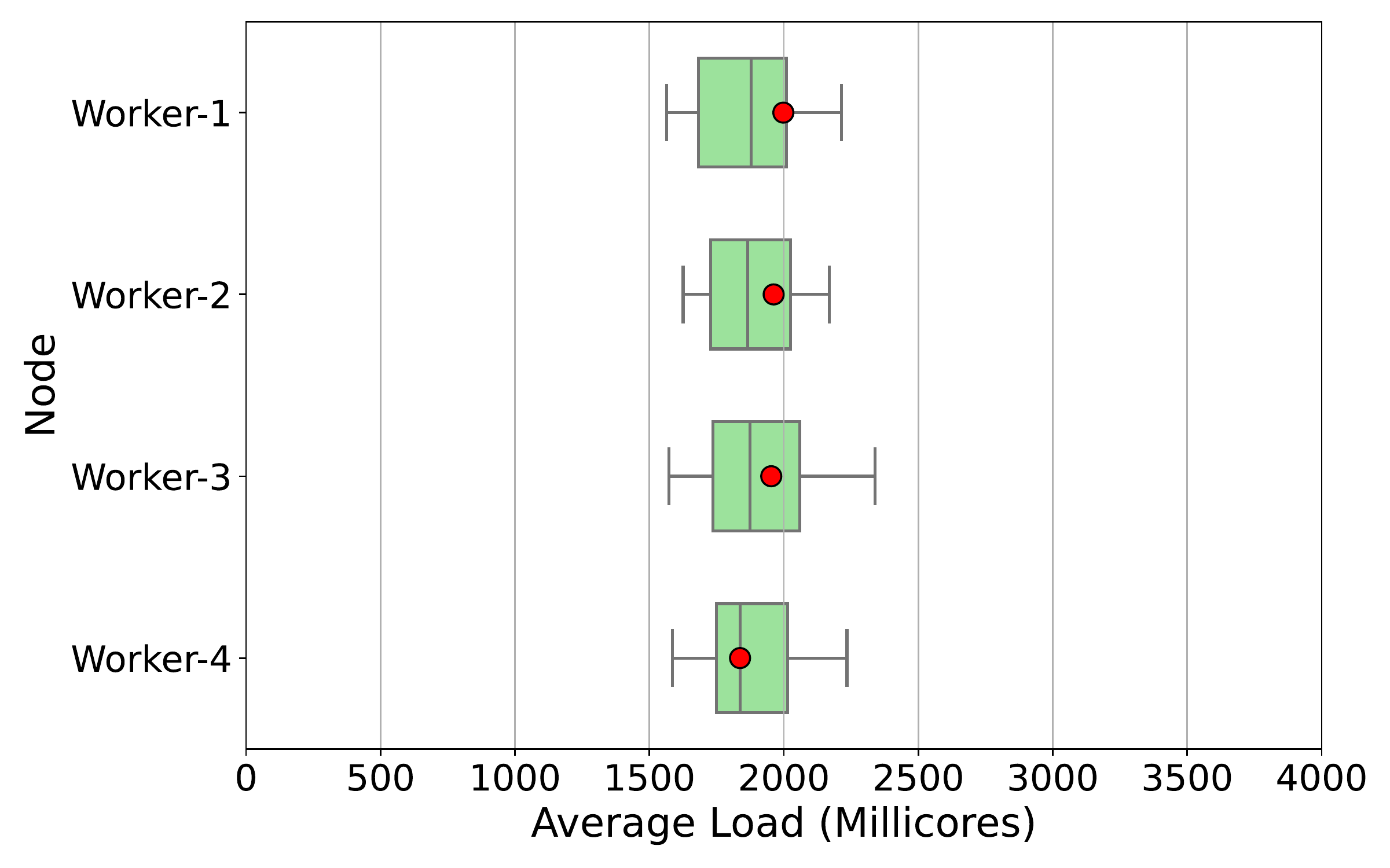}\\
Workflow P1
  \end{minipage}
  \hfill
  \begin{minipage}{0.2344\linewidth}\centering
    \includegraphics[width=\linewidth,trim={4.1cm .4cm .3cm .3cm},clip]{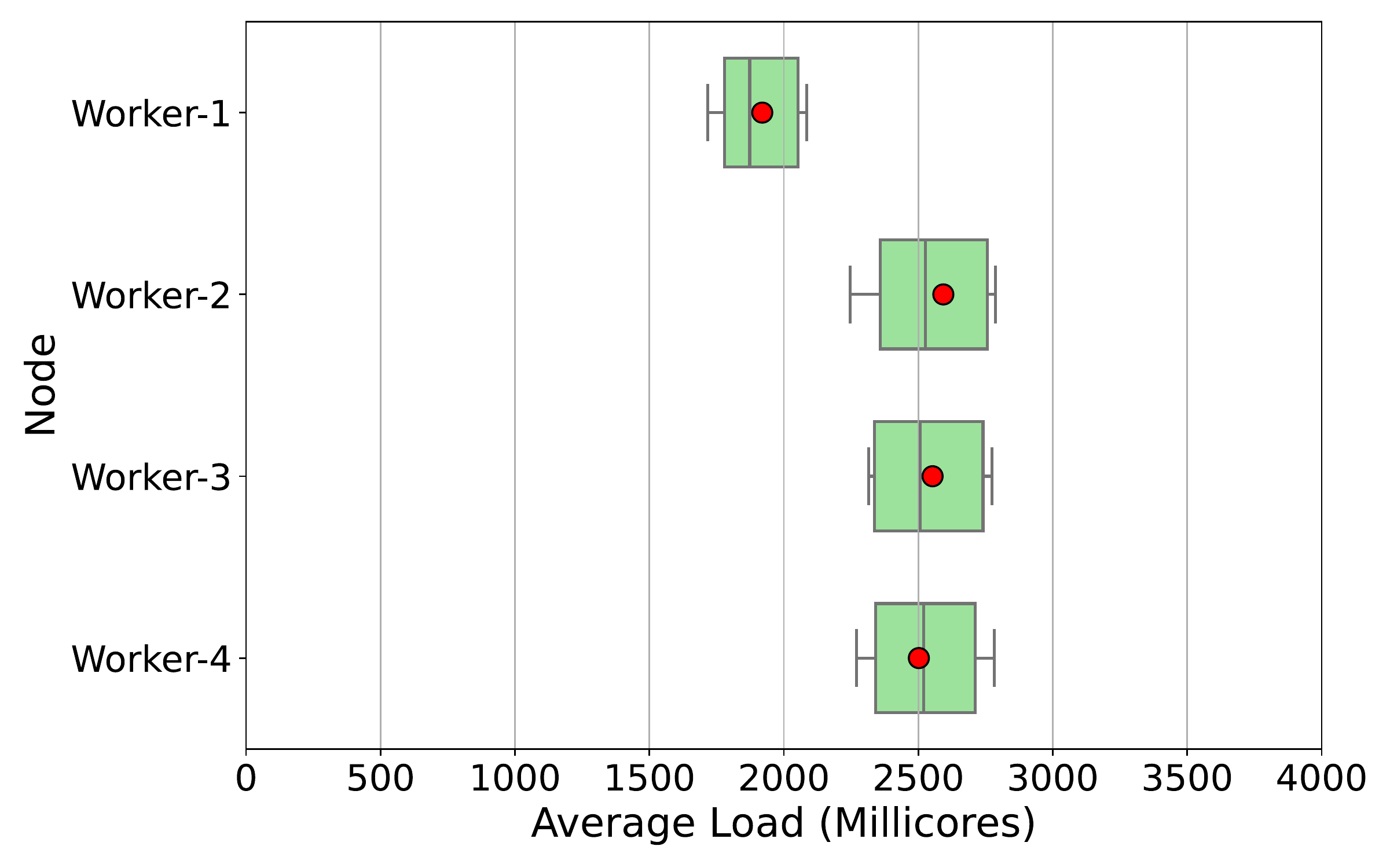}\\
Workflow P2
  \end{minipage}
  \hfill
  \begin{minipage}{0.2344\linewidth}\centering
    \includegraphics[width=\linewidth,trim={4.1cm .4cm .3cm .3cm},clip]{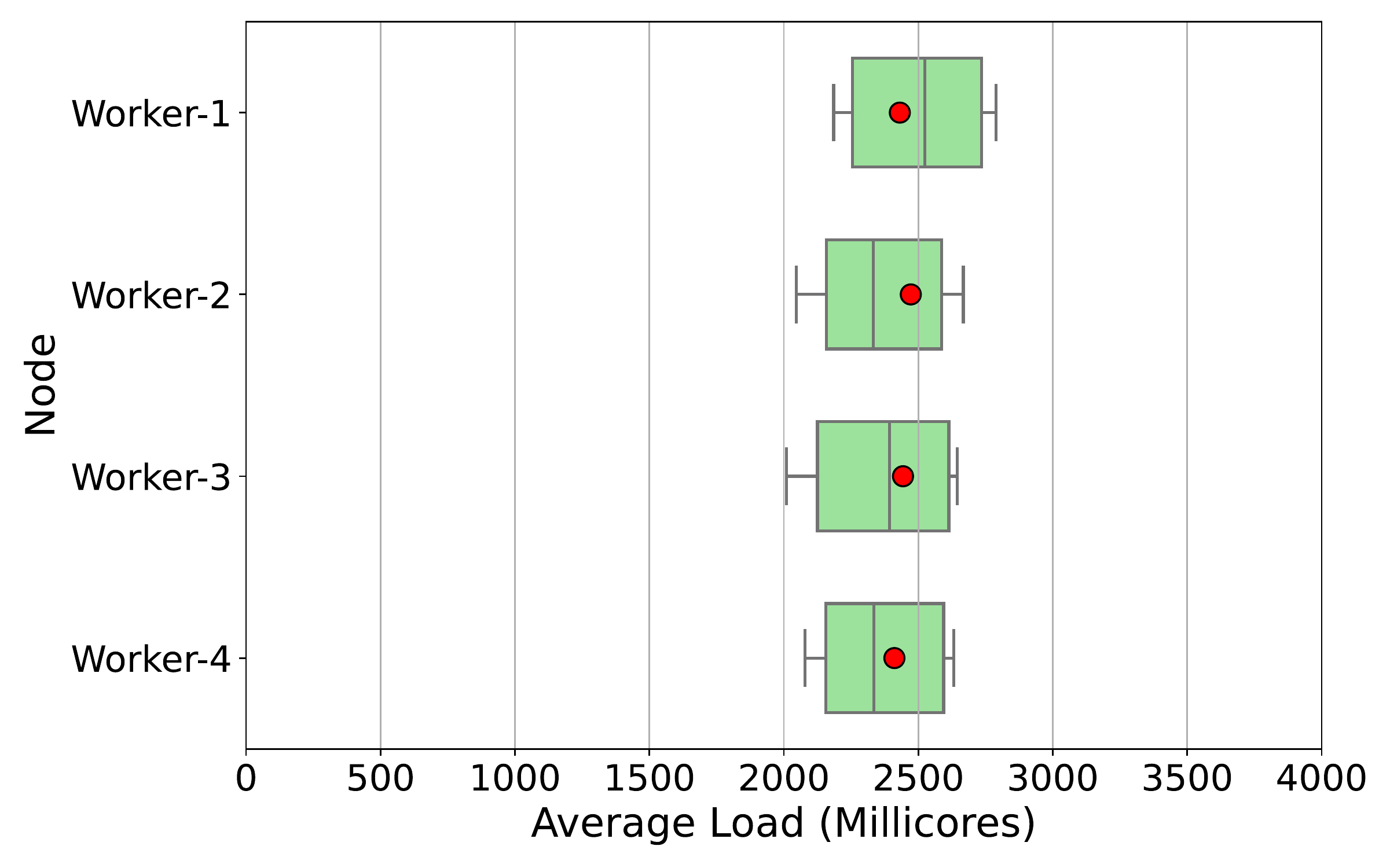}\\
Workflow P3
  \end{minipage}
  \hfill
  \begin{minipage}{0.2344\linewidth}\centering
    \includegraphics[width=\linewidth,trim={4.1cm .4cm .3cm .3cm},clip]{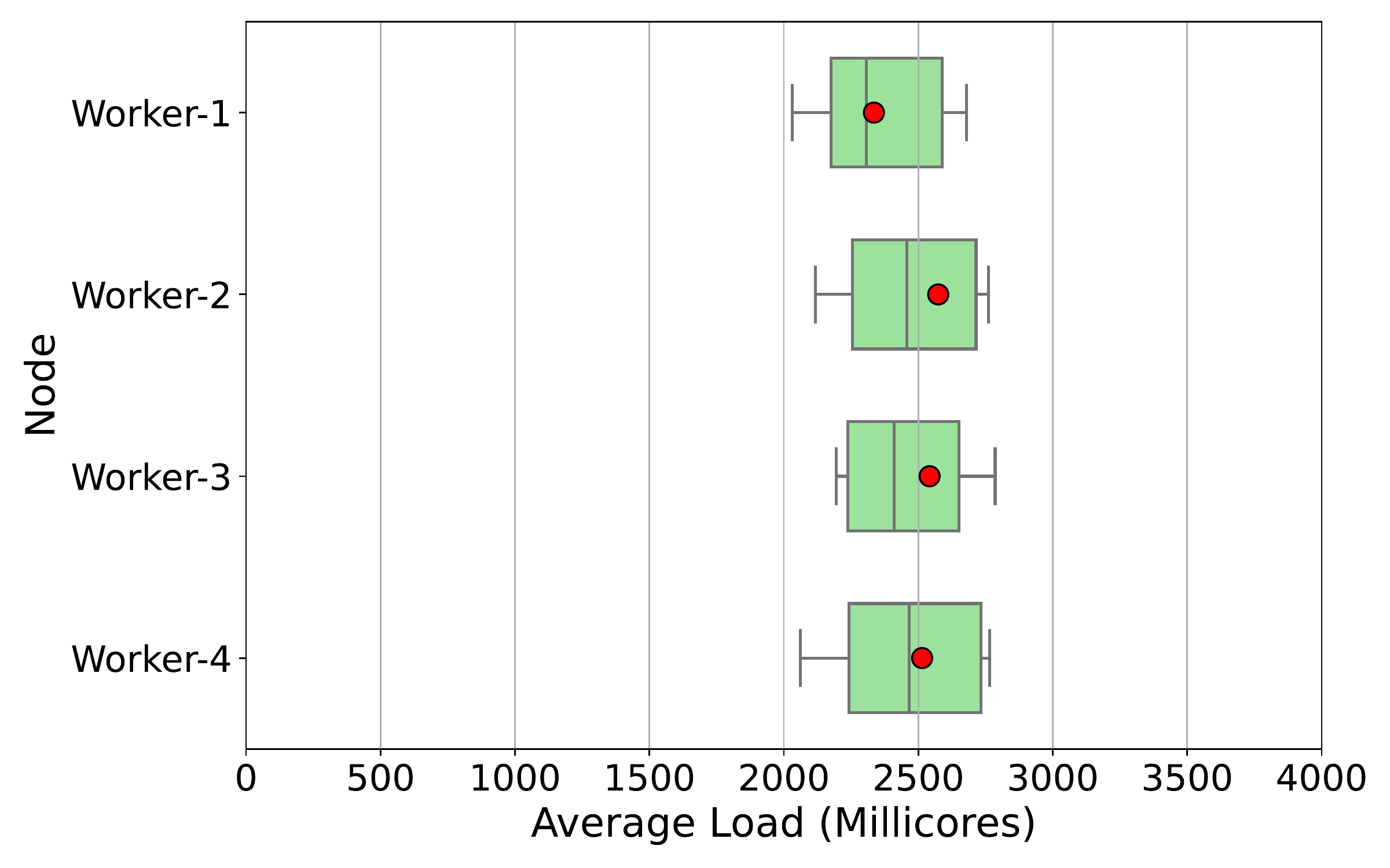}\\
Workflow P4
  \end{minipage}

\medskip

  \begin{minipage}{0.268\linewidth}\centering
    \includegraphics[width=\linewidth,trim={1.2cm .4cm .3cm .3cm},clip]{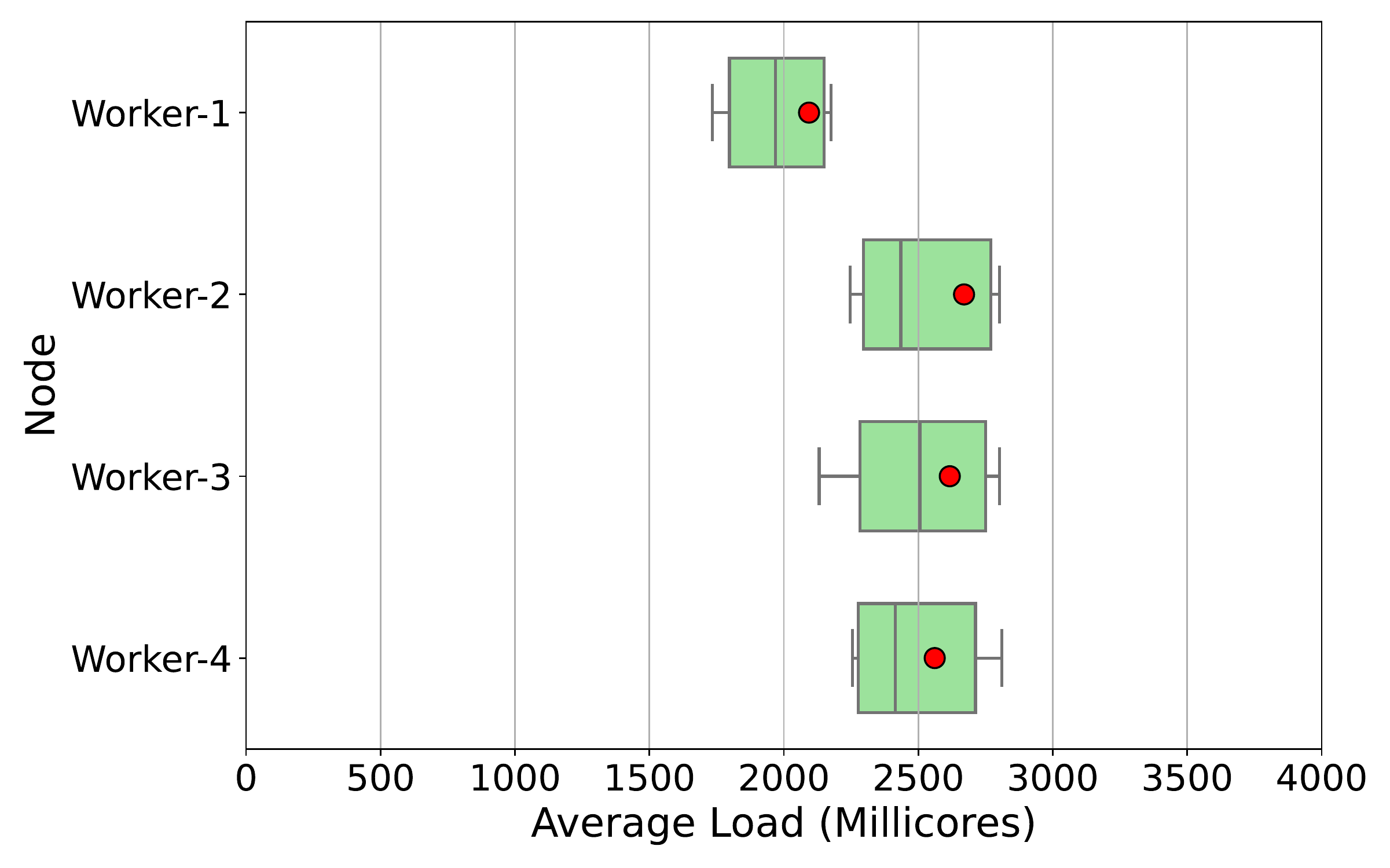}\\
Workflow T1
  \end{minipage}
  \hfill
  \begin{minipage}{0.2344\linewidth}\centering
    \includegraphics[width=\linewidth,trim={4.1cm .4cm .3cm .3cm},clip]{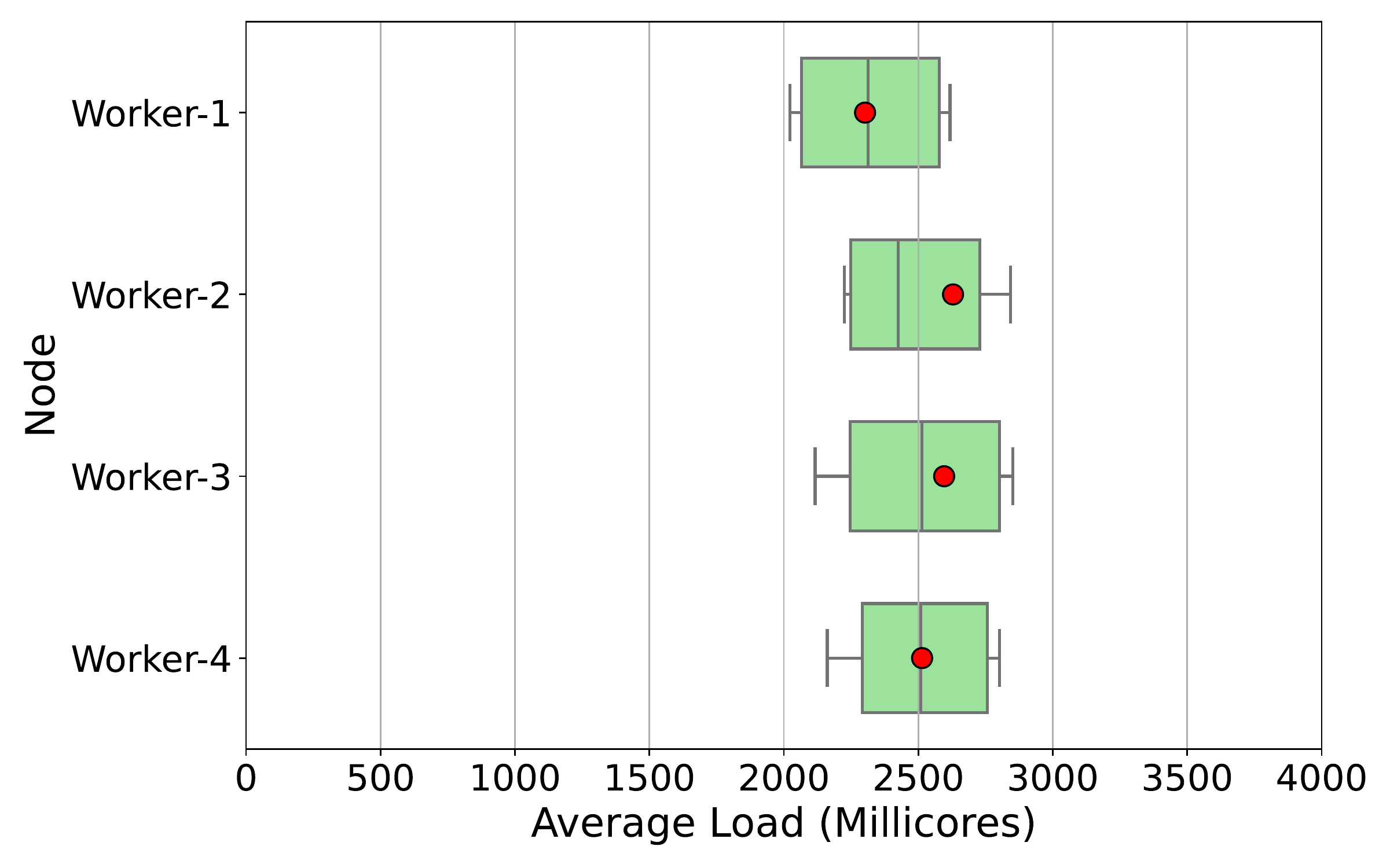}\\
Workflow T2
  \end{minipage}
  \hfill
  \begin{minipage}{0.2344\linewidth}\centering
    \includegraphics[width=\linewidth,trim={4.1cm .4cm .3cm .3cm},clip]{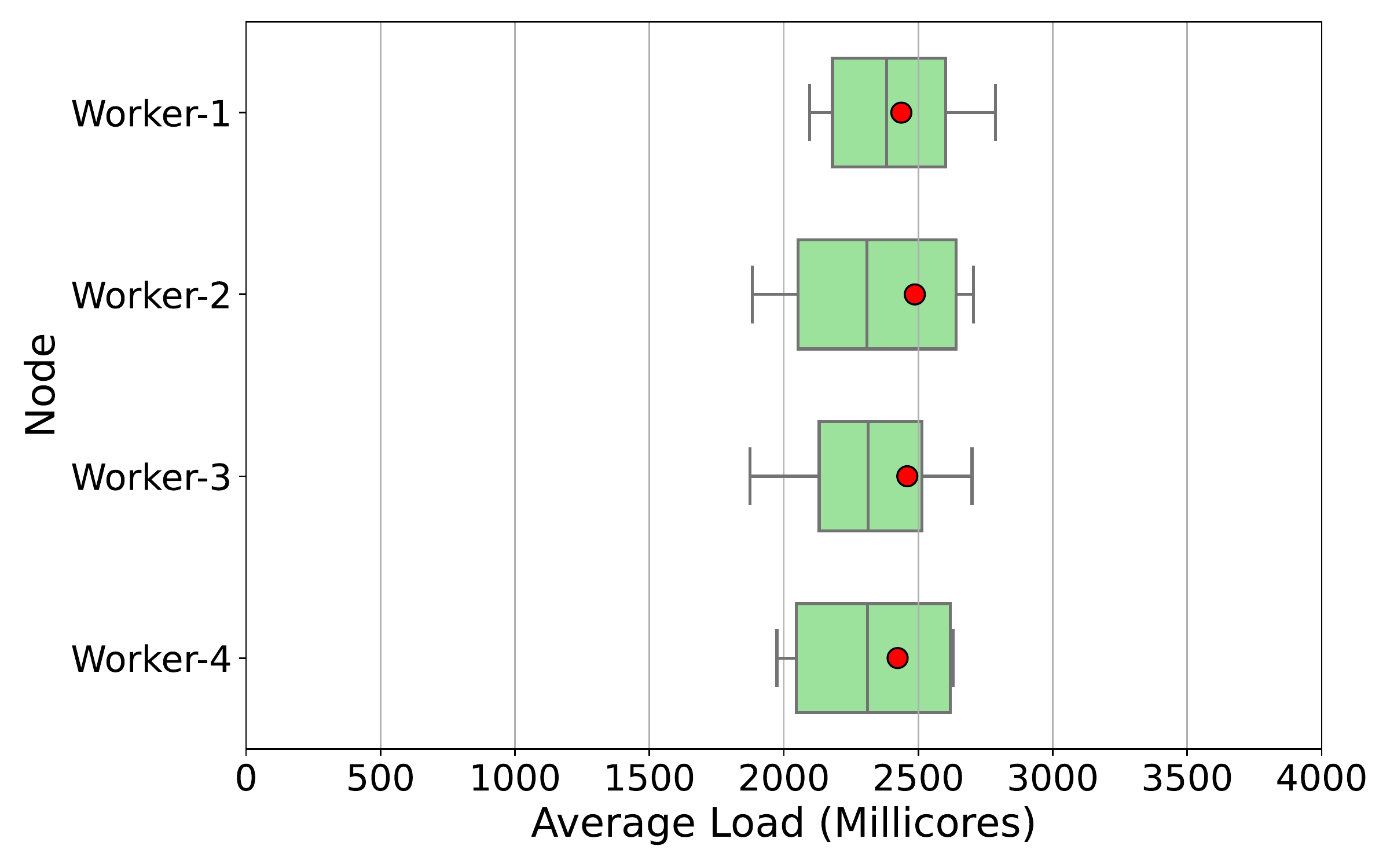}\\
Workflow T3
  \end{minipage}
  \hfill
  \begin{minipage}{0.2344\linewidth}\centering
    \includegraphics[width=\linewidth,trim={4.1cm .4cm .3cm .3cm},clip]{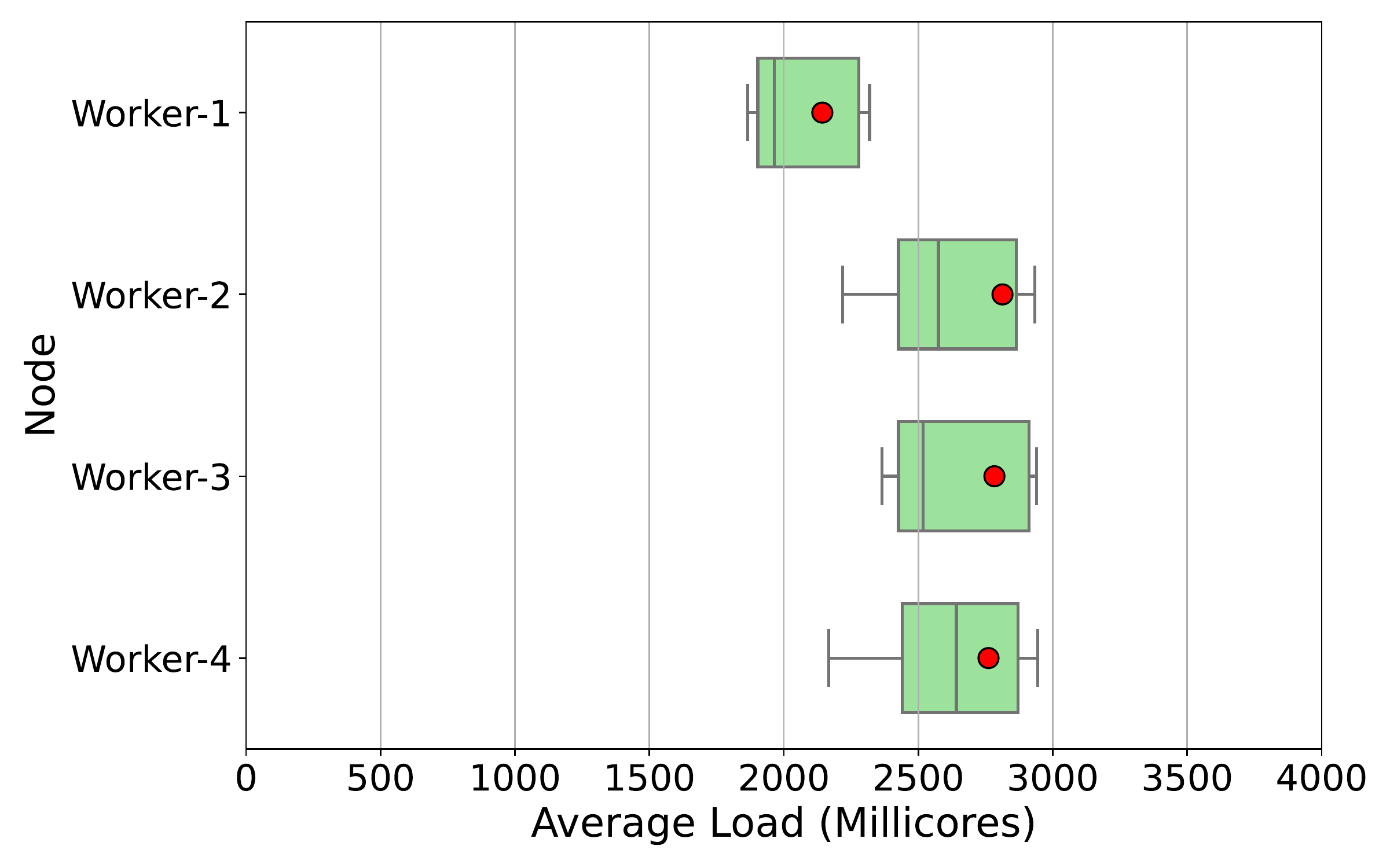}\\
Workflow T4
  \end{minipage}
  \caption{\label{fig:3b1e}Mixed workflow measurements for cluster
    configuration 3B1C. The first line of plots reports consumption by
    service, the second line by node.}
\end{figure}

\begin{figure}[t]
\centering\footnotesize
\begin{minipage}[t]{0.268\linewidth}\centering
    \includegraphics[width=\linewidth,trim={1.2cm .4cm .3cm .3cm},clip]{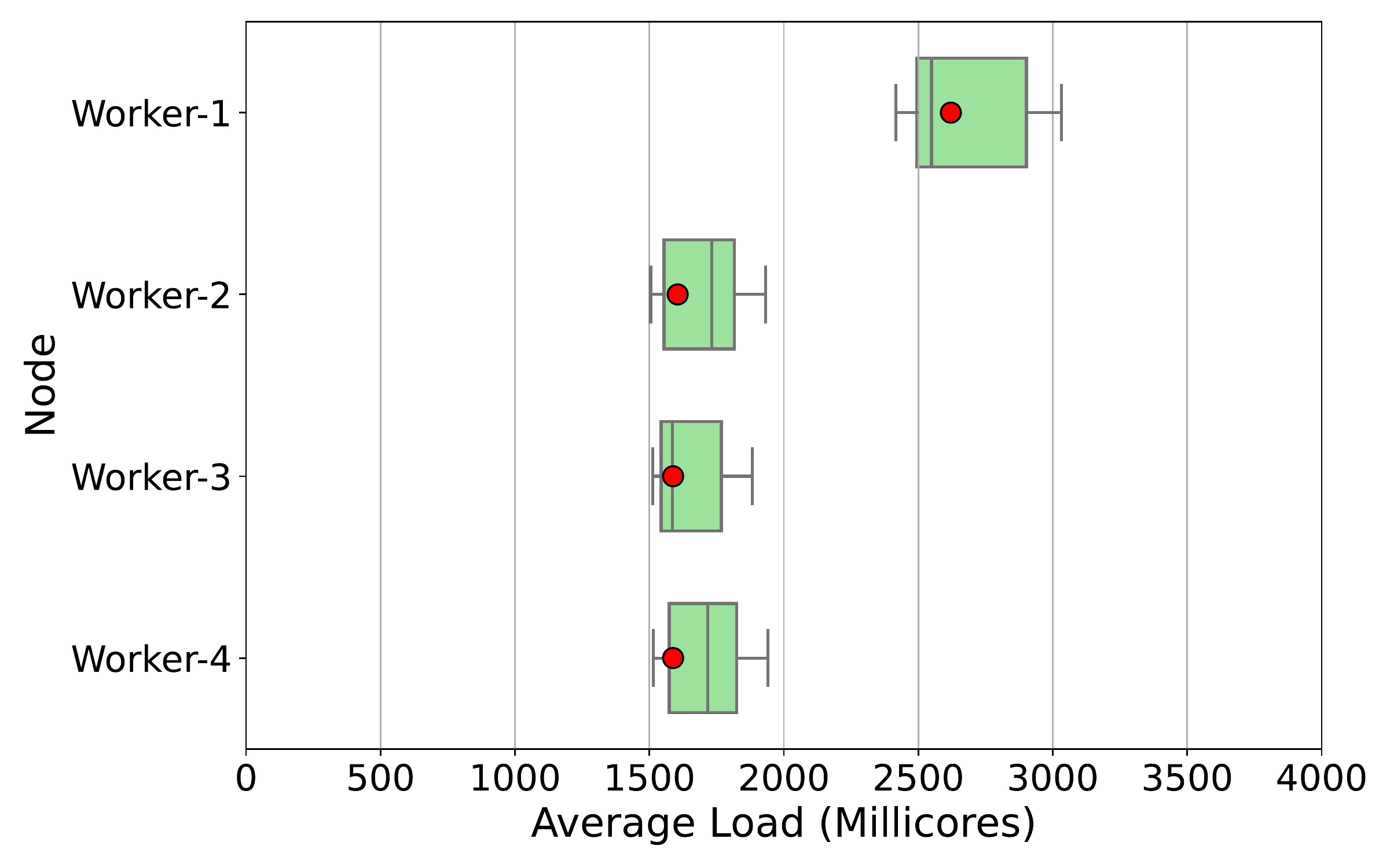}\\
    Workflow P1
  \end{minipage}
  \hfill
  \begin{minipage}[t]{0.2344\linewidth}\centering
    \includegraphics[width=\linewidth,trim={4.1cm .4cm .3cm .3cm},clip]{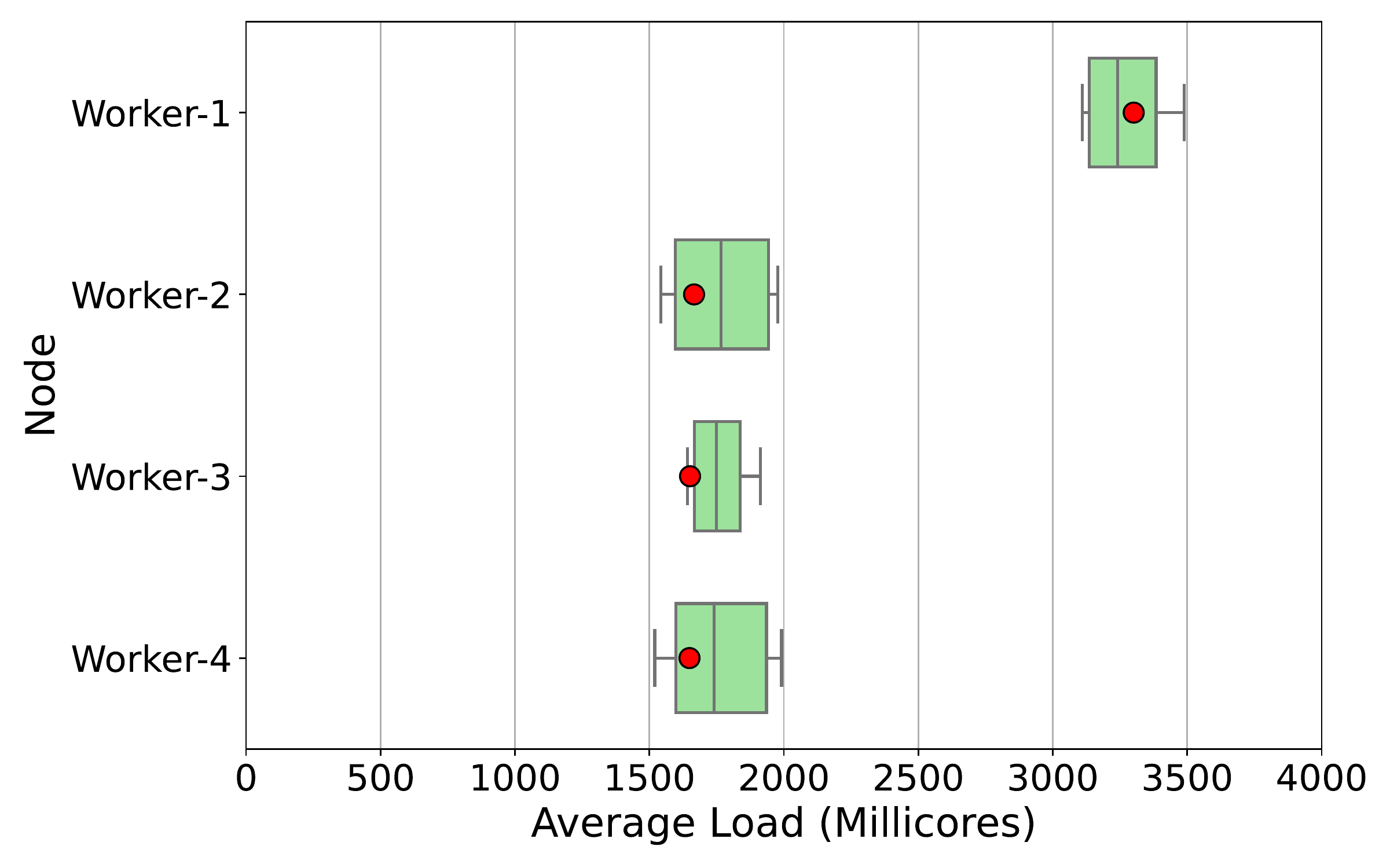}\\
    Workflow P2
  \end{minipage}
  \hfill
  \begin{minipage}[t]{0.2344\linewidth}\centering
    \includegraphics[width=\linewidth,trim={4.1cm .4cm .3cm .3cm},clip]{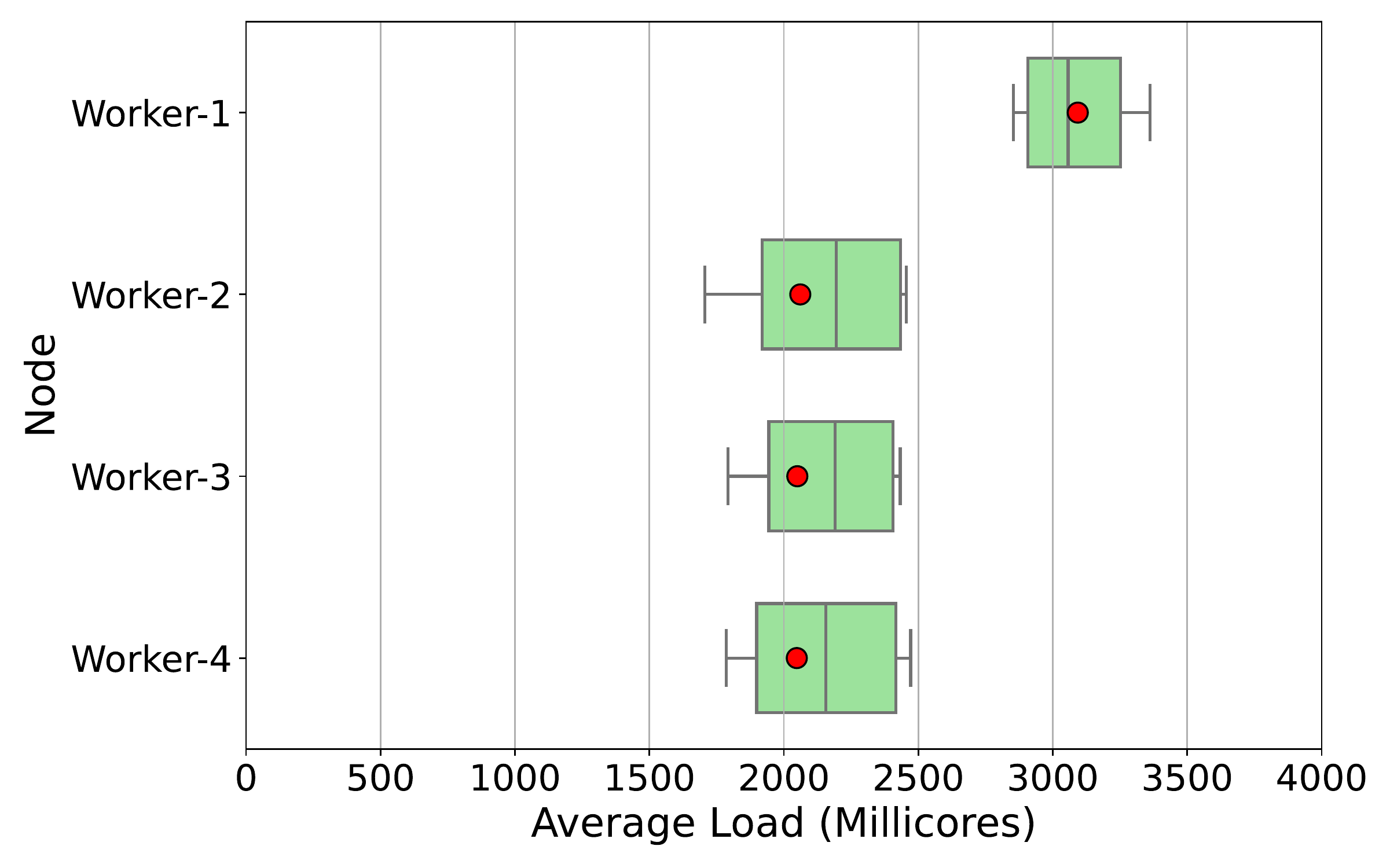} \\
    Workflow P3 
  \end{minipage}
  \hfill
  \begin{minipage}[t]{0.2344\linewidth}\centering
    \includegraphics[width=\linewidth,trim={4.1cm .4cm .3cm .3cm},clip]{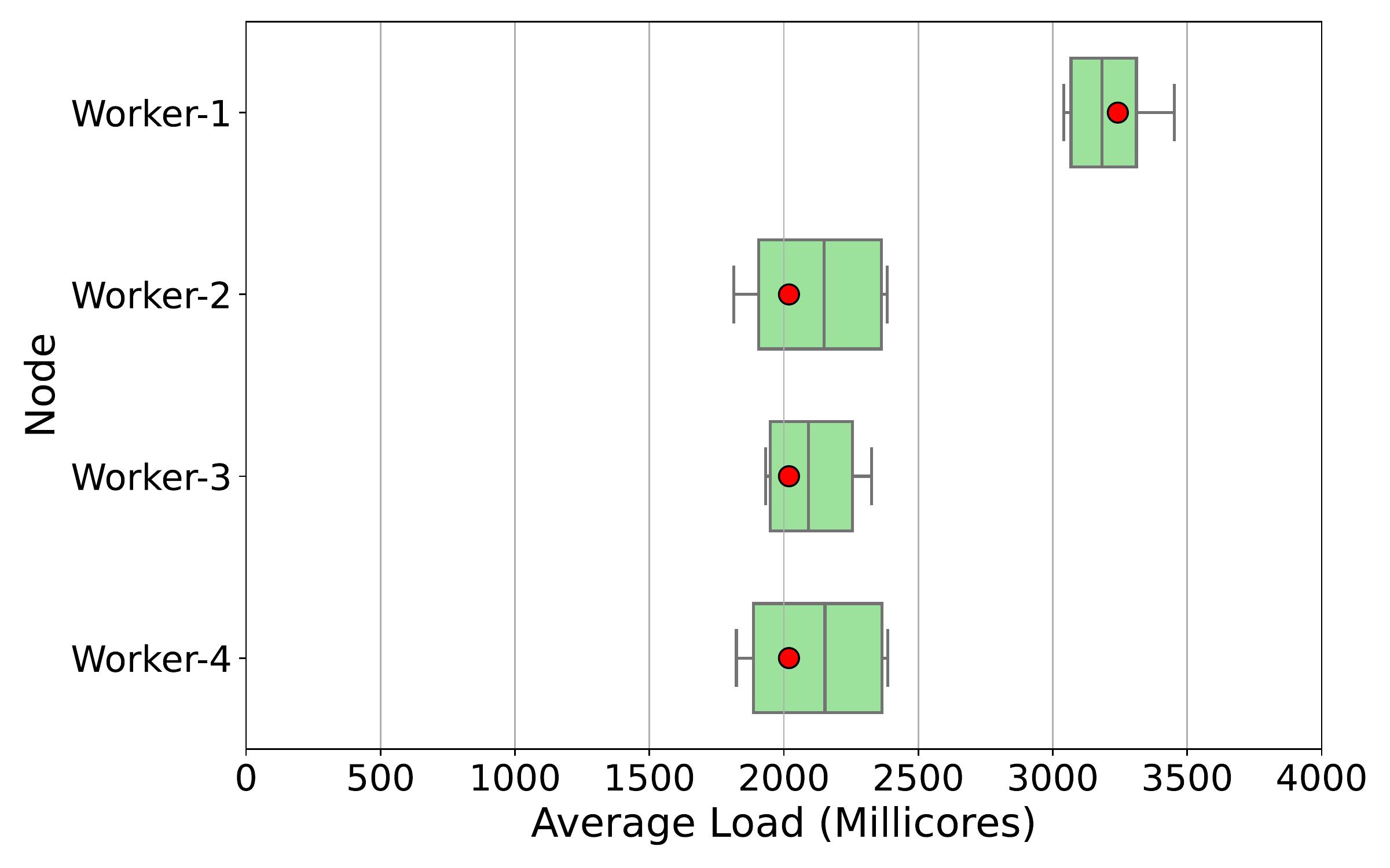} \\
    Workflow P4
  \end{minipage}

\medskip

  \begin{minipage}[t]{0.268\linewidth}\centering
    \includegraphics[width=\linewidth,trim={1.2cm .4cm .3cm .3cm},clip]{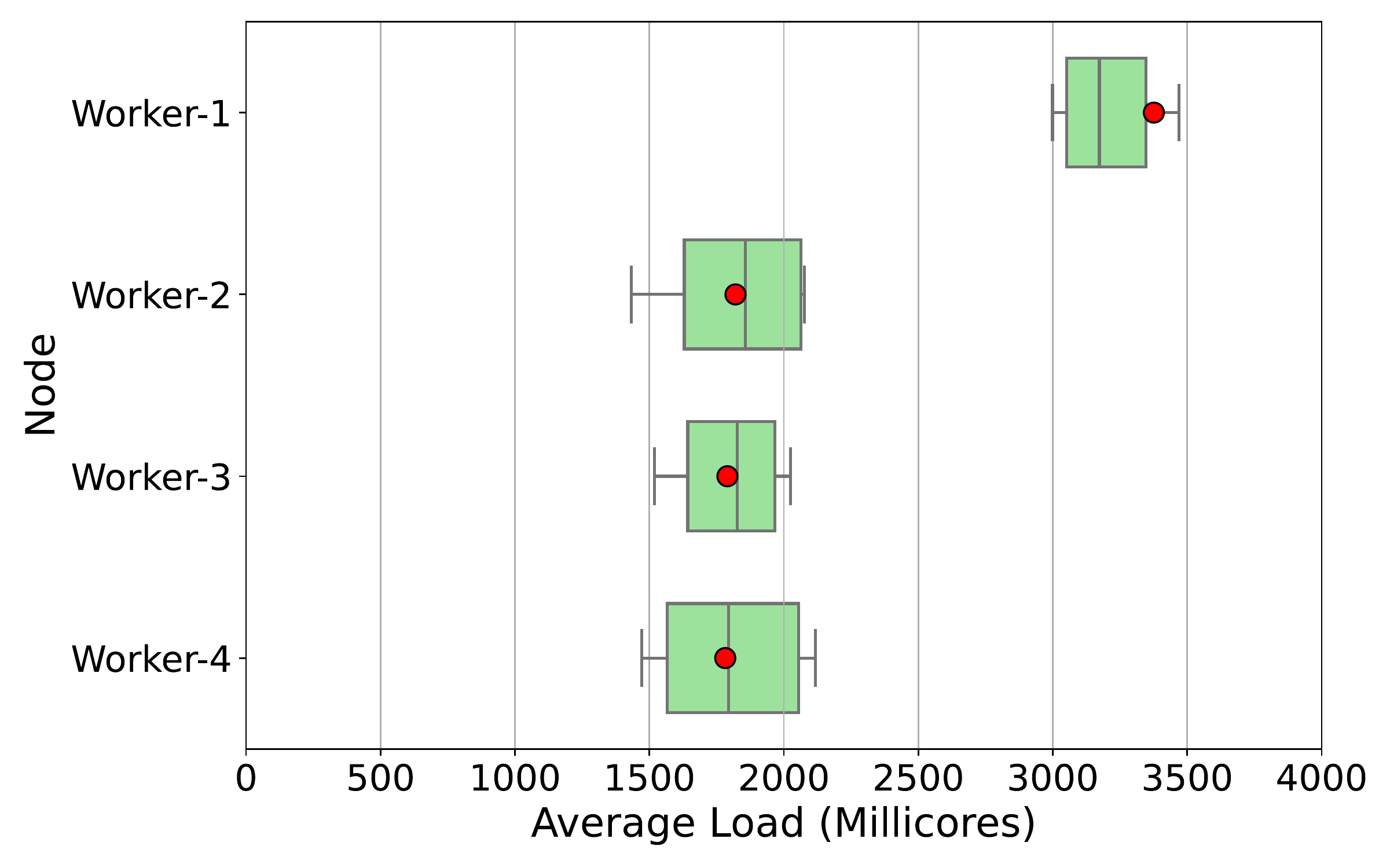}\\
    Workflow T1
  \end{minipage}
  \hfill
  \begin{minipage}[t]{0.2344\linewidth}\centering
    \includegraphics[width=\linewidth,trim={4.1cm .4cm .3cm .3cm},clip]{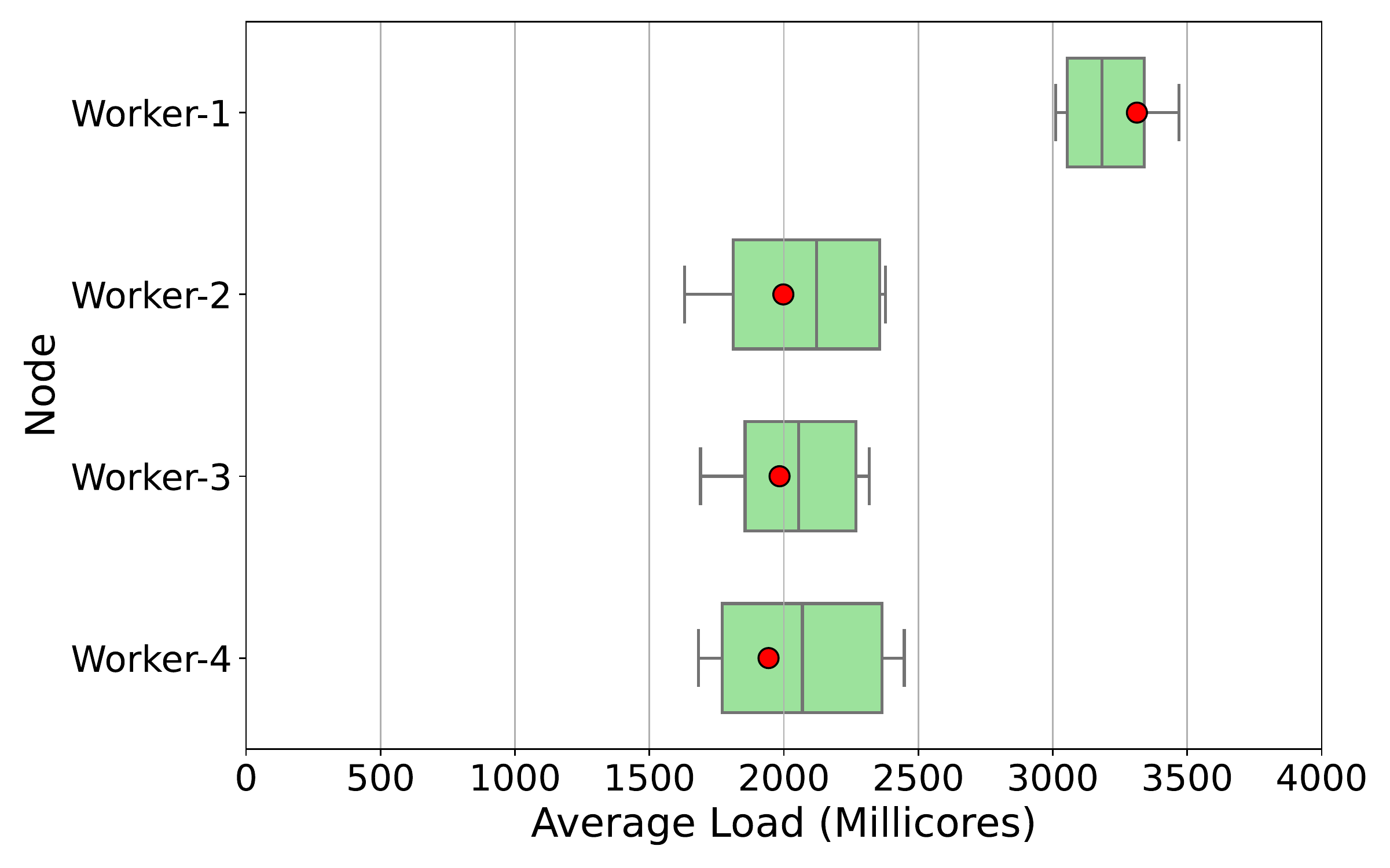} \\
    Workflow T2
  \end{minipage}
  \hfill
  \begin{minipage}[t]{0.2344\linewidth}\centering
    \includegraphics[width=\linewidth,trim={4.1cm .4cm .3cm .3cm},clip]{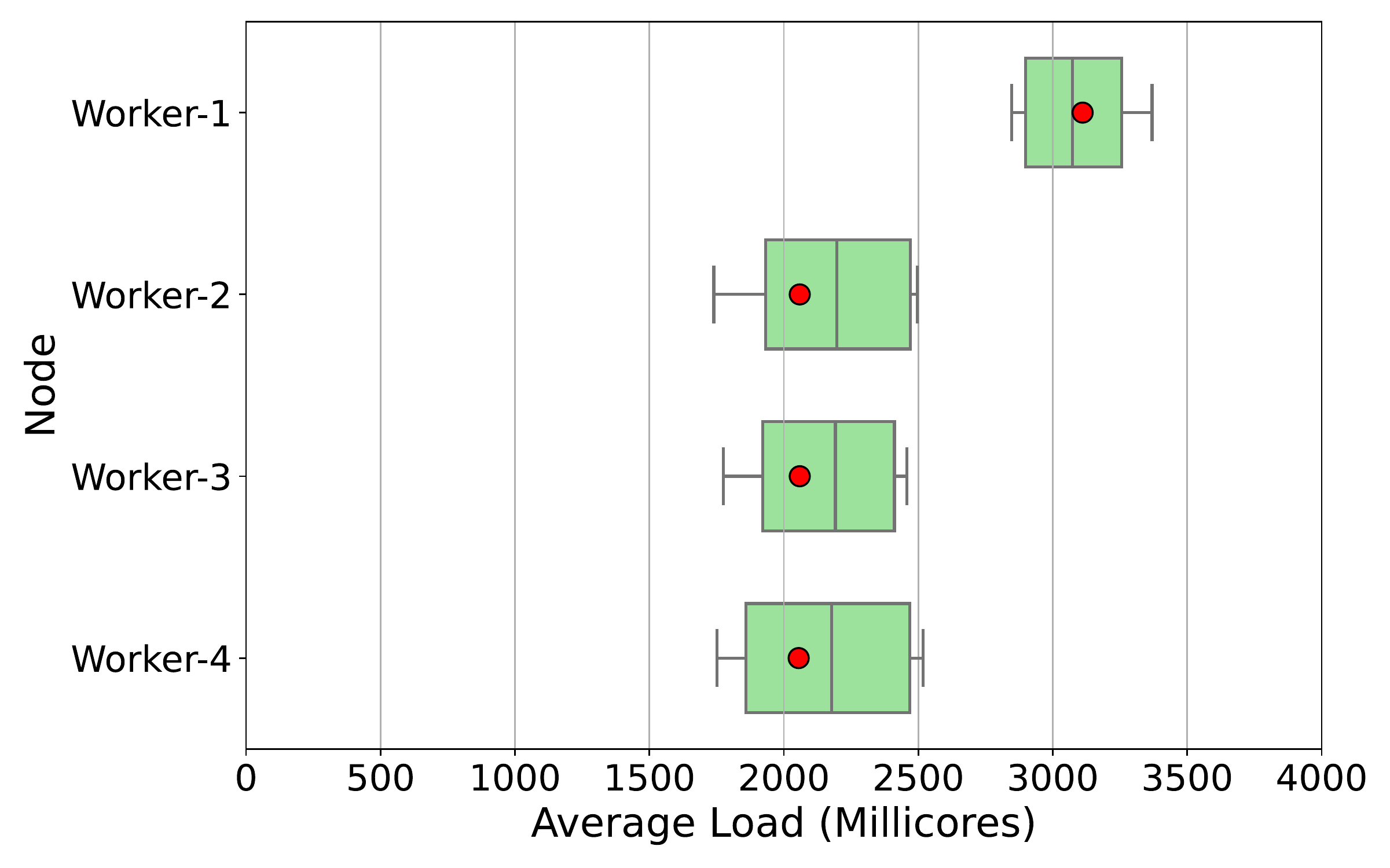}\\
    Workflow T3 
  \end{minipage}
  \hfill
  \begin{minipage}[t]{0.2344\linewidth}\centering
    \includegraphics[width=\linewidth,trim={4.1cm .4cm .3cm
      .3cm},clip]{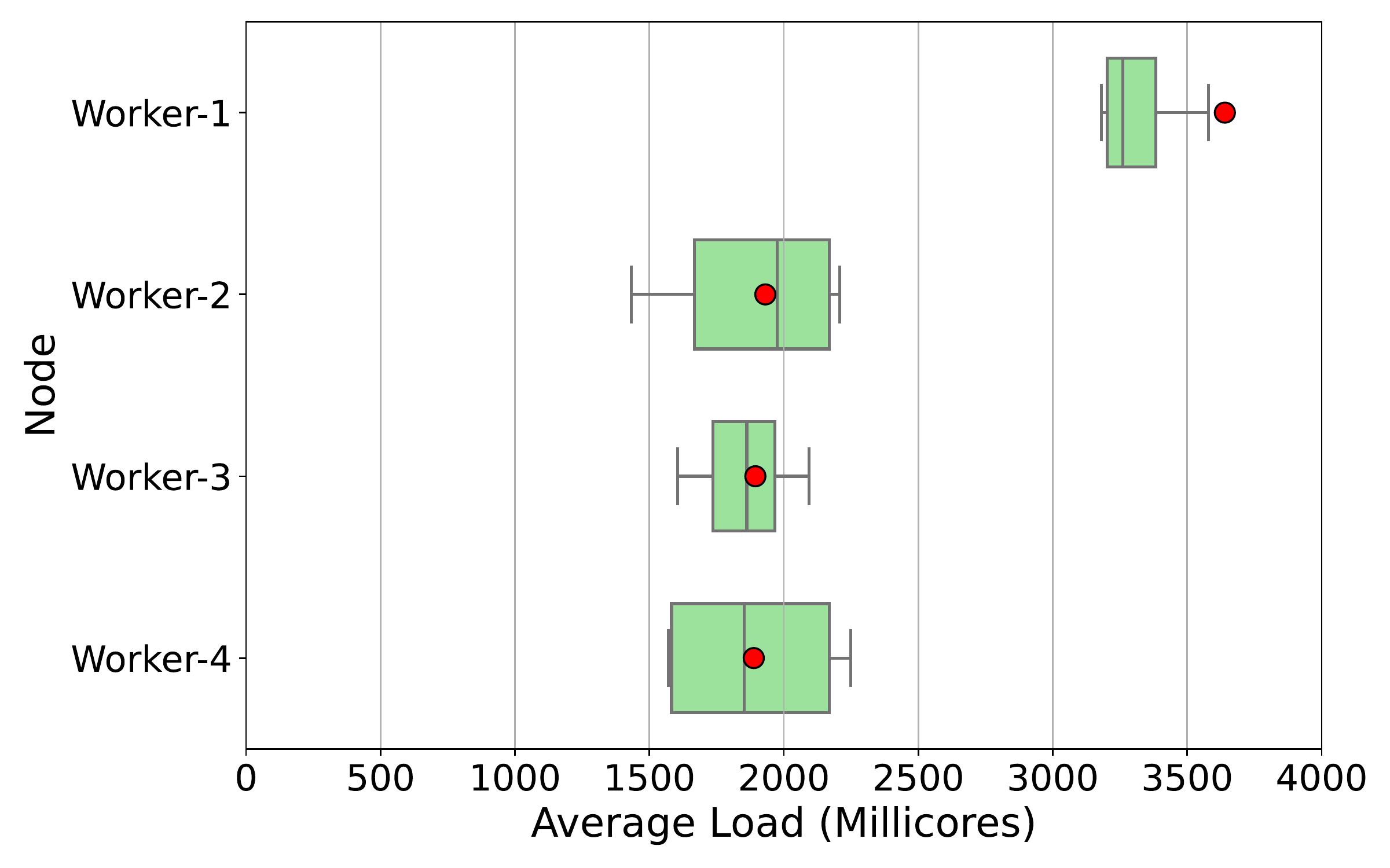}\\
    Workflow T4
  \end{minipage}
  \caption{\label{fig:1b3e}Mixed workflow measurements for cluster
    configuration 1B3C. The first line of plots reports consumption by
    service, the second line by node.}
\end{figure}

Figures~\ref{fig:2b2e}--\ref{fig:1b3e} compare measurements for the
twelve iterations of the stress tests to the corresponding model
predictions for the mixed workflows specified in Table~\ref{tab:rq2},
using node turnover in the cluster experiments.
The consumption recorded during the experiments is presented as box
plots for all cluster configurations.  In particular,
Figure~\ref{fig:2b2e} shows the results of the four pair workflows P1,
P2, P3 and P4 (first row) and four triplet workflows T1, T2, T3 and T4
(second row) for the cluster configuration 2B2C (see
Table~\ref{tab:nodes}), and Figures~\ref{fig:3b1e} and~\ref{fig:1b3e}
show the corresponding results for the cluster configurations 3B1C and
configuration 1B3C, respectively. The green area of the box plots
cover 50\% of the observations, and the two brackets span to the
minimum and maximum value observed. The red dot depicts the expected
consumption from the calibrated model.

In these experiments, we observe that the expected resource
consumption from the model corresponds well to the observed
consumption measured on the cluster during the turnover stress
tests. In our experiments, the best balanced cluster load is
obtained with cluster configuration 3B1C (shown in
Figure~\ref{fig:3b1e}) and we see that this can be detected from the
model predictions.
When the system is overloaded, the model can predict that some
failures may occur in a given scenario, but it cannot predict how such
failures will impact resource consumption. This is because effects
such as requests lost due to oversaturated queues are not reflected in
the model.
For example, in Figure~\ref{fig:1b3e} (second row) the workflows tend
to systematically overload Worker~1 (which is clearly not a desirable
scheduling) and the predicted consumption from the model is less
accurate than for the non-overloaded scenarios.

\begin{figure}[t]
  \centering
    \includegraphics[width=\linewidth]{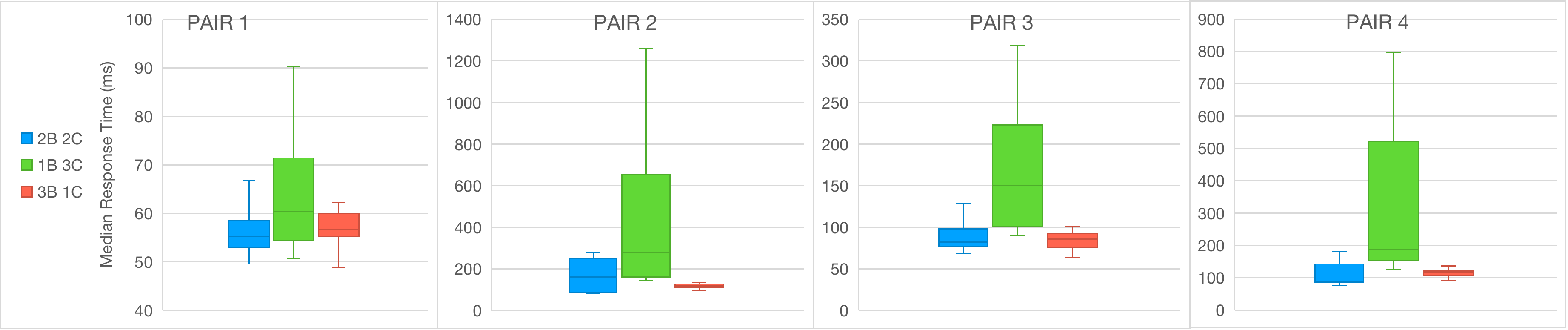}\\
    \includegraphics[width=\linewidth]{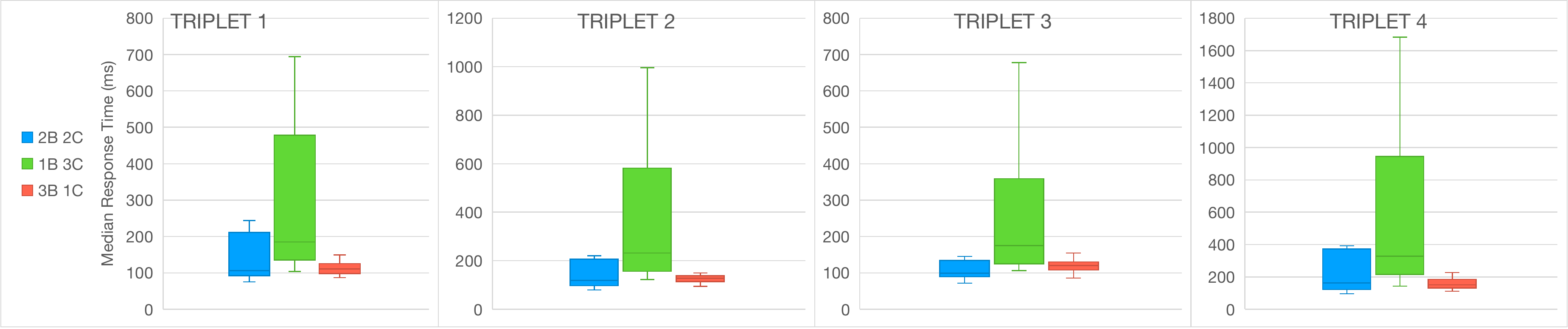}
    \caption{\label{fig:rq3median}Comparison of median response time
      between multiple cluster settings for pairs (top) and triplets
      (bottom).}
\end{figure}

\paragraph{RQ3}
Figure~\ref{fig:rq3median} shows the time of the median request in the
stress tests for each workflow and cluster configuration. In the
plots, the $Y$-axis depicts a time scale in milliseconds and the
$X$-axis depicts the different cluster configurations. The three
different cluster settings are identified by the three colours --- the
top-down order in the legend reflects the left-to-right order of boxes
in the plots.

In these experiments, we can observe that cluster configuration 3B1C
(the rightmost, red column of each plot in Figure~\ref{fig:rq3median})
was the most resilient and performant under the different workloads.
We can further observe that in the experiments in which the load is
fairly distributed among the nodes,  cluster configuration 2B2C is
likewise performant. Nevertheless, there are cases (P2 and T4) where
the unbalance seems impairing for performance.

\subsection{Threats to Validity}

Our experiments reduce the configuration space in terms of the number
of node configurations and workflows, and by only considering static
workflows.  First, in principle there could be a combinatory explosion
of node configurations on a cluster. However, in practice, we believe
that configurations do not vary too much due to deployment constraints
that prevent many configurations from being used. In the proposed
methodology, we consider a set of calibrated nodes with fixed
workloads since containers and pods are not independent. Containers
and pods affect each other's consumption and performance when running
on the same machine~\cite{Zhao2017}. Thus, the single pod consumption,
stressed by the same demand, can differ from one node configuration to
another.  Second, cloud-native applications with a huge number of
workflows have not been considered in the experiments. This potential
limitation of the methodology could be addressed by a fully automated
sampling process as an extension to the current work (see
Section~\ref{sec:discussion}). Third, our experiments have not
considered dynamic workflows. We do not believe that this is a major
limitation of the proposed methodology because a dynamic workflow can
be treated as a set of static workflows, one for each workflow
variation in accordance to a parameter change.




\section{Discussion}
\label{sec:discussion}
We complement the presentation of the modeling framework by discussing
two perspectives on its applicability: the integration of the modeling
framework in a Continuous Integration/Continuous Delivery (CI/CD)
pipeline and its generalization beyond Kubernetes and CPU resources.

\subsection{From a Modeling Framework to a Full-fledged Tool}


Nowadays, cloud-native applications are built, packaged, tested and
deployed automatically by mean of CI/CD pipelines. These pipelines can
be seen as sequences of stages where each stage runs a set of scripts.
Companies tend to split \emph{staging} and \emph{development}
environments from \emph{production} to prevent that problems spread
from the applications under development and testing to the production
environment.  Having a sandbox in which applications can be properly
tested and possibly calibrated before they move to the production
environment, can be considered as part of today's best practices in
software engineering.

The modeling framework presented in this paper enables the further
development of an automated calibration stage in such a sandbox, which
would be an additional stage in the pipeline for building an
application, before it moves into production.  The modeling framework
generates resource consumption plots for services and nodes as the
outcome of simulations.  The comparison of different plots for
different simulations, helps in finding suitable configurations. A
next step towards fully integrating the modeling framework in a CI/CD
calibration stage could be to, e.g., configure multiple simulation
scenarios ahead of time, run the simulations for the resulting models
and automatically generate comparison plots for the considered
deployments.

To configure a simulation scenario, the model must specify a set of
clients calling each workflow with a certain demand (i.e., RPS,
intensity) and the services involved for each workflow. We believe
this could be significantly simplified by developing a proper GUI and
automating the corresponding model generation alongside the
calibration process.  In the evaluation of our proposed methodology
for instantiating the modeling framework
(Section~\ref{ssec:casestudies}), we considered an application
exposing an API of about 10 different workflows.  Among these, we
identified the three most common and resource demanding workflows that
were responsible for the most of the application's CPU consumption.
We believe this scenario, with a few workflows that dominate the
resource consumption of an application, is fairly common and covers a
wide range of applications deployed in Kubernetes. An interesting line
of future work would be to test a larger number of workflows together,
especially from several applications.  Although specifying a large
number or workflows in the model can be demanding, the additional
manual work in a CI/CD setting would in fact be fairly limited; the
configuration of URLs to access the different endpoints exposed via
the API in order to build the cost tables for calibration, would also
be needed for integration testing.

The modeling framework is well-suited to quickly discover bottlenecks
in a configuration: When resource consumption remains within the
thresholds, the outcome of the simulations can be used to explore
configurations.  Although our models can be used to detect hazardous
scenarios in which consumption goes beyond the thresholds, they cannot
be used to fully explain the consequences of these scenarios.  For
example, if a service pod is overloaded because it reaches its CPU
limit or its host node is saturated, the pod will trigger a cascade of
request failures as messages are dropped from the queues (see
Section~\ref{sec:results}).  It would significantly increase the
complexity of the modeling framework to capture how such chaotic
failures may impact the cluster.  In this paper, we have opted to
abstract away the dependencies between tasks in a workflow and model
them as a set of tasks that execute in parallel, we believe this
abstraction from service dependencies does not induce a significant
loss of precision for the analysis of resource consumption, because
the pods processing requests form a pipeline, microservices process
requests asynchronously and consumption falls within the
thresholds. However, workflows with dependencies can be modeled in ABS
when needed (see, e.g., \cite{johnsen17softcom,lin16fase}).

\subsection{Generalizing the Modeling Framework  Beyond Kubernetes and CPU Consumption}


The presented methodology is currently tailored to containerized
applications, thus, we believe it can be applied to other cloud
deployment architectures. Some parts of the implemented framework can
be easily reused; for example, the chosen monitoring tools were used
to monitor custom cloud deployments before the Kubernetes platform was
introduced. Other parts of the framework will require further
investigation and careful changes; for example, the prediction model
would need to be adjusted for different deployment platforms. The
current model calculates CPU consumption based on the load balancing
strategy used on the cluster. Other platforms may offer different load
balancing strategies and may interact with containers in different
ways; e.g., evicting or killing containers that exceed their memory
limits.

Resources generally fall into two categories, counting semaphores
(like memory) and temporal (like CPU). Both categories are already
covered in our modeling framework.  For the evaluation framework, we
have focused on how to derive resource models empirically for CPU
resources.  For memory-intensive applications, memory usage will also
be of interest, and the cost tables should include memory usage.  The
memory usage of a node can easily be obtained in the calibration
process from the memory usage of the hosted containers.  In contrast
to CPU resources, memory is acquired and released rather than consumed
over time.  The simulations check that no pod or node exceeds its
memory limit and output memory consumption by services and nodes (see
Section~\ref{sec:nodesModel}).  Other resources such as disk I/O,
network bandwidth, and energy consumption can also in principle be
monitored and targeted by our methodology. Proper instrumentation
would then be required to monitor the chosen metric and build the
appropriate cost tables.  In this case, the resource model should be
extended to include the provisioning of the new resource and the
simulation model to capture how the resource is acquired and released.


\section{Related work}
\label{sec:related}
We position our contribution with respect to related work concerning the modeling of cloud systems, the optimisation of microservice management and tools for improving Kubernetes deployment.

\paragraph{Resource Models for Cloud-based Applications}
Whereas there are many cloud modeling languages (see, e.g., Bergmayr
\etal\cite{bergmayr18csur}), the majority of them deals with the
description of cloud deployment configurations. In contrast, this
paper is part of a line of work on formal modeling of virtualized
systems in ABS \cite{johnsen10fmco}, a concurrent, executable modeling
language. The perspective on virtualized systems taken in this line of
work, is to focus on resource provisioning and quality-of-service,
which typically affects the timing behavior of systems on the
cloud. The underlying technical idea is to introduce a separation of
concerns between the resource needs of different computational tasks,
and resource provisioning in the infrastructure
\cite{johnsen10icfem,johnsen12icfem,johnsen15jlamp}. This
approach has been successfully applied to different kinds of
virtualization infrastructure, including Amazon
AWS~\cite{johnsen16isola}, Hadoop YARN~\cite{lin16fase} and Hadoop
Spark Streaming~\cite{lin20ijguc}. The concurrency model of ABS, based
on actors, has also been used for the verification of industrial case
studies in a DevOps setting~\cite{ABHJSTW13}, for the analysis of
worst-case memory bounds~\cite{albert11fm} and for parallel cost
analysis~\cite{albert18tocl}, a novel static analysis method related
to parallelism and maximal span. The formal model of Kubernetes
presented in this paper differs from previous work in its
\emph{nested} virtualization; i.e., the containerization of
microservices leads to two levels of book-keeping in the
resource-sensitive architecture, corresponding to the pods and nodes
of the Kubernetes framework. Furthermore, the lack of isolation that
we observed for the containers led to a generalization of the cost
models used in the discussed work, from cost expressions to cost
tables.  An early case study of the nested
model~\cite{turin2020formal} did not account for this lack of
isolation and the associated cost tables. To the best of our
knowledge, our paper introduces the first concrete methodology for
instantiating cost models for Kubernetes deployments; we outline a
methodology for instantiating cost models for Kubernetes deployments
and provide a concrete example of how the methodology can be applied.

\paragraph{Optimization of Microservice Management}
General techniques for resource provisioning on the cloud are surveyed
by Zhang \etal\cite{zhang2016}. These techniques are not specific to
Kubernetes, and include algorithms aiming to improve autoscaling
\cite{jindal2018}, performance \cite{boza2019,rodriguez2018} and
energy efficiency~\cite{zhu2017}. Methods for optimizing microservices
include model-driven optimization techniques such as
\cite{gouda2014,samini2016}; these are also not specific to Kubernetes
systems.

It has been shown that deployment management can be formalized as
finite state machines, such as the Aeolus \cite{dicosmo14ic} and
TOSCA-compliant deployment models \cite{brogi15esocc}, which have been
adapted to formally reason about the static deployment of
microservices in Kubernetes \cite{chareonsuk2021}. For example, the
static deployment of microservices can be encoded as a constraint
problem \cite{Bravetti-et-al:LNCS-2019}. This work, which is based on
Aeolus, takes an ABS model as its starting point. In contrast to our
work, their focus is on how to solve the logical grouping of
microservices on nodes and the resource consumption of the deployed
microservices has not been considered.

\paragraph{Tools for Improving Kubernetes Deployments}
Several approaches have been proposed to improve resource allocation
for Kubernetes systems. For example, Ramos \etal \cite{ramos2021}
propose a machine learning model for detection of Docker-based app
overbooking on Kubernetes and RLSK \cite{huang2020rlsk} is a deep
Reinforcement Learning Scheduler for Kubernetes that uses
reinforcement learning to refine deployment heuristics. To improve
resource distribution, Zhang \etal \cite{wei2018research} proposed to
combine ant colony and particle swarm optimization algorithms. Li
\etal \cite{li2020dynamic} introduced a dynamic Input/Output sensing
scheduler for Kubernetes. The scheduler considers the disk pressure in
the scheduling process and tries to balance the node disk I/O usage
across the cluster dynamically. Similarly, Gaia \cite{song2018gaia} is
a scheduler specifically designed to improve load distribution on
GPUs, treating GPU resources in the same way as Kubernetes treats
CPUs. Townend \etal \cite{townend2019improving} and Wang \etal
\cite{wang2020icsoc} studied schedulers to reduce energy consumption
and heat waste. These schedulers need to generalise over the kind of
services that are being instantiated, so even an optimal
deployment~\cite{lebesbye2021} that has been statically decided, may
turn out to be poor and benefit from being refined after collecting
some data. For a recent survey on scheduling approaches for
Kubernetes, see~\cite{carrion2022k8sschedsurvey}.

Closer to our work, Medel \etal\cite{medel2016} propose a model-based
approach to predict performance and resource management for Kubernetes
systems. Their work focuses on simulating the lifecycle behaviour of
containers in a Kubernetes deployment, using timed Petri nets
\cite{tpnbook}. While their work targets pod and container lifecycle
management, our work has focused on resource consumption and load
balancing, and we model nodes in order to address bigger clusters with
multiple nodes and services. In contrast to our work, they do not
propose a specific methodology to leverage the model to estimate
performance and resource consumption of specific cloud-native
applications.

Interestingly, Mendel \etal\ point out that according to their experiments two containers that are in the same pod perform better than if they are deployed on different pods.  This seems to be the case also for identical pods deployed on the same node rather than deployed on two nodes.  In our experiments, we have also observed that pods are rarely independent and that modeling pods in isolation leaves open the problem of how to calculate resource consumption.  By focusing on resource consumption, our model complements the work of Mendel \etal; in fact, they point to resource contention for containers as a direction for future work \cite{medel2018}, in order to investigate the behavior of different resource management policies, which is what our model achieves.

In contrast to the above mentioned work on model-based analyses of Kubernetes deployment, our work proposes a methodology for applying the proposed modeling framework to concrete cloud-native applications, and validates the methodology on a concrete use case.


\section{Conclusion and Future Work}
\label{sec:conclusion}

The problem of predicting resource-efficient cluster configurations in
a complex industrial scenario quickly becomes challenging for the
human administrator. In this paper, we propose and evaluate a modeling
framework and an associated model-based methodology which can be
integrated in a Continuous Integration/Continuous Delivery (CI/CD)
pipeline to address this problem. The proposed methodology aims to
reduce a continuous space of possible cluster configurations to a
finite number of experiments, in order to instantiate the modeling
framework for cloud-native applications deployed on a Kubernetes
cluster.  The resulting model-based analysis can be used to predict
the resource load of different nodes in the cluster for different
scenarios of stress.

A particular challenge that we encountered in developing the modeling
framework for Kubernetes clusters, was a lack of isolation between
pods (due to reuse in the underlying system).  To address this
challenge, we propose the use of node images and cost tables in the
resource model, rather than uniform cost expressions as used in
previous work. The derivation of these cost tables was handled in the
associated methodology by a cost sampling strategy. The granularity of
the sampling strategy determines how precisely the model reflects the
resource consumption of the real system on the cluster.  Since
resource consumption need not be linear, only sampling resource
consumption for a finite number of points in the domain of RPS levels
that can be processed, will always lead to an approximation of the
continuum.  In future work, we plan to investigate the cost benefit
trade-off of sampling strategies with different granularities
following the proposed methodology. Obviously, the more time is
invested in the sampling process, the more accurate the resulting
model.

The proposed methodology is not meant to investigate \emph{critical}
service loads.  With very high service demands, the stressed pods and
nodes are no longer able to guarantee a high success rate. In this
case, errors cause exceptions that detour the execution such that the
resource consumption reflects the handling of exceptions rather than
the handling of requests. We did not aim to capture such erroneous
behavior in the cost tables of our model.

To evaluate the proposed methodology, we instantiated, calibrated and
explored a model of the cloud-native application \emph{Online
  Boutique} with stress tests under different mixed workflow
scenarios. The resulting model reflects the different workflows and
how they stress their associated services, the load balancing of the
services, and the composition of pods and nodes. By simulating mixed
workflow scenarios, we were able to understand which nodes will be
overloaded on the cluster. In fact, the best configuration in terms of
performance turned out to be the best balanced configuration with
respect to resource consumption in the majority of the cases.  The
results show that the expected load calculated by the model is close
to the average load observed on the real Kubernetes cluster.

In the experimental part of this paper, we used stress tests that
generated requests with uniformly distributed delays, both for the
sampling to derive the model and for the evaluation of derived model
with mixed stress tests. An interesting line of future work is to
compare the results obtained using these uniformly distributed delays
to non-uniformly distributed delays such as bursts of requests
according to a non-uniform distribution, both with respect to the
accuracy of predictions and the granularity of the required sampling
strategy.  It would also be interesting to investigate whether other
metrics than the node loads could be predicted with equally satisfying
accuracy.

Our experiments demonstrate that the proposed modeling framework and
methodology can already be used to derive models that can facilitate
deployment decision making. Another line of future work is to enhance
the usability of the approach by automating the sampling process and
the derivation of resource models for Kubernetes deployed cloud-native
applications, resulting in a tool capable of discovering the cluster
settings and generating the simulation module automatically after the
sampling is completed. Many Kubernetes plugins already implement such
automatic configuration retrieval.

\paragraph{Acknowledgments}
Our experiments on real Kubernetes deployment configurations were
performed on the HPC4AI~\cite{DBLP:conf/cf/AldinucciRPSVDG18} and
NREC\footnote{\href{https://www.nrec.no}{www.nrec.no}} clusters, two
academic platforms for deploying high-performance applications. We are
grateful for this support, as well as for feedback on our work, from
Marco Aldinucci at the University of Turin, and from Geir Horn, Rudi
Schlatte, and the SIRIUS center at the University of Oslo. We thank
the anonymous reviewers for their constructive comments that helped
improve the clarity and presentation in the paper.


\end{document}